%% file: paper.tex
\documentclass[twocolumn,showpacs,aps,prd]{revtex4}

\usepackage{graphicx} 
\usepackage{rotating}

\usepackage{graphicx}
\usepackage{dcolumn}
\usepackage{amsmath}
\usepackage{subfigure}

\RequirePackage{xspace}

\def\CP{\ensuremath{C\!P}\xspace}
\def\CPT  {\ensuremath{C\!PT}\xspace}

\def\Babar{\slshape B\kern-0.1em{\footnotesize A}\kern-0.1em B\kern-0.10em{\footnotesize A\kern-0.20em R}} 
\def\babar{\slshape B\kern-0.1em{\scriptsize A}\kern-0.1em B\kern-0.10em{\scriptsize A\kern-0.20em R}} 
\def\Bbar{\kern 0.18em\overline{\kern -0.18em B}{}} 
\def\BbarSubscr{\kern 0.25em\overline{\kern -0.18em B}{}} 
\def\Dbar{\kern 0.20em\overline{\kern -0.20em D}{}}

\def\gev{\mathrm{\,Ge\kern -0.1em V}} 
\def\mev{\mathrm{\,Me\kern -0.1em V}} 
\def\gevc{\mathrm{\,Ge\kern -0.1em V\!/}c} 
\def\gevcc{\mathrm{\,Ge\kern -0.1em V\!/}c^2} 
\def\mevcc{\mathrm{\,Me\kern -0.1em V\!/}c^2} 
\def\gevccsq{\mathrm{\,Ge\kern -0.1em V^2\!/}c^4} 
\def\ubar{\kern 0.10em \overline{\kern-0.10em u}\kern 0.05em{}} 
\def\dbar{\kern 0.15em \overline{\kern-0.15em d}\kern 0.05em{}} 
\def\sbar{\kern 0.10em \overline{\kern-0.10em s}\kern 0.05em{}} 
\def\cbar{\kern 0.10em \overline{\kern-0.10em c}\kern 0.05em{}} 
\def\bbar{\kern 0.10em \overline{\kern-0.10em b}\kern 0.05em{}} 
\def\nbar{\kern 0.10em \overline{\kern-0.10em n}\kern 0.05em{}} 
\def\pbar{\kern 0.10em \overline{\kern-0.10em p}\kern 0.05em{}} 
\def\Nbar{\kern 0.25em \overline{\kern-0.25em N}\kern 0.05em{}}

\def\BzBzb   {\ensuremath{\Bz {\kern -0.16em \Bzb}}\xspace}

\long\def\inst#1{\par\nobreak\kern 4pt\nobreak {\it #1}\par\vskip 10pt plus 3pt minus 3pt} 

\begin{document} 

\widetext 

\begin{flushleft}
    SLAC-PUB-15425 \\
    {\Babar}-PUB-12/026
\end{flushleft}

\title{
    \boldmath Measurement of  $\CP$-violating asymmetries in $B^0 \to \left( \rho \pi \right)^0$ decays using a time-dependent Dalitz plot analysis
}

\input authors_sep2012.tex

\input abstract.tex

\end{abstract}
\maketitle

\section{INTRODUCTION} 
\label{sec:introduction} 

\noindent

Within the standard model (SM) of particle physics, $\CP$ violation in the quark sector is described by the Cabibbo-Kobayashi-Maskawa (CKM) quark-mixing matrix. Physics beyond the SM may result in measured values of observables that deviate from the values expected based on other CKM parameter measurements and the SM.

The decay $B^0 \rightarrow \pi^+ \pi^- \pi^0$ \cite{ref:conj} is well suited to the study of $\CP$ violation and has been previously explored by both the {\Babar}~\cite{ref:matt2007} and Belle~\cite{ref:belle2007} Collaborations. Early studies of this mode were ``quasi-two-body'' (Q2B) analyses that treated each $\rho$ resonance separately in the decays $B^0\rightarrow\rho^0\pi^0(\rho^0\rightarrow\pi^+\pi^-)$ and $B^0\rightarrow\rho^\pm\pi^\mp(\rho^\pm\rightarrow\pi^\pm\pi^0)$. However, as first noted by Snyder and Quinn~\cite{ref:snyder}, a complete time-dependent Dalitz plot (DP) analysis is sensitive to the interference between the strong and weak amplitudes in the regions where the $\rho^+$, $\rho^-$, and $\rho^0$ resonances overlap.
This interference allows the unambiguous extraction of the strong and weak relative phases, and of the $\CP$-violating parameter $\alpha \equiv {\rm arg}[-V^{\phantom *}_{td}V^{*}_{tb}/(V^{\phantom *}_{ud}V^{*}_{ub})]$, where $V_{q q^\prime}$ are components of the CKM matrix. A precision measurement of $\alpha$ is of interest because it serves to further test the SM and constrain new physics that may contribute to loops in Feynman diagrams.

In this paper, we present an update of an earlier {\Babar} analysis. We use the full {\Babar} dataset collected at the $\Upsilon(4S)$ resonance, corresponding to an increase of 25\% in the number of $B$ meson decays, and include a number of improvements to both the reconstruction and selection procedures. Among these are improved charged-particle tracking and particle identification (PID), and a reoptimized multivariate discriminator, used both for event selection and as a variable in the final fit.

Section~\ref{sec:theorysec} contains an introduction to the theory behind this analysis and the formalism used. We proceed to descriptions of the detector (Sec.~\ref{sec:detectorsec}), the datasets (Sec.~\ref{sec:datasetsec}), and the event selection procedures (Sec.~\ref{sec:selectionsec}). This is followed by a presentation of the fitting procedure (Sec.~\ref{sec:fitsec}) and of the systematic studies (Sec.~\ref{sec:systematicsec}). Finally, we present the fit results (Sec.~\ref{sec:resultssec}) and a conclusion (Sec.~\ref{sec:conclusion}). An overview of robustness studies is provided in an Appendix.

\section{\boldmath THEORY OVERVIEW} 
\label{sec:theorysec}

\subsection{Time-Independent Probability Distribution} 

The time-independent amplitudes for $B^0$ and ${\overline B}^0$ decays to $\pi^+\pi^-\pi^0$ are given by
\begin{eqnarray}
\label{timeIndepEq}
A_{3\pi} & =& f_{+}A^{+} + f_{-}A^{-} + f_{0}A^{0},  \nonumber \\
{\overline A}_{3\pi} & =&  f_{+}{\overline A}^{+} + f_{-}{\overline A}^{-} + f_{0}{\overline A}^{0},
\end{eqnarray}
\noindent respectively, where $A^\kappa$ and ${\overline A}^\kappa$ with $\kappa \in \{ +,-,0 \}$ are complex amplitudes associated with the $\rho^+$, $\rho^-$, and $\rho^0$ resonances, respectively, and $f_{\kappa}=f_{\kappa}(m,\theta_{\kappa})$ are defined in terms of modified relativistic Breit-Wigner resonances~\cite{ref:modRelBreitWigner} modeling the three $\rho$ resonances. The angle $\theta_{\kappa}$ is the helicity angle for the resonance,  defined as the angle between the $\pi^0$ ($\pi^-$) momentum and the negative of the momentum of the recoiling $\pi^-$ ($\pi^+$) for the $\rho^+$ ($\rho^-$), and as the angle between the $\pi^+$ momentum and the negative of the momentum of the recoiling $\pi^0$ for the $\rho^0$. All helicity angles are calculated in the $\rho$ rest frame. In the fit, we include the $\rho(770)$ as well as its radial excitation, the $\rho(1450)$; therefore, each $f_{\kappa}$ is a sum of modified relativistic Breit-Wigner resonances, $F$, for the $\rho^\kappa(770)$ and $\rho^\kappa(1450)$:
\begin{align}
f_{\kappa}(m,\theta_\kappa) \propto F_{\rho(770)}(m,\theta_\kappa) + a_{\rho^\prime} e^{i \phi_{\rho^\prime}} F_{\rho(1450)}(m,\theta_\kappa),
\end{align}
\noindent where $a_{\rho^\prime}$ and $\phi_{\rho^\prime}$ are the magnitude and phase of the $\rho(1450)$ resonance relative to the $\rho(770)$. We include systematic uncertainties, described in Sec.~\ref{sec:exclusionof1700}, to account for possible contributions from the $\rho(1700)$.

\subsection{Time-Dependent Probability Distribution} 
\label{timeDepSec}

Using the time-independent amplitudes $A_{3\pi}$ and ${\overline A}_{3\pi}$, we can express the full time-dependent probability for a meson that is a $B^0$ ($\mathcal{A}^-_{3\pi}$) or ${\overline B^0}$ ($\mathcal{A}^+_{3\pi}$) at the time the other $B$ meson decays, to decay to $\pi^+\pi^-\pi^0$ as
\begin{eqnarray}
	|\mathcal{A}^\pm_{3\pi}(\Delta t)|^2 & & = \frac{e^{-|\Delta t|/\tau_{B^0}}}{4 \tau_{B^0}} \left( |A_{3\pi}|^2 + |{\overline A}_{3\pi}|^2 \phantom{ \left[ \frac{q}{p} {\overline A}_{3\pi} A^*_{3\pi} \right]} \right.  \nonumber \\
	& & \mp \left( |A_{3\pi}|^2 - |{\overline A}_{3\pi}|^2 \right) \cos (\Delta m_d \Delta t) \phantom{ \left[ \frac{q}{p} {\overline A}_{3\pi} A^*_{3\pi} \right]}  \nonumber \\
	& & \left. \pm 2\ {\rm Im} \left[ \frac{q}{p} {\overline A}_{3\pi} A^*_{3\pi} \right] \sin(\Delta m_d \Delta t) \right),
	\label{timeDepEq}
\end{eqnarray}
where $\tau_{B^0}$ is the mean neutral $B$ lifetime, $\Delta m_d$ is the mass difference between the heavy and light neutral $B$ mass eigenstates, $p$ and $q$ are the complex parameters in the definitions of the neutral mass eigenstates $p|B^0\rangle \pm q|{\overline B}^0\rangle$, and $\Delta t$ is the time difference between the decays of the fully reconstructed $B$ meson ($B_{3\pi}$) and the $B$ meson used to determine the $B$ flavor ($B_{\rm tag}$). In Eq.~\eqref{timeDepEq}, as in the fit, we assume that the heavy and light mass eigenstates have the same lifetime, that there is no $\CP$ violation in $B^0{\overline B^0}$ mixing ($|q/p| = 1$), and that $\CPT$ is conserved.

\subsection{Square Dalitz Plot Formalism}

\begin{figure*}[!htb]
	\begin{center}
		\includegraphics[height=6cm]{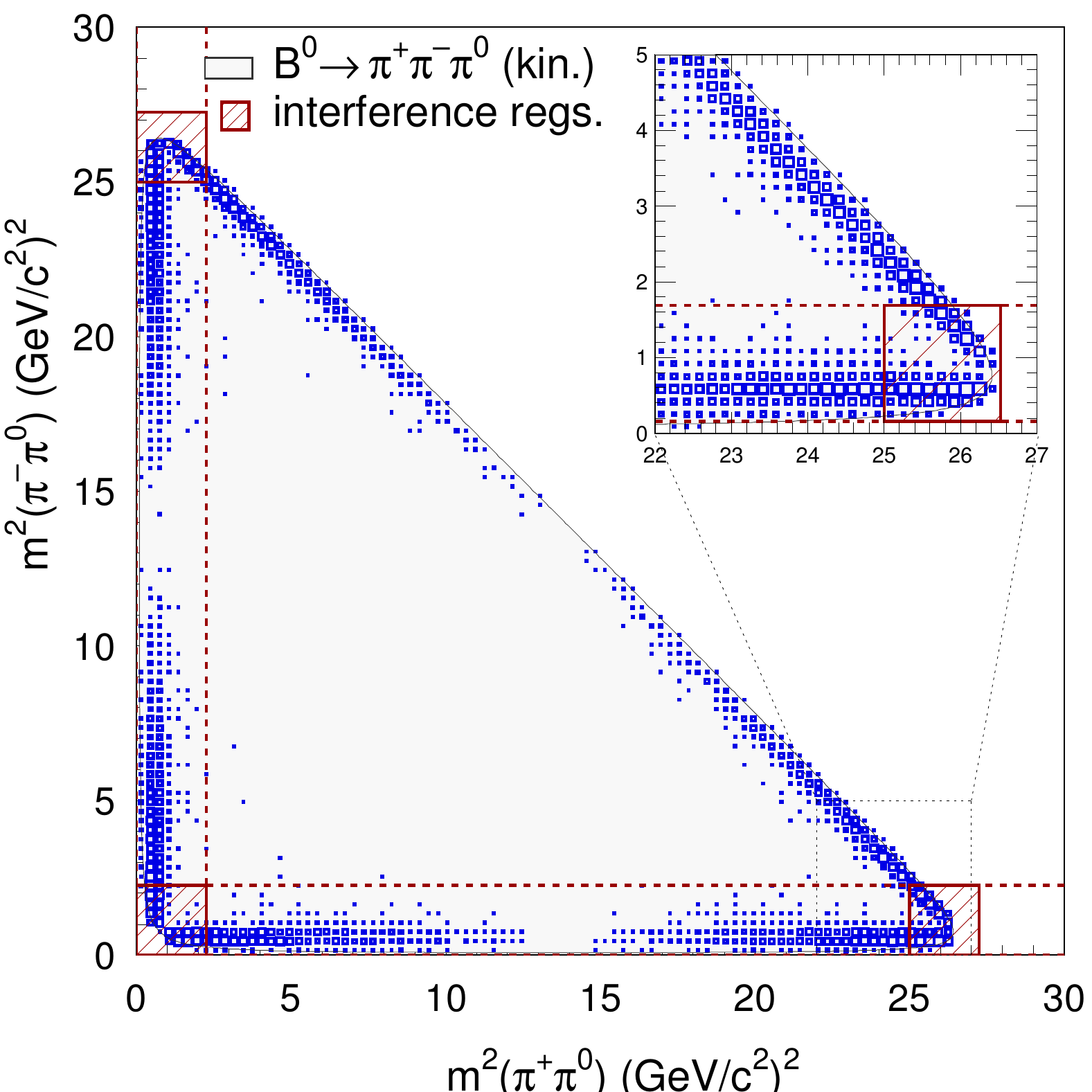}
		\includegraphics[height=4.6cm]{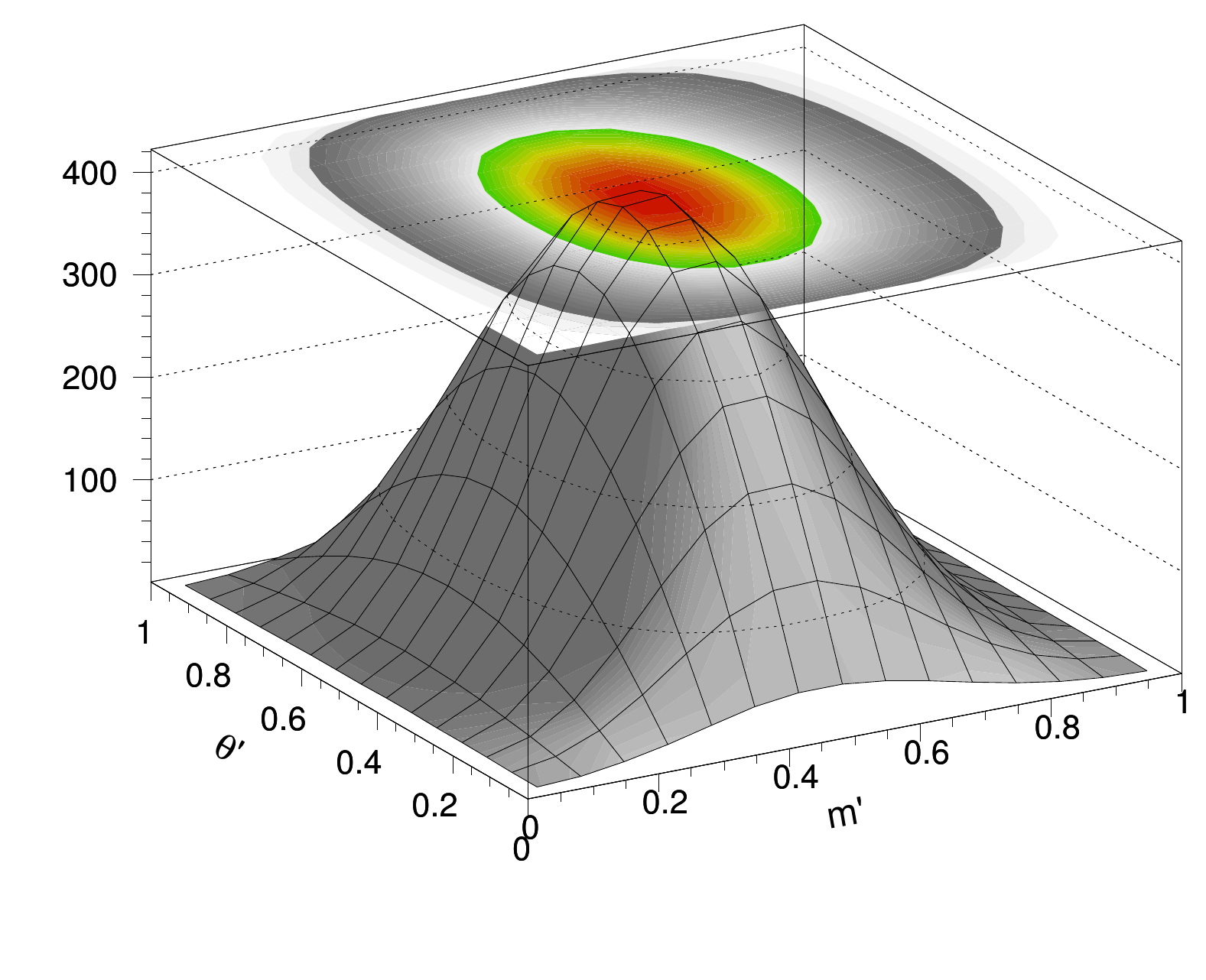}
		\includegraphics[height=6cm]{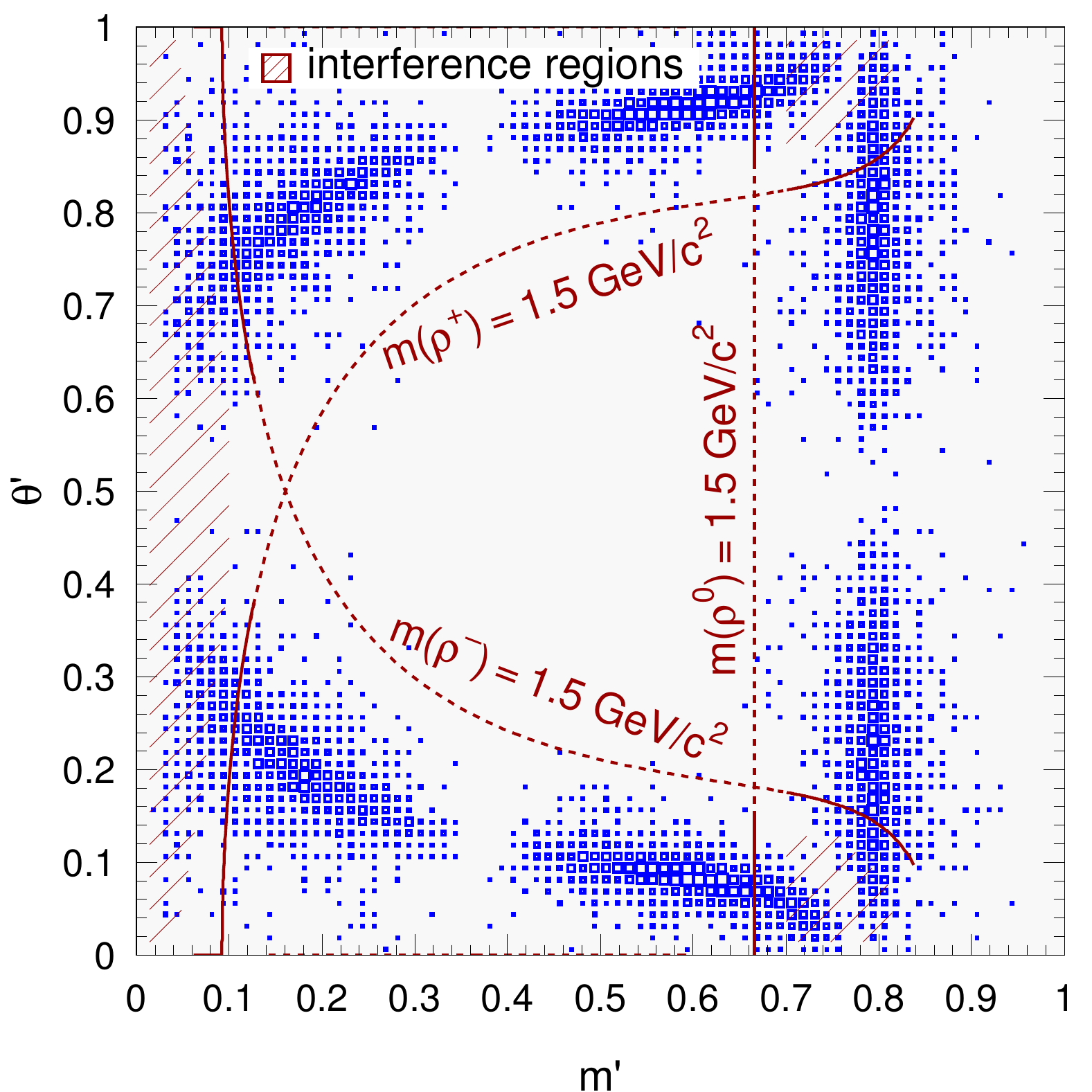}
		\caption{(color online) Nominal (left) and square (right) $B^0 \rightarrow \rho\pi$ Dalitz plots obtained from Monte Carlo generated events without detector simulation~\cite{ref:matt2007}. The amplitudes in Eq.~\eqref{timeIndepEq} are generated with values $A^+=A^-=A^0=1$ so that they interfere destructively for equal $\pi\pi$ masses. The hatched areas indicate the main overlap regions between the different $\rho$ bands. The dashed lines in the square Dalitz plot correspond to $m(\rho^{+,-,0})=1.5\gevcc$. The middle plot depicts the Jacobian determinant of the transformation and shows the distribution in the square Dalitz plot for uniformly distributed events in the nominal Dalitz plot.}
		\label{fig:sdp}
	\end{center}
\end{figure*}

While nonresonant phase-space decays uniformly populate the kinematically allowed region of a DP, signal $\rho \pi$ events populate the boundaries of this region due to the low mass of the $\rho$ resonances relative to the $B$ mass. In particular, the interference regions of the signal DP, which provide sensitivity to the relative phases of the $\rho$ resonances, are confined to small regions in the three corners of the DP. In order to expand these regions of interest and avoid the use of bins of variable size, we perform a transformation of the DP that maps the kinematically allowed region onto a dimensionless unit square. The transformation is described by
\begin{align}
	dm_+ dm_- \rightarrow |{\rm det} J| dm^\prime d\theta^\prime,
\end{align}
\noindent with the square Dalitz plot (SDP) coordinates
\begin{align}
	m^\prime &\equiv \frac{1}{\pi} \arccos \left( 2\frac{m_0 - m_0^{\rm min}}{m_0^{\rm max} - m_0^{\rm min}} - 1 \right), \\
	\theta^\prime &\equiv \frac{1}{\pi} \theta_0,
\end{align}
\noindent where $m_{\pm}$ is the invariant mass of the $\pi^\pm\pi^0$ system, $m_0$ is the invariant mass of the two charged pion candidates, $\theta_0$ is the $\rho^0$ helicity angle defined earlier, $m_0^{\rm max} = m_{B^0} - m_{\pi^0}$ and $m_0^{\rm min} = 2 m_{\pi^+}$ are the kinematic limits of the $m_0$ mass, and $J$ is the Jacobian of the transformation. The determinant of the Jacobian is given by
\begin{align}
	| {\rm det} J | = 4 |{\bf p}_+^* | | {\bf p}_0^*| m_0 \frac{\partial m_0}{\partial m^\prime} \frac{\partial \cos \theta_0}{\partial \theta^\prime},
\end{align}
\noindent where
\begin{align}
|{\bf p}_+^* | &= \sqrt{(E_+^*)^2 - m_{\pi^+}^2}, \\
|{\bf p}_0^* | &= \sqrt{(E_0^*)^2 - m_{\pi^0}^2},
\end{align}
and the energies $E_+^*$ and $E_0^*$ of the $\pi^+$ and $\pi^0$ are defined in the $\pi^+\pi^-$ center-of-mass (CM) frame. Figure~\ref{fig:sdp} shows an example of a standard DP (left) and its transformed SDP counterpart (right), plotted using simulated $B^0 \rightarrow \rho \pi$ decays, where the three $\rho$ resonances are assumed to have the same amplitude.

\subsection{$U/I$ Formalism}

If one explicitly inserts Eq.~\eqref{timeIndepEq} into Eq.~\eqref{timeDepEq}, the full time-dependent amplitude for a $B^0$ or ${\overline B^0}$ meson to decay to $\pi^+\pi^-\pi^0$ can be written in terms of
\begin{eqnarray}
	|\mathcal{A}_{3\pi}|^2 \pm |\mathcal{\overline A}_{3\pi}|^2 &=& \sum_{\kappa \in \left[ +,-,0 \right]} |f_\kappa|^2 U^\pm_\kappa  \nonumber \\
	&& + 2 \sum_{\kappa < \sigma \in \left[ +,-,0\right]} \left( {\rm Re} \left[ f_\kappa f_\sigma^* \right]U_{\kappa \sigma}^{\pm,{\rm Re}} \right.  \nonumber \\
	&& \left. - {\rm Im} \left[ f_\kappa f_\sigma^* \right]U_{\kappa \sigma}^{\pm,{\rm Im}} \right),
\end{eqnarray}
\noindent and
\begin{eqnarray}
		{\rm Im} \left[ \frac{q}{p} \mathcal{\overline A}_{3\pi} \mathcal{A}^*_{3\pi} \right] &=& \sum_{\kappa \in \left[ +,-,0 \right]} |f_\kappa|^2 I_\kappa \nonumber \\
		&& + \sum_{\kappa < \sigma \in \left[ +,-,0\right]} \left( {\rm Re} \left[ f_\kappa f_\sigma^* \right] I_{\kappa \sigma}^{\rm Im} \right.  \nonumber \\
		&& \left. + {\rm Im} \left[ f_\kappa f_\sigma^* \right] I_{\kappa \sigma}^{\rm Re} \right),
\end{eqnarray}
\noindent with
\begin{eqnarray}
	\label{UsIsDefFirst}
	U_{\kappa}^{\pm} &=& |A^\kappa|^2 \pm |{\overline A}^\kappa|^2 , \\
	U_{\kappa \sigma}^{\pm, {\rm Re(Im)}} &=& {\rm Re(Im)} \left[ A^\kappa A^{\sigma *} \pm {\overline A}^\kappa {\overline A}^{\sigma *} \right] , \\
	I_\kappa &=& {\rm Im} \left[ {\overline A}^\kappa A^{\kappa *} \right] , \\
	I_{\kappa \sigma}^{\rm Re} &=& {\rm Re} \left[ {\overline A}^\kappa A^{\sigma *} - {\overline A}^\sigma A^{\kappa *} \right] , \\
	I_{\kappa \sigma}^{\rm Im} &=& {\rm Im} \left[ {\overline A}^\kappa A^{\sigma *} + {\overline A}^\sigma A^{\kappa *} \right].
	\label{UsIsDefLast}
\end{eqnarray}
\noindent The 27 real-valued $U$ and $I$ coefficients provide an alternative parameterization to tree and penguin amplitudes (as well as $\alpha$) or to the amplitudes $A^\kappa$ and ${\overline A}^\kappa$~\cite{ref:snyderquinnusis}. The $U$ and $I$ parameters can also be directly related to the Q2B ${\mathcal C}$ and ${\mathcal S}$ parameters often used in $\CP$-violation analyses~\cite{ref:cpv}, where ${\mathcal C}$ parameterizes direct $\CP$ violation, and ${\mathcal S}$ parameterizes mixing-induced $\CP$ violation (involving the angle $\alpha$ in this analysis). The related parameter $\Delta {\mathcal C}$ describes the asymmetry between the rates $\Gamma (B^0 \rightarrow \rho^+ \pi^-) + \Gamma ( {\overline B}^0 \rightarrow \rho^- \pi^+) $ and $\Gamma (B^0 \rightarrow \rho^- \pi^+) + \Gamma ( {\overline B}^0 \rightarrow \rho^+ \pi^-) $, while $\Delta {\mathcal S}$ is related to the strong phase difference between the different amplitudes describing the decay $B^0 \rightarrow \rho \pi$. The $U$ and $I$ parameters are related to the ${\mathcal C}$ and ${\mathcal S}$ parameters through the relations
\begin{align}
	{\mathcal C}^+ = \frac{U^-_+}{U^+_+},\ {\mathcal C}^- = \frac{U^-_-}{U^+_-},\ {\mathcal S}^+ = \frac{2 I_+}{U^+_+},\ {\mathcal S}^- = \frac{2 I_-}{U^+_-},
	\label{q2bequation}
\end{align}
\noindent\ and
\begin{align}
	\mathcal{A}_{\rho \pi} = \frac{U^+_+-U^+_-}{U^+_+ + U^+_-},
\end{align}
\noindent where
\begin{align}
	\label{firstQ2B}
	{\mathcal C} &= ({\mathcal C}^+ + {\mathcal C}^-)/2, \\
	\Delta {\mathcal C} &= ({\mathcal C}^+ - {\mathcal C}^-)/2, \\
	{\mathcal S} &= ({\mathcal S}^+ + {\mathcal S}^-)/2, \\
	\Delta {\mathcal S} &= ({\mathcal S}^+ - {\mathcal S}^-)/2.
	\label{lastQ2B}
\end{align}
Note that while ${\mathcal C}$, ${\mathcal S}$, $\Delta {\mathcal C}$, and $\Delta {\mathcal S}$ do not depend on interference effects between the $\rho$ resonances, the $U$ and $I$ parameter formalism accounts for these features and is thus appropriate for a full DP analysis. While some degree of physical intuition is lost when using the $U$ and $I$ parameters instead of the standard complex amplitudes and phases, there are several practical motivating factors for their adoption in the fit:

\begin{itemize}
\item{Whereas there is a two-fold ambiguity in $\alpha$ ($\alpha$ versus $90^\circ-\alpha$), there is a unique solution in a fit to the $U$ and $I$ parameters, which encompasses both solutions for $\alpha$).}

\item{The $U$ and $I$ parameterization results in uncertainties that are more Gaussian than those in a standard amplitude and phase fit.}

\item{It is simpler to average $U$ and $I$ fit results from different measurements or experiments that publish the full covariance matrix.}
\end{itemize}

For physical solutions, there are constraints between the $U$ and $I$ parameters. In the case of three $\rho$ resonances ($\rho^+$, $\rho^-$, and $\rho^0$), a fit to the complex tree and penguin amplitudes as well as the weak phase $\alpha$ involves 11 unknown parameters, which reduce to 10 parameters when the arbitrary global phase is removed. A $U$ and $I$ fit is equivalent to such a fit, but involves many more parameters. However, when the $\rho^0\pi^0$ amplitude is small, as is observed in nature, the values of 11 of the 27 $U$ and $I$ parameters become unimportant. Due to the high degree of correlation between the various $U$ and $I$ fit parameters, the 27 parameters actually represent only 12 independent parameters. Neglecting the arbitrary phase and the overall normalization, this reduces to 10, and once isospin relations are taken into account, only 9 independent parameters remain.

Because the $U$ and $I$ formalism is used in the final fit without any constraints on the parameters (aside from fixing $U^+_+=1$ to set the overall normalization), it is possible for the free parameters to take on unphysical values that do not correspond to any physical set of $\rho$ amplitudes. The final fit values from the 2007 {\babar} analysis \cite{ref:matt2007} are one such unphysical set. We determined that no biases are introduced due to the fitted values of the parameters being unphysical.

\section{THE {\Babar} DETECTOR AND EXPERIMENT}
\label{sec:detectorsec}

The data used in this analysis were collected with the {\babar} detector at the PEP-II asymmetric-energy $e^+e^-$ storage ring at SLAC. A detailed description of the {\babar} detector is presented in Ref.~\cite{ref:detector}. The tracking system used for track and vertex reconstruction has two components: a 5-layer silicon vertex tracker and a drift chamber, both operating within a $1.5\ {\rm T}$ magnetic field generated by a superconducting solenoidal magnet. A detector of internally reflected Cherenkov light associates Cherenkov photons with tracks for particle identification. The energies of photons and electrons are determined from the measured light produced in electromagnetic showers inside a CsI(Tl) crystal electromagnetic calorimeter. Muon candidates are identified with the use of the instrumented flux return of the solenoid. The flux return instrumentation initially consisted of resistive plate chambers and was later modified to consist of a mixture of resistive plate chambers and limited streamer tubes.

\section{DATA SAMPLE AND MC SIMULATION}
\label{sec:datasetsec}

\subsection{Data Samples}

For the final fit, we use the full ``on-resonance'' {\babar} dataset of $431.0\ {\rm fb}^{-1}$ collected at the $\Upsilon(4S)$ resonance energy (${\sqrt s} = 10.58\gevcc$). When optimizing background suppression criteria, $44.6\ {\rm fb}^{-1}$ of ``off-resonance'' data, collected $40 \mev$ below the $\Upsilon(4S)$ resonance, are used to model ``continuum''  $e^+e^-\rightarrow q {\overline q}\ (q=u,d,s,c)$ background.

\begin{table*}[tbp!] 
\begin{centering}
\caption{$B$ background decay modes included in the final fit. Modes generated taking into account interference effects in the Dalitz plot are indicated by the ``Dalitz'' label. Longitudinal polarization is indicated by ``[longitudinal]''.}
\renewcommand{\arraystretch}{1.2}
\begin{tabular}{ c | l c c  }
Class	& Decay mode						                                                                           & $\mathcal{B}$ [$10^{-6}$]	& \# of expected events \\ \hline \hline

0	  	& $B^+ \rightarrow \rho^+ \rho^0\ {\rm [longitudinal]} $                                                         & $24.0 \pm 1.9$		& $ 129 \pm 10 $ \\
0	  	& $B^+ \rightarrow a_{1}^{+} (\rightarrow (\rho \pi)^+) \pi^0 $                                  & $26 \pm 7$		& $ 53 \pm 14 $ \\
0	  	& $B^+ \rightarrow a_{1}^{0} (\rightarrow \rho^{\pm} \pi^{\mp}) \pi^+ $		 & $20 \pm 6$		& $ 37 \pm 11 $ \\

1		& $B^+ \rightarrow \pi^+ K_{s}^0 (\rightarrow \pi^+ \pi^-) $                                       & $7.99 \pm 0.35$	& $ 6.96 \pm 0.31 $ \\

2	  	& $B^+ \rightarrow K^+ \pi^+ \pi^- \ {\rm Dalitz}$						 & $51.0 \pm 2.9$		& $ 34.8 \pm 2.0 $ \\
2	  	& $B^+ \rightarrow \pi^+ \pi^+ \pi^- \ {\rm Dalitz}$    	                    & $16.2 \pm 1.5$	& $ 203 \pm 19 $ \\

3	  	& $B^+ \rightarrow \pi^0 \rho^+ $									 & $10.9 \pm 1.4$	&  $ 120 \pm 15 $ \\
3	  	& $B^+ \rightarrow \pi^+ K_{s}^0 (\rightarrow \pi^0 \pi^0) $				 & $3.54 \pm 0.15$	&  $ 24.2 \pm 1.1 $ \\

4	  	& $B^+ \rightarrow \pi^+ \pi^0 $										 & $5.7 \pm 0.5$	&  $ 38.6 \pm 3.4 $ \\
4	  	& $B^+ \rightarrow K^+ \pi^0 $										 & $12.9 \pm 0.6$	&  $ 18.6 \pm 0.9 $ \\

5	  	& $B^0 \rightarrow K_{s}^{0} \pi^+ \pi^- \ {\rm Dalitz} $  & $44.8 \pm 2.6$	& $ 15.7 \pm 0.9 $ \\

6	  	& $B^0 \rightarrow \rho^+ \rho^- {\rm [longitudinal]} $				 			 & $24.2 \pm 3.1$	& $ 122 \pm 16 $ \\
6	  	& $B^0 \rightarrow a_1^- \pi^+$						  			 & $33 \pm 5$	& $ 61 \pm 9 $ \\
6	  	& $B^0 \rightarrow a_1^0 \pi^0$						  			 & $11.0 \pm 1.7$	& $ 22.8 \pm 3.5 $ \\

7	  	& $B^0 \rightarrow K^+ \pi^- $						  				 & $19.4 \pm 0.6$	& $ 21.6 \pm 0.7 $ \\

8	  	& $B^0 \rightarrow K^+ \pi^- \pi^0\ {\rm Dalitz} $                       & $35.9 \pm 2.6$       & $ 398 \pm 29 $ \\

9	  	& $B^0 \rightarrow \gamma K^{{\rm *}0} (892) (\rightarrow K^+ \pi^-) $                 & $40.1 \pm 2.0$	& $ 31.8 \pm 1.6 $ \\
9	  	& $B^0 \rightarrow \gamma K^{{\rm *}0} (1430) (\rightarrow K^+ \pi^-) $              & $12.4 \pm 2.4$	& $ 3.2 \pm 0.6 $ \\
9	  	& $B^0 \rightarrow \eta^{\prime} (\rightarrow \rho^0 \gamma) \pi^0 $			 & $0.35 \pm 0.18$	& $ 4.2 \pm 2.1 $ \\

10	  	& $B^0 \rightarrow \pi^0 K_{s}^{0} (\rightarrow \pi^+ \pi^-) $			  	 & $3.39 \pm 0.21$	& $ 21.8 \pm 1.4 $ \\
11	  	& $B^0 \rightarrow D^- (\rightarrow \pi^- \pi^0) \pi^+ $					 & $3.35 \pm 0.27$	& $ 399 \pm 32 $ \\
12	  	& $B^0 \rightarrow {\overline D}^0 (\rightarrow K^+ \pi^-,\ K^+ \pi^- \pi^0) \pi^0 $  & $46.7 \pm 4.5$	& $ 124 \pm 12 $ \\
13	  	& $B^0 \rightarrow {\overline D}^0 (\rightarrow \pi^+ \pi^-) \pi^0	 $	 		 & $0.367 \pm 0.034$	& $ 48 \pm 5 $ \\
14	  	& $B^0 \rightarrow J/\psi (\rightarrow e^+e^-,\mu^+\mu^-) \pi^0	 $			 & $2.09 \pm 0.19$	& $ 153 \pm 14 $ \\ \hline
15	  	& $B^0 \rightarrow {\rm neutral\ generic\ } b \rightarrow c{\rm \ decays} $		 &				& $466 \pm 14$ \\
16	  	& $B^+ \rightarrow {\rm charged\ generic\ } b \rightarrow c{\rm \ decays} $	 &				          & $921 \pm 21$ \\ \hline \hline
		& Total									                                                &                                  & $3478 \pm 65$
\label{tab:bbkgtab}
\end{tabular}
\end{centering}
\end{table*}

\subsection{Monte Carlo Samples}

Event simulation based on the Monte Carlo (MC) method is used to evaluate backgrounds and to determine signal-event reconstruction efficiencies. For all MC samples, detector response is accounted for using the {\tt GEANT4} package \cite{ref:geant4} in a full {\babar} detector simulation.

$B^0\rightarrow\rho^+\pi^-$ and $B^0\rightarrow\rho^0\pi^0$ signal decays are simulated in separate MC datasets, but the $B^0\rightarrow\rho^0\pi^0$ sample is not used to determine background selection criteria or to model signal distributions since the nominal branching fraction for $B^0 \rightarrow \rho^{\pm} \pi^{\mp}$ is $11.5 \pm 3.1$ times larger than the branching fraction for $B^0 \rightarrow \rho^0 \pi^0$~\cite{ref:rpp2012}.

$B$-decay backgrounds are modeled using MC samples consisting of $B$ decays to specific final states as well as ``generic'' MC samples consisting of charged and neutral $B$ decays to unconstrained final states. In the generic MC, dominant branching fractions are fixed to the results of recent measurements~\cite{ref:rpp2011}. Due to the uncertainty on branching fractions for charmless decays, all charmless events are removed from the generic MC and charmless modes of interest are explicitly included among the $B$ background samples consisting of decays to specific final states. We use a total of 24 $B$-decay MC samples corresponding to 29 different final states  (Table~\ref{tab:bbkgtab}) as well as MC samples of generic charged and neutral $B$ decays.

The expected number of events for each charmless $B$ background is calculated according to
\begin{align}
	n_{\rm exp} = 2 n_{BB} {\mathcal B} {\mathcal B}_{\rm mode}  \epsilon,
\end{align}
\noindent where $n_{BB}$ is the number of produced $B{\overline B}$ pairs, ${\mathcal B}$ is the branching fraction~\cite{ref:rpp2011} (approximately $1/2$) for an $\Upsilon(4S)$ to decay to a charged or neutral $B{\overline B}$ pair (whichever is appropriate for the mode in question), ${\mathcal B}_{\rm mode}$ is the branching fraction for the $B$ decay mode, and $\epsilon$ is the efficiency for reconstructing events in the mode, determined from MC. The factor of 2 is included because either of the $B$ mesons in a given event may decay to the mode of interest.

In the case of the charged and neutral generic $B$ backgrounds, the number of events expected for each mode is
\begin{align}
	n_{\rm exp} = n_{\rm MC} \frac{n_{BB} {\mathcal B}}{n_{\rm gen}},
\end{align}
\noindent where $n_{\rm gen}$ is the number of generated charged (neutral) MC events and ${n_{\rm MC}}$ is the number of charged (neutral) generic MC events that remain after all selection criteria have been applied.

An additional simulated dataset, which is used for validation studies, consists of DP-parameterized $B^0\rightarrow\rho\pi$ decays. This dataset is used to verify flavor-tagging conventions.

\section{EVENT SELECTION AND BACKGROUND SUPPRESSION}
\label{sec:selectionsec}

\subsection{Event Preselection}

The kinematics of $B$ meson decays that are fully reconstructed at {\babar} can be characterized by two variables: $m_{\rm ES}$ and $\Delta E$. The beam-energy-substituted mass $m_{\rm ES}$ is the invariant mass of the reconstructed $B$ candidate calculated under the assumption that its energy in the $e^+e^-$ CM frame is half the total beam energy. We define
\begin{align}
m_{\rm ES} = \sqrt{\left[(s/2 + {\vec p_i} \cdot {\vec p_B})/E_i\right]^2 - |{\vec p_{B}}|^2},
\end{align}
where $\sqrt{s}$ is the total beam energy in the $e^+e^-$ CM frame, $(E_i,{\vec p}_i)$ is the four-momentum of the $e^+e^-$ system in the laboratory frame, and ${\vec p}_B$ is the $B$-candidate momentum in the laboratory frame. The second kinematic variable is defined by
\begin{align}
\Delta E = E^{*}_{B} - \frac{1}{2}\sqrt{s},
\end{align}
where $E^{*}_{B}$ is the measured energy of the $B$ candidate in the $e^+e^-$ CM frame.

Pairs of oppositely charged tracks are combined with $\pi^0 \to \gamma \gamma$ candidates to construct $B^0\rightarrow\pi^+\pi^-\pi^0$ candidates. During a preselection stage, we require $E^{*}_{B}$ to lie between $4.99$ and $5.59\gev$. For the charged tracks corresponding to the $\pi^{\pm}$ candidates, we require a maximum momentum of $10 \gevc$, a minimum of 12 hits in the drift chamber, a maximum distance-of-closest-approach (DOCA) relative to the beamspot center in the $x$-$y$ plane of 1.5 cm, and a DOCA along the $z$ axis between $-10$ and $+10$ cm. We require the two photon candidates that form the $\pi^0$ candidate to have an energy in the laboratory frame between $30 \mev$ and $10 \gev$, and lateral moments in the electromagnetic calorimeter less than 0.8. We require a $\pi^0\rightarrow\gamma\gamma$ mass between 100 and $160 \mevcc$, and a total $\gamma\gamma$ laboratory energy greater than $200 \mev$.

\subsection{Primary Selection Criteria}

Following preselection, we require $m_{\rm ES}$ to lie between $5.200$ and $5.288 \gevcc$. For the charged tracks corresponding to the $\pi^{\pm}$ candidates, we require a minimum transverse momentum of $0.1 \gevc$. The lateral moments of the two photons from the $\pi^0$ candidate are each required to lie between 0.01 and 0.60, while the laboratory-frame energies in the electromagnetic calorimeter are required to exceed $50 \mev$. The mass of the $\pi^0$ candidate ($m_{\gamma\gamma}$) is required to satisfy $110 < m_{\pi^0} < 160 \mevcc$.

We calculate $\Delta t$ by measuring the distance along the beam axis between the $B_{3\pi}$ and $B_{\rm tag}$ decay vertices and using the boost ($\beta\gamma \approx 0.56$) of the $e^+e^-$ system. The time difference $\Delta t$ and its estimated uncertainty are required to satisfy $|\Delta t| < 20\ {\rm ps}$ and $\sigma(\Delta t)<2.5\ {\rm ps}$.

The average number of $B$ candidates (measured as the total number of candidates that pass all the preceding selection criteria divided by the total number of events with at least one candidate) is 1.45. To retain only one $B$-decay candidate in each event, we select the candidate that has a $\pi^0\rightarrow\gamma\gamma$ candidate mass closest to the world average value of the $\pi^0$ mass. In the case of multiple $B$ candidates reconstructed with the same $\pi^0$ candidate, we select one $B$ candidate arbitrarily. A tighter requirement of $5.272<m_{\rm ES}<5.288 \gevcc$ is applied after selecting a single candidate in each event.

We use a lower $m_{\rm ES}$ sideband of on-resonance data, $5.215 < m_{\rm ES} < 5.270 \gevcc$, as well as off-resonance data, to model the distribution of continuum events. 

\subsection{Transformed $\Delta E^{\prime}$ Definition and Selection Criterion }

Because the width of the $\Delta E$ distribution is highly correlated with the $\pi^+\pi^-$ mass and hence varies across the DP, we introduce the dimensionless transformed variable $\Delta E^{\prime}=(2\Delta E - \Delta E_{+} - \Delta E_{-})/(\Delta E_{+} - \Delta E_{-})$, where $\Delta E_{\pm}(m_{\pi^+\pi^-}) = c_{\pm} - (c_{\pm} \mp \overline{c})(m_{\pi^+\pi^-}/m_{\pi^+\pi^-}^{\rm max})^2$, $m_{\pi^+\pi^-}$ is the measured $\pi^+\pi^-$ mass, and the parameter $m_{\pi^+\pi^-}^{\rm max}$ is fixed at  $5.0 \gevcc$ (corresponding roughly to the maximum observed value of $m_{\pi^+\pi^-}$). The $\overline{c}$ and $c_{\pm}$ parameters are calculated from fits to $B^0 \to \rho^+\pi^-$ signal MC. The dimensionless quantity $\Delta E^{\prime}$ serves to reduce the degree of correlation with $m_{\pi^+\pi^-}$ and is included as an input variable in the final fit.

To calculate the three $c$ parameters, signal MC events are divided into seven equal-sized bins in $m_{\pi^+\pi^-}$ from 0 to $5.143\gevcc$. In each mass bin, the peak in $\Delta E$ is fit with the sum of two Gaussians, and the mean ($\mu$) and width ($\sigma$) of the narrower Gaussian are extracted. From these parameters, two sets of datapoints are constructed: one consisting of the values of $\mu + 3\sigma$ for each mass bin, and the other consisting of the values of $\mu - 3\sigma$ for each mass bin. The first set is fit with the quadratic function corresponding to $\Delta E_{+}(m_{\pi^+\pi^-})$, which yields values for $c_{+}$ and $\overline{c}$. A similar fit is performed on the second set using the function $\Delta E_{-}(m_{\pi^+\pi^-})$, which yields values for $c_{-}$ and $\overline{c}$. The $\overline{c}$ value used in the final transformation is obtained by averaging the $\overline{c}$ values from the separate fits. For the final calculation of $\Delta E^{\prime}$, we use the parameter values $\overline{c} =  0.0792 \gev$, $c_{-} = -0.1433 \gev$, and $c_{+} = 0.1093 \gev$. Candidates are selected if $\Delta E^{\prime}$ is between $-1$ and $+1$, which is roughly a $\pm 3\sigma$ criterion on $\Delta E$.

\subsection{ Particle Identification Selection Criteria }
\label{sec:ultimatePID}

In the previous {\babar} analysis~\cite{ref:matt2007}, the charged pion candidates were required to be inconsistent with muon, kaon, proton, and electron hypotheses. Improvements in the {\babar} reconstruction software now provide decreased false-positive rates for a given signal efficiency. We apply a PID selection criterion corresponding to a signal-to-square-root-of-background ($S/\sqrt{B}$) ratio (calculated using off-resonance data to model background, and correctly reconstructed $B^0\rightarrow\rho^\pm\pi^\mp$ signal MC to model signal) of 1.26 (scaled relative to all previous selection criteria). For comparison, application of the PID criteria of the previous analysis achieves a scaled $S/\sqrt{B}$ ratio of 1.14. The new PID criterion selects 54.6\% of off-resonance data and 93.2\% of correctly reconstructed signal MC events relative to the previous selection criteria.

\subsection{ Multivariate Discriminator Selection }

Continuum $q\overline{q}$ events are the primary source of background in this analysis. To improve discrimination between continuum background and signal events, we train a neural network (NN) on off-resonance data and signal MC using the ``MLP'' (multi-layer perceptron) NN implementation in the TMVA software package~\cite{ref:tmva}. The NN is trained using 10,000 events representative of signal, 7,500 events representative of background, and 200 training iterations. Validation is performed using an independent sample of 20,000 signal and 7,500 background events. The NN is configured to use two hidden layers with 6 and 5 nodes, respectively. The NN uses four event-shape variables that help distinguish between the roughly isotropic shape of $B$ decays and the more jet-like shape of continuum events. The training variables include the Legendre moment {\tt L0} of the rest of the event, defined as the sum of the magnitudes of the momenta of all charged particle candidates not belonging to the reconstructed $B$ candidate; the ratio of the Legendre moment {\tt L2} to the Legendre moment {\tt L0}, where {\tt L2} is defined as the sum over all tracks and neutral clusters not belonging to the $B$ candidate of $p \cdot (3 \cos^2{\theta} - 1)/2$ where $p$ is the magnitude of the momentum of each track or cluster and $\theta$ is the angle between the $B$ thrust axis and the momentum corresponding to the track or cluster; the cosine of the angle between the $B$ candidate momentum and the beam axis; and the cosine of the angle between the thrust axis of the $B$ candidate and the beam axis. All of these input variables are calculated in the CM frame.

While optimizing the NN discriminator, we studied whether some improvement in performance might be achieved by training the discriminator separately in each of seven $B$-flavor tagging categories (each of which has a different flavor tagging purity), or applying NN selection criteria separately in each of the $B$-flavor tagging categories. These approaches were found to yield negligible improvements in signal efficiency for a given degree of background rejection.

\begin{figure}[tbp!]
\begin{center}
\includegraphics[width=0.48\textwidth]{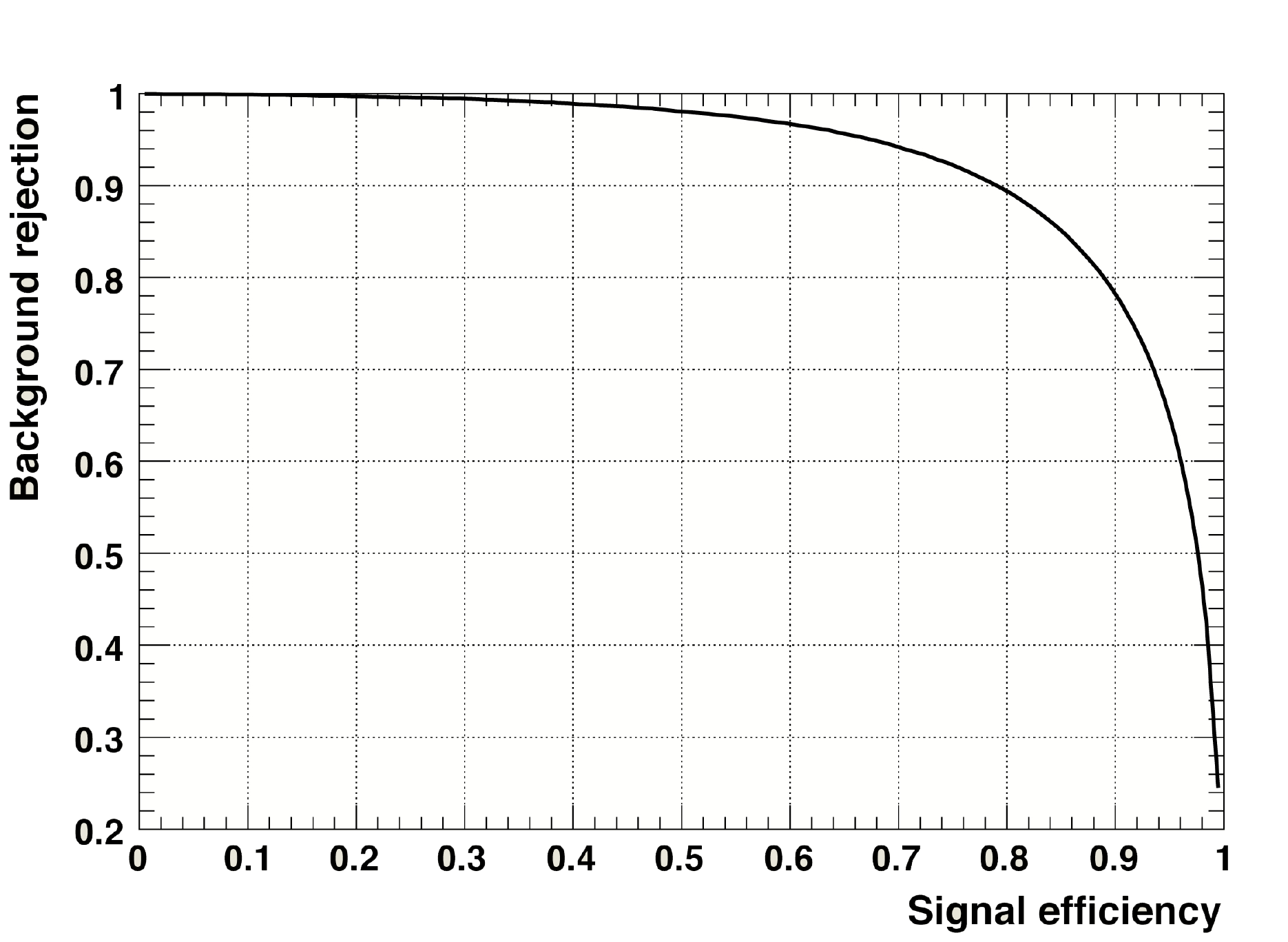}
\caption{(color online) $B^0 \rightarrow \rho^\pm \pi^\mp$ continuum background rejection versus signal efficiency for final neural network implementation, with continuum background represented by off-resonance data, and signal represented by signal MC. Points on the curve correspond to different requirements on the neural network output. Our selection criterion retains $75\%$ of signal events.}
\label{fig:tmvaeff}
\end{center}
\end{figure}

\begin{figure}[tbp!]
\begin{center}
\includegraphics[width=0.48\textwidth]{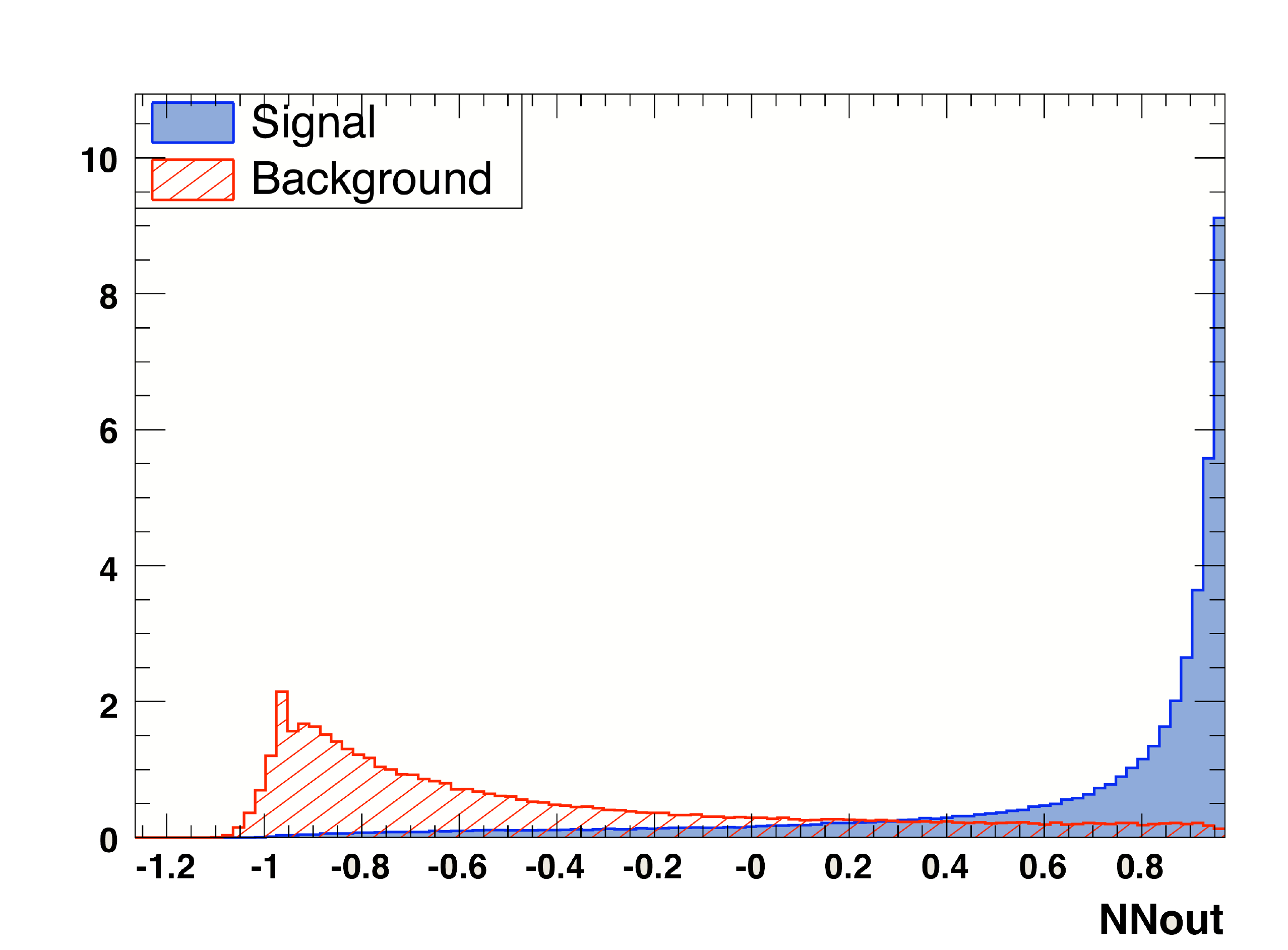}
\caption{(color online) NN output distributions for signal $B^0 \rightarrow \rho^\pm \pi^\mp$ (solid blue) and continuum background (hatched red) with an arbitrary vertical scale. Continuum background is represented by off-resonance data while signal is represented by signal MC. Our selection criterion requires ${\rm NN}>0.58$.}
\label{fig:separation}
\end{center}
\end{figure}

The final performance of the NN discriminator is shown in Fig.~\ref{fig:tmvaeff}, while Fig.~\ref{fig:separation} shows the separation achieved between signal and background samples during training. The NN is used to apply a loose selection criterion that retains 75\% of the signal events remaining after all previous selection criteria. The NN output is also used as an input variable in the final fit.

\subsection{Selection Performance}

Table~\ref{tab:stepbystepeff} provides a summary of the signal efficiencies for each step of the selection process for simulated $B^0\rightarrow\rho^+\pi^-$, $B^0\rightarrow\rho^0\pi^0$, and nonresonant (three-body phase space) $B^0\rightarrow\pi^+\pi^-\pi^0$ events.

\begin{table}
\caption{Selection efficiencies, $\epsilon^{\rm MC}_{X}$, relative to previous criteria for simulated $B^0\rightarrow X$ events ($X=\rho^+\pi^-$, $\rho^0\pi^0$, nonresonant $\pi^+\pi^-\pi^0$). Statistical uncertainties on all efficiencies are less than $0.001$.}

\begin{center}
\begin{tabular}{ c | c l c l c  }
Cuts											    & $\epsilon^{\rm MC}_{\rho^+\pi^-}$ & $\epsilon^{\rm MC}_{\rho^0\pi^0}$ & $\epsilon^{\rm MC}_{\pi^+\pi^-\pi^0}$ \\ \hline \hline 
Preselection									    & $0.650$			&$0.582$                       	 & $0.547$ \\
$p_T > 0.1\gev$								    & $0.994$		          & $0.986$                          & $0.999$ \\
$|\Delta t| < 20\ {\rm ps}\ \&\ \sigma(\Delta t)<2.5\ {\rm ps}$    &  $0.966$  	         	          & $0.957$                          & $0.960$              \\
Photon energy and lateral mom.		                                & $0.928$                        & $0.960$                          & $0.943$     \\
$110<m(\gamma\gamma)<160 \mevcc$                                  & $0.982$	                   & $0.983$                          & $0.983$  \\
{\tt PID}	                                                     			             & $0.920$                      & $0.936$                          & $0.928$    \\
$5.200<m_{\rm ES}<5.288 \gevcc$                                           & $0.994$                    & $0.995$                        & $0.996$ \\
$-1<\Delta E^{\prime}<1$                                                            & $0.819$                      & $0.863$                   	 & $0.871$ 	\\
${\rm NN} > 0.58$ 		                                                           & $0.749$                      & $0.768$                           & $0.762$   \\
$5.272<m_{\rm ES}<5.288 \gevcc$                                           & $0.798$                      & $0.826$			 & $0.937$   \\ \hline
Cumulative efficiency				                               & $0.255$                      & $0.265$                           & $0.279$	
\label{tab:stepbystepeff}
\end{tabular}
\end{center}

\end{table}

\section{\boldmath MAXIMUM LIKELIHOOD FIT}
\label{sec:fitsec}

We perform an unbinned extended maximum likelihood fit to the selected events in order to extract the event yields and the $U$ and $I$ parameters. The input variables are $m_{\rm ES}$, $\Delta E^{\prime}$, the NN output, and the three time-dependent-SDP variables $m^{\prime}$, $\theta^{\prime}$, and $\Delta t$. We also use $\sigma_{\Delta t}$ (the per-event uncertainty on $\Delta t$) as a scale factor in the signal $\Delta t$ resolution function. The likelihood function used in the fit consists of separate components for signal, continuum background, charged $B$ backgrounds, and neutral $B$ backgrounds. The signal PDF is subdivided into components describing correctly reconstructed or ``truth-matched'' (TM), and misreconstructed or self-cross-feed (SCF) candidates. Truth-matched events are identified using MC generator information and contain correctly reconstructed $\pi^+$, $\pi^-$, and $\pi^0$ candidates, two of which were produced in a $\rho$ decay. It is also required that the $\rho$ and the remaining $\pi$ come from the same $B$-meson parent. For nonresonant MC, we instead require that all three pions come directly from the same $B$ parent with no intermediate $\rho$ resonance. All signal events that are not TM are SCF. In $B^0\rightarrow\rho^\pm\pi^\mp$ MC, $22\%$ of reconstructed $B$ candidates are SCF while in $B^0\rightarrow\rho^0\pi^0$ MC, $14\%$ are SCF.

The signal and $B$-background components are further divided by $B$-flavor tagging category and the flavor of the tag-side $B$. The $B_{\rm tag}$ flavor is estimated using multivariate discriminator techniques and is classified as belonging to one of seven flavor-tagging categories corresponding to different degrees of probability that the flavor has been correctly determined \cite{ref:flavor}.

Additionally, prior to the final fit, a transformation is applied to the NN variable. This has the effect of broadening the peak in the signal NN distribution, and transforming the continuum NN distribution to make fitting with the smoothing algorithm (see Sec.~\ref{sec:sigpar}) more effective. The transformation is defined by
\begin{align}
	\label{transformNN}
	{\rm NN}_{\rm new} = 1 - \arccos ({\rm NN}_{\rm old} + \beta),
\end{align}
\noindent where $\beta=0.029$ is an offset approximately equal to 1 minus the maximum of the original NN distribution.

\subsection{Likelihood Function}

The PDF ${\mathcal P}^c_i$ for an event $i$ in tagging category $c$ is the sum of the probability densities of all the signal and background components, namely
\begin{eqnarray}
	\label{likelihood}
	{\mathcal P}^c_i & \equiv & N_{3\pi}f^c_{3\pi} \left[ (1-{\overline f}^c_{\rm SCF}) {\mathcal P}^c_{3\pi{\rm -TM},i} + {\overline f}^c_{{\rm SCF}} {\mathcal P}^c_{3\pi{\rm -SCF},i}  \right] \nonumber \\
	& & + N_{q {\overline q}}^c \frac{1}{2} (1+q_{{\rm tag},i} A_{q {\overline q}, {\rm tag}}) {\mathcal P}^c_{q {\overline q}, i} \nonumber \\
	& & + \sum^{N^{B^+}_{\rm class}}_{j=1} N_{B^+ j} f^c_{B^+j} \frac{1}{2} (1+q_{{\rm tag},i} A_{B^+,{\rm tag},j}) {\mathcal P}^c_{B^+,ij} \nonumber \\
	& & + \sum^{N^{B^0}_{\rm class}}_{j=1} N_{B^0 j} f^c_{B^0 j} {\mathcal P}^c_{B^0 ,ij},
\end{eqnarray}
\noindent where

\begin{itemize}
	\item	{$N_{3\pi}$ is the total number of $\pi^+\pi^-\pi^0$ signal events in the data sample (both TM and SCF);}
	\item {$f^c_{3\pi}$ is the fraction of signal events (TM and SCF) in flavor tagging category $c$;}
	\item {${\overline f}^c_{\rm SCF}$ is the fraction of misreconstructed signal events (SCF) in tagging category $c$;}
	\item {${\mathcal P}^c_{3\pi{\rm -TM},i}$ and ${\mathcal P}^c_{3\pi{\rm -SCF},i}$ are the products of PDFs of the discriminating variables used in tagging category $c$ for TM and SCF events, respectively;}
	\item {$N^c_{q {\overline q}}$ is the number of continuum events in flavor tagging category $c$;}
	\item {$q_{{\rm tag},i}$ is the tag flavor of the event where we use the convention that $q_{{\rm tag},i}=1$ for $B_{\rm tag}=B^0$ and $q_{{\rm tag},i}=-1$ for a $B_{\rm tag}={\overline B}^0$;}
	\item {$A_{q{\overline q},{\rm tag}}$ is the flavor tag asymmetry, parameterizing possible charge asymmetry in continuum events;}
	\item {${\mathcal P}^c_{q {\overline q},i}$ is the continuum PDF for tagging category $c$;}
	\item {$N^{B^+}_{\rm class}$ ($N^{B^0}_{\rm class}$) is the number of charged (neutral) $B$-background classes included in the fit (including generic modes); see Table~\ref{tab:bbkgtab};}
	\item {$N_{B^+ j}$ ($N_{B^0 j}$) is the number of expected events in the charged (neutral) $B$-background class $j$;}
	\item {$f^c_{B^+ j}$ ($f^c_{B^0 j}$) is the fraction of charged (neutral) $B$-background events of class $j$ that are in flavor tagging category $c$;}
	\item {$A_{B^+,{\rm tag},j}$ is the flavor tag asymmetry of charged $B$-background class $j$;}
	\item {${\mathcal P}^c_{B^+ , ij}$ is the $B^+$-background PDF for tagging category $c$ and class $j$;}
	\item {${\mathcal P}^c_{B^0 , ij}$ is the $B^0$-background PDF for tagging category $c$ and class $j$.}
\end{itemize}

The PDFs ${\mathcal P}^c_X$ are the product of PDFs for the discriminating variables $x_1 = m_{\rm ES}$, $x_2 = \Delta E^\prime$, $x_3 = {\rm NN}_{\rm new}$, and the three time-dependent-SDP variables $x_4 = \{ m^\prime, \theta^\prime, \Delta t \}$:
\begin{align}
	{\mathcal P}^c_{X,i(j)} \equiv \prod^{4}_{k=1} {\mathcal P}^c_{X,i(j)}(x_k).
\end{align}
The extended likelihood function including all tagging categories is given by
\begin{align}
	{\mathcal L}  \equiv \prod^{7}_{c=1} e^{ - {\overline N}^c } \prod^{N^c}_{i=1} {\mathcal P}^{c}_{i},
\end{align}
where ${\overline N}^c$ is the total number of events expected in tagging category $c$ and $N^c$ is the observed number.

\subsection{Signal Parameterization}
\label{sec:sigpar}

The effect of experimental resolution for $\Delta t$ in signal events is taken into account by convolving the PDF describing the true $\Delta t$ distribution with a sum of three Gaussians.

\begin{figure}[!htb]
\begin{center}
\includegraphics[width=0.48\textwidth,clip=true,trim=70 70 125 70]{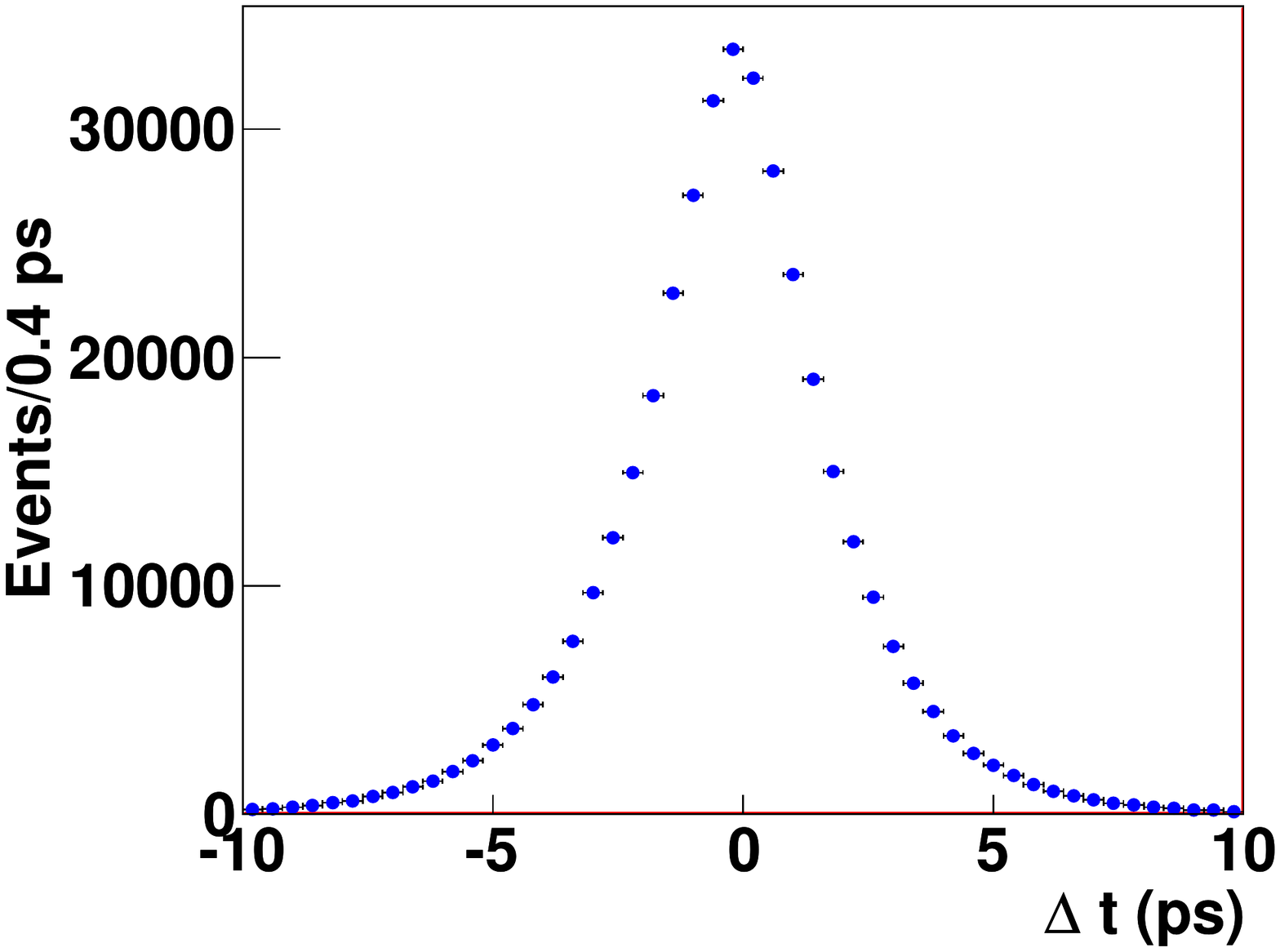}
\includegraphics[width=0.48\textwidth,clip=true,trim=70 70 125 70]{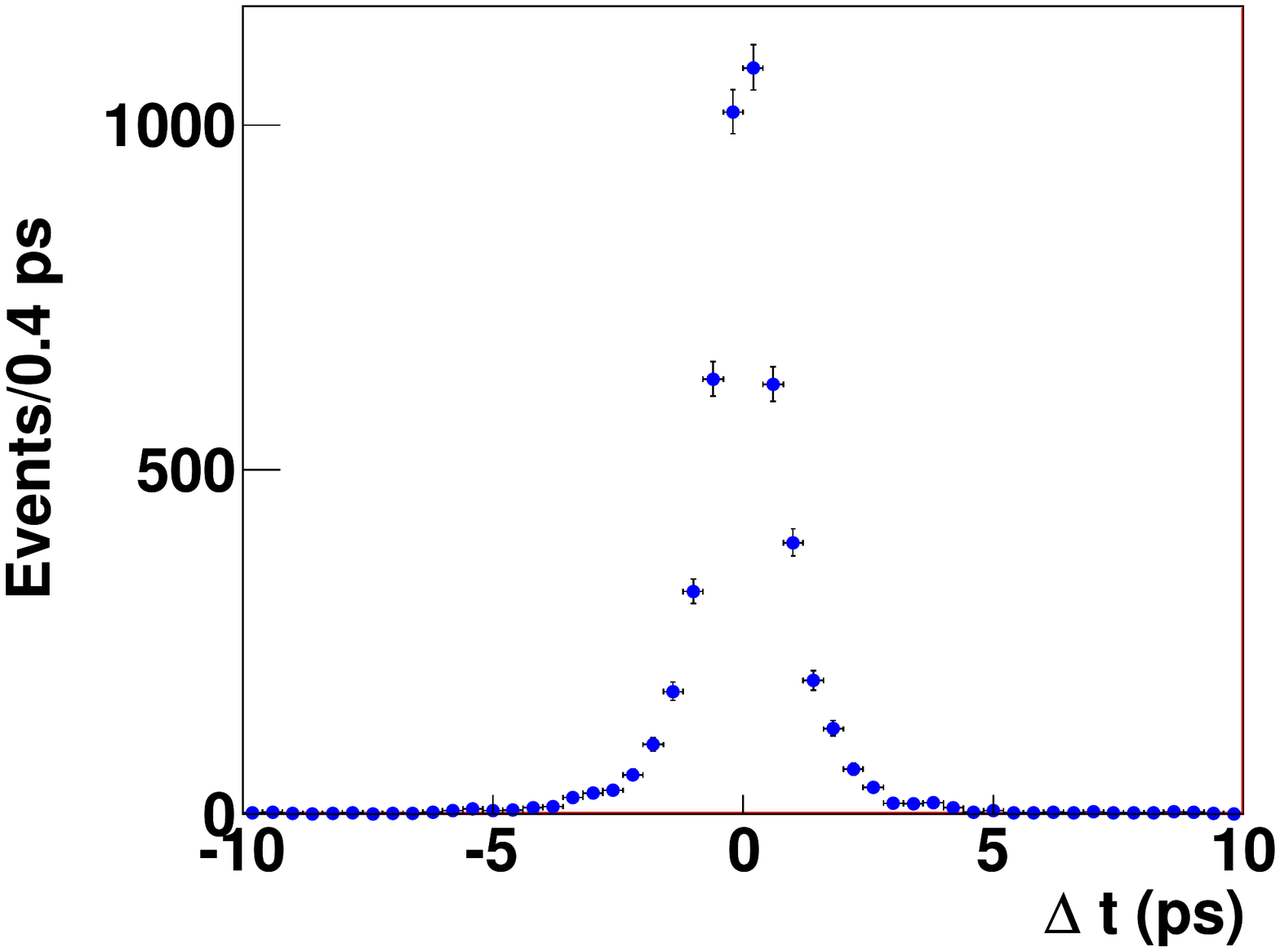}
\caption{Distribution of $\Delta t$ in simulated $B^0\rightarrow\rho^\pm\pi^\mp$ signal events (top) and in off-resonance data (bottom).}
\label{fig:deltatfigure}
\end{center}
\end{figure}

The triple-Gaussian resolution function is constructed using a narrow ``core'' Gaussian, a slightly wider ``tail'' Gaussian, and a very wide ``outlier'' Gaussian. In the final fit to on-resonance data, all signal $\Delta t$ parameters are fixed to values obtained from fits to fully reconstructed $B$ decays. In the fits to fully reconstructed $B$ decays, the mean and width of the outlier Gaussian are fixed to 0 and 8 ps, respectively. Similarly, the width of the tail Gaussian is fixed to 3 ps, but its mean is allowed to vary. Finally, both the mean and width of the core Gaussian are allowed to vary in the fit, and the means of the core Gaussian are allowed to take on different values in each $B$-flavor tagging category. The means and widths of the core and tail Gaussians are scaled by the per-event uncertainty on $\Delta t$ ($\sigma_{\Delta t}$):
\begin{eqnarray}
	{\mathcal R}_{\rm sig}(\Delta t, \sigma_{\Delta t}) &=& (1 - f_{\rm tail} - f_{\rm out}) \nonumber \\
										   & & \times G(\Delta t; s^b_{\rm core} \sigma_{\Delta t}, s^{\sigma}_{\rm core} \sigma_{\Delta t}) \nonumber \\
										   & & + f_{\rm tail} G(\Delta t,s^b_{\rm tail} \sigma_{\Delta t}; s^{\sigma}_{\rm tail} \sigma_{\Delta t}) \nonumber \\
										   & & + f_{\rm out} G(\Delta t;s^b_{\rm out}, \sigma_{\rm out}),
\end{eqnarray}
\noindent where $G(x;x_0,\sigma)$ is a Gaussian with mean $x_0$ and width $\sigma$. See Fig.~\ref{fig:deltatfigure} (upper plot) for the distribution of $\Delta t$ in signal MC events.

The $\Delta E^{\prime}$ signal distribution is modeled using the sum of two Gaussians where all five free parameters depend linearly on $m^2(\pi^+\pi^-)$ in order to account for residual dependence on the mass. The $m_{\rm ES}$ signal distribution is parameterized using a bifurcated Crystal Ball function, which is composed of a one-sided Gaussian and a Crystal Ball function:
\begin{align}
f(x)=\left\{ \begin{array}{l}
	C e^{-(x-m)^2/2s^2_R} {\rm \ for\ } (x-m)>0, \\
	C e^{-(x-m)^2/2s^2_L}  {\rm \ for\ } -A<\frac{x-m}{s_L}<0, \\
	C (\frac{b}{A})^b e^{-\frac{A^2}{2}} (\frac{b}{A}-A-\frac{x-m}{s_L})^{-b} {\rm \ for\ } \frac{x-m}{s_L}<-A.
\end{array} \right.
\end{align}
\noindent The parameters describing both these lineshapes are extracted from fits to signal MC events. In the final fit, the parameter $m$ is free to vary for $m_{\rm ES}$ while the core Gaussian mean and width, and the slope of the mean (i.e., the dependence of the mean on $m^2(\pi^+\pi^-)$) are free to vary for $\Delta E^\prime$.

The ${\rm NN}_{\rm new}$ signal distribution is modeled by nonparametric histograms generated by smoothing the ${\rm NN}_{\rm new}$ distribution in signal MC.

The $\Delta E^\prime$ distribution for SCF signal candidates is described using a single Gaussian with mean and width fixed to the values extracted
from a fit to SCF signal MC.

The distributions of $m_{\rm ES}$ and ${\rm NN}_{\rm new}$ for SCF signal events are modeled with nonparametric histograms generated by smoothing the appropriate one-dimensional distributions in SCF signal MC. The nonparametric PDFs for the ${\rm NN}_{\rm new}$ distributions in both TM and SCF signal are generated separately for each flavor-tagging category.

While the SDP distribution for TM signal events is parameterized by the full time-dependent decay probability, the SCF distribution is parameterized by modifying this distribution. Using signal MC, we create a binned map in the SDP that contains the probability for an event generated in each bin of the SDP to be reconstructed in the same bin, or each of the other bins. This map is convolved with the time-dependent decay PDF to generate the SCF signal PDF in the SDP.

\subsection{Continuum Background Parameterization}

The $\Delta t$ distribution for continuum background is modeled using a sum of three Gaussians. The parameters of this triple Gaussian are obtained from a fit to off-resonance data in which all parameters are allowed to vary. See Fig.~\ref{fig:deltatfigure} (lower plot) for the distribution of $\Delta t$ in off-resonance data.

The $\Delta E^{\prime}$ distribution for continuum background is modeled using a second-order polynomial with parameter values extracted from a fit to the $m_{\rm ES}$ sideband in on-resonance data.

The distribution of $m_{\rm ES}$ for continuum background is modeled using an Argus \cite{ref:argus} function with shape and endpoint parameters that are allowed to vary in the final fit.

To account for residual dependence on DP position, the distribution of ${\rm NN}_{\rm new}$ for continuum background is modeled using a second-order polynomial function in which each coefficient depends linearly on $m_{\rm dist}$, the minimum invariant mass of any $\pi\pi$ combination in the $B^0\rightarrow\pi^+\pi^-\pi^0$ candidate, which acts as a measure of the distance from the edge of the DP. In addition, the polynomial is multiplied by $(1-{\rm NN}_{\rm new})^a$ where $a$ is a linear function of $m_{\rm dist}$. All the polynomial parameters are free to vary in the final fit.

The two-dimensional SDP distribution for continuum events is obtained by applying Gaussian kernel smoothing algorithms to the SDP distribution for on-resonance $m_{\rm ES}$ sideband data and generating a two-dimensional histogram from the resulting PDF, which serves as a nonparametric PDF in the fit. Histogram PDFs are generated separately for each flavor-tagging category and, within each category, for both $B$-flavor tags. Bins in these histograms are mirrored across the $\theta^\prime=0.5$ axis so that the distributions are symmetric in $\theta^\prime$. A number of parameters are allowed to vary in the fit to allow for an asymmetry.

\subsection{$B$-Background Parameterization}

The functional form of the $\Delta t$ resolution functions for the $B$ backgrounds is the same as that for signal. Parameter values are obtained from separate fits to fully simulated MC data samples representative of each $B$-background class. 

Simulated samples for each of the $B$-background classes are used to generate nonparametric PDFs for use in the final fit. One-dimensional PDFs are used for $\Delta E^{\prime}$, $m_{\rm ES}$, and ${\rm NN}_{\rm new}$, without any splitting by flavor-tagging category. Two-dimensional SDP PDFs are generated for each $B$-background class and each $B$-flavor tag within that class.

\subsection{Dalitz-Plot-Dependent Selection Efficiency}

\begin{figure}[!htb]
\begin{center}
\begin{turn}{0} \includegraphics[clip=true,trim=110 60 75 110,width=0.48\textwidth]{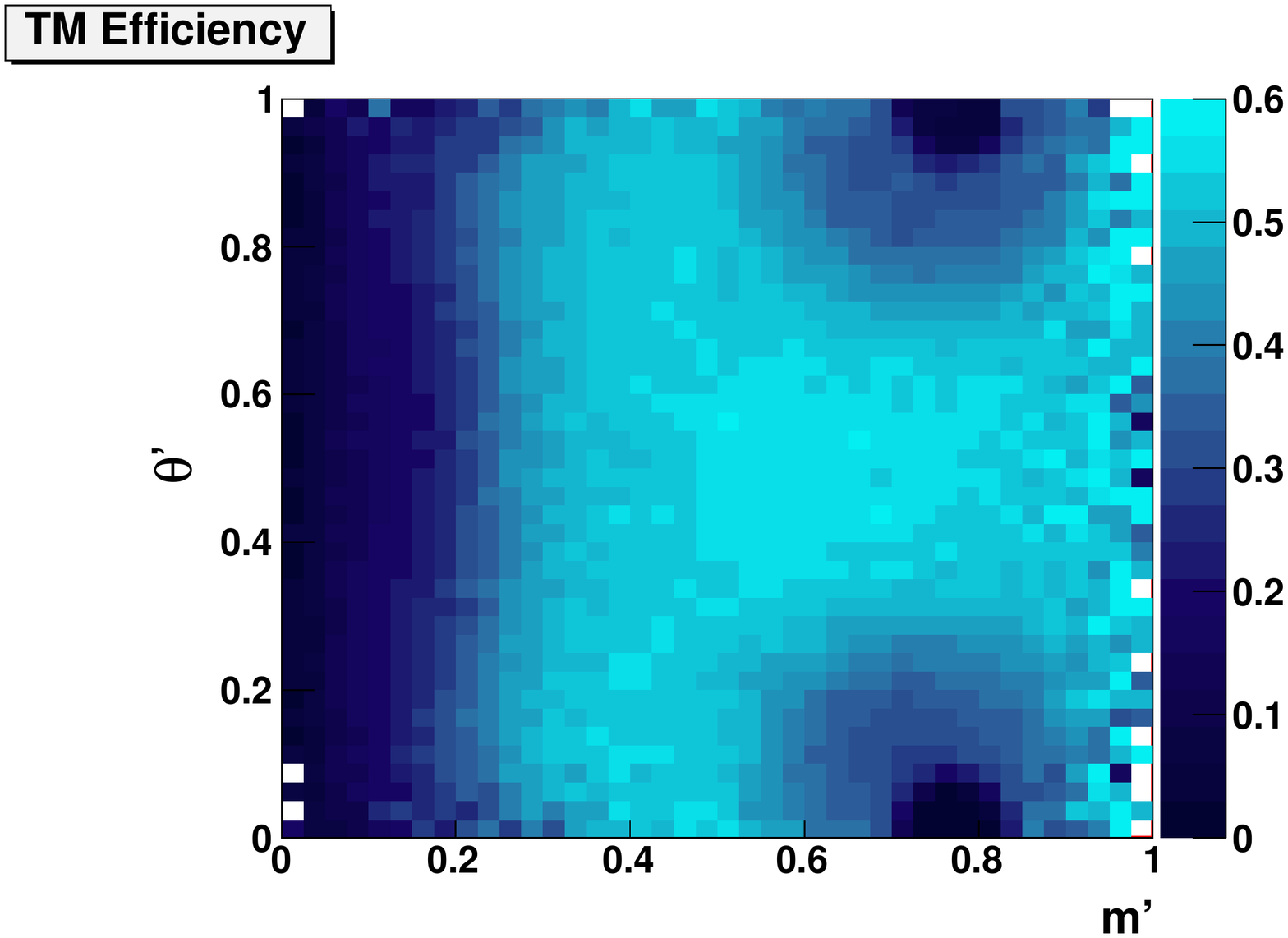} \end{turn}
\begin{turn}{0} \includegraphics[clip=true,trim=110 60 75 110,width=0.48\textwidth]{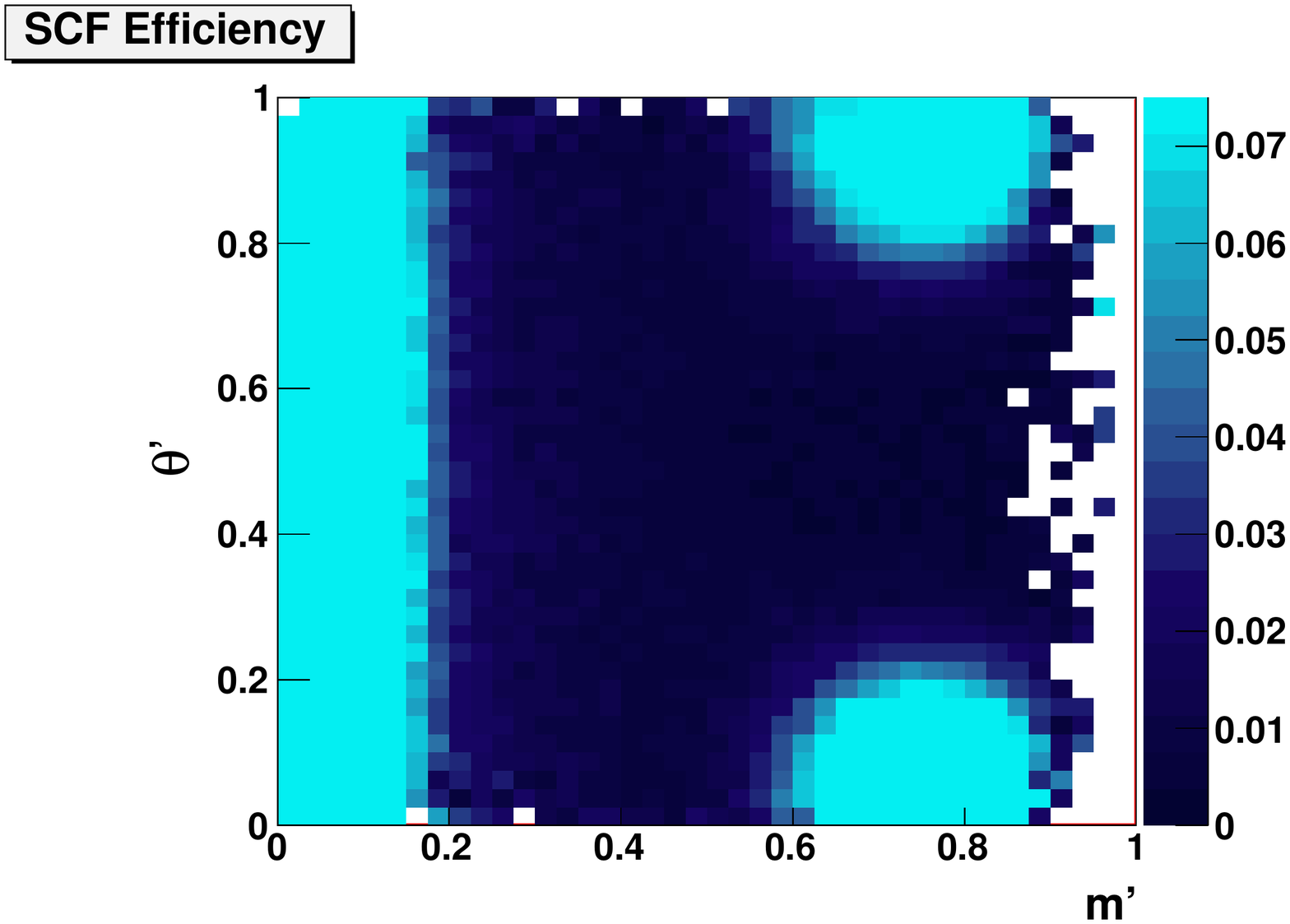} \end{turn}

\caption{(color online) Square Dalitz plot map of TM (top) and SCF (bottom) selection efficiencies. The color scale on the TM plot has a maximum of 0.60 while that on the SCF plot has a maximum of 0.075.}

\label{fig:dpeff}
\end{center}
\end{figure}

Selection efficiencies across the SDP are calculated from a combination of all available nonresonant ($B^0 \rightarrow \pi^+\pi^-\pi^0$), $B^0 \rightarrow \rho^\pm \pi^\mp$, and $B^0 \rightarrow \rho^0 \pi^0$ MC samples. We divide the SDP into a 40 by 40 grid and, for each bin, calculate both the fraction of events generated in that bin that are correctly reconstructed (TM), and the fraction of events generated in that bin that are misreconstructed (SCF). From this, we generate tables of efficiencies that are used as inputs to the fit. Histograms of the TM and SCF selection efficiencies are provided in Fig.~\ref{fig:dpeff}.

\section{SYSTEMATIC STUDIES}
\label{sec:systematicsec}

\subsection{The Effect of the $\rho(1700)$}
\label{sec:exclusionof1700}

We include the $\rho(1450)$ in the final fit with an assumption that the relative magnitudes and phases between the three $\rho(1450)$ resonances are the same as for the $\rho(770)$. Whereas there is reasonable motivation for this assumption in the case of the $\rho(1450)$ since the $\rho(770)$ and $\rho(1450)$ have the same quantum numbers, the $\rho(1700)$ does not share these quantum numbers ($\ell=2$ instead of $0$). Since the $\rho(1700)$ is not expected to provide a large contribution, we exclude the $\rho(1700)$ from the fit and associate a systematic uncertainty with this omission.

Naively, one might calculate this systematic uncertainty by fitting the full dataset with and without the $\rho(1700)$ and calculating a single change in the $U$ and $I$ parameters, but this approach suffers from statistical uncertainties due to the fact that there is only one dataset available. Nonetheless, it is still useful to calculate a covariance matrix using this approach in order to estimate the magnitude of the systematic uncertainty. As a first order assessment of the systematic uncertainty, we calculate a covariance matrix with elements given by
\begin{eqnarray}
	C_{i,j} \equiv \delta_i \delta_j,
\end{eqnarray}
\noindent where $\delta_i$ is the difference between the value of the $i{\rm th}$ $U$ or $I$ parameter in fits with and without the $\rho(1700)$. When the three $\rho(1700)$ resonances are included in the fit, their magnitudes and phases are all allowed to vary independently. The square roots of the diagonal elements of the matrix (in other words, the absolute value of the change in each $U$ or $I$ parameter) are given in Table~\ref{tab:full1700Cov} along with the ratio relative to the statistical uncertainty on each parameter in the fit without the $\rho(1700)$. The ratios are generally less than 1; only 4 of the 26 ratios are greater than 1, indicating that the changes in the $U$ and $I$ parameters resulting from the inclusion of a $\rho(1700)$ in the Dalitz model are mostly small.

To assess the uncertainties on the changes in the $U$ and $I$ parameters, we employ the bootstrap technique introduced by Efron~\cite{ref:efron}. This approach allows us to calculate a covariance matrix associated with the uncertainty on the $U$ and $I$ parameters extracted from the fits to the full dataset by fitting a large number of datasets generated by sampling events with replacement from the full dataset. Each of the approximately 1000 resampled datasets generated in this manner has the same number of events as the original dataset and is fit with and without the $\rho(1700)$ component. For each pair of fits, the change in the $U$ and $I$ parameters is calculated and a covariance matrix is generated by determining the covariances of the changes in the $U$ and $I$ parameters in all these pairs of fits. As demonstrated by Efron, the covariance matrix obtained in this manner is an estimate of the covariance matrix associated with the uncertainty on the changes in the $U$ and $I$ parameters. Therefore, the square roots of the diagonal elements of this matrix are estimates of the uncertainty on the change in each $U$ and $I$ parameter. These uncertainties, as well as the ratio of the mean change of each $U$ and $I$ across all the bootstrapped fits to its estimated uncertainty (from the bootstrap), are given in the last column in Table~\ref{tab:full1700Cov}. As the ratios in this table demonstrate, the mean changes in the $U$ and $I$ parameters are consistent with 0. Given these results, we use the covariance matrix obtained by bootstrapping to characterize the systematic uncertainty associated with excluding the $\rho(1700)$.

\begin{table}
\caption{Magnitude of changes in $U$ and $I$ parameters between fits with and without the $\rho(1700)$, the ratio of these magnitudes relative to statistical uncertainties from the fit without the $\rho(1700)$, the bootstrapped estimate of the uncertainty on changes in $U$ and $I$ parameters between fits with and without the $\rho(1700)$, and the ratio of the mean change in the $U$ and $I$ parameters across all bootstrapped fits to their bootstrap-estimated uncertainties.}
\begin{center}
\begin{tabular}{ l | l c l c l l }
Parameter & $|\Delta U|$ & $ |\Delta U| / \sigma_{\rm stat}\quad$       & $\sigma_{\Delta U}^{\rm bs}$ & $\left<\Delta U\right> / \sigma_{\Delta U}^{\rm bs}$ \\ \hline
$I_0$                 &                 0.017           &   0.46      &      0.020           &       $ -0.36  $ \\
$I_-$                 &                 0.015           &   0.24       &            0.022          &       $  {\phantom -}0.41   $ \\
$I_{-0}^{\rm Im}$     &                 0.40            &   0.92      &      0.42           &       $ -0.70  $ \\
$I_{-0}^{\rm Re}$     &                 0.39            &   0.67      &      0.7           &       $  {\phantom -}0.28  $ \\
$I_+$                 &                 0.0013          &   0.02       &           0.023          &       $  {\phantom -}0.18  $ \\
$I_{+0}^{\rm Im}$     &                 0.024           &   0.07     &      0.42           &       $ -0.11  $ \\
$I_{+0}^{\rm Re}$     &                 0.18            &   0.40      &     0.7            &       $  {\phantom -}0.26  $ \\
$I_{+-}^{\rm Im}$     &                 1.34             &   1.81       &            0.9           &       $ -0.95  $ \\
$I_{+-}^{\rm Re}$     &                 0.68            &   0.90      &      0.9           &       $ -0.59  $ \\
$U_0^-$               &                 0.012           &   0.23      &     0.029          &       $  {\phantom -}0.29   $ \\
$U_0^+$               &                 0.006           &   0.21      &     0.017          &       $  {\phantom -}0.54  $ \\
$U_{-0}^{-,{\rm Im}}$ &                 0.31            &   0.70      &      0.5           &       $  {\phantom -}0.76  $ \\
$U_{-0}^{-,{\rm Re}}$ &                 0.31            &   0.89      &      0.34           &       $  {\phantom -}0.49  $ \\
$U_{-0}^{+,{\rm Im}}$ &                 0.14            &   0.70      &      0.21           &       $ -0.76  $ \\
$U_{-0}^{+,{\rm Re}}$ &                 0.25            &   1.47       &     0.17           &       $ -0.77  $ \\
$U_-^-$               &                 0.00041           &   0.00    &           0.034          &       $  {\phantom -}0.19  $ \\
$U_-^+$               &                 0.015            &   0.21      &            0.05          &       $  {\phantom -}0.15  $ \\
$U_{+0}^{-,{\rm Im}}$ &                 0.18            &   0.61      &      0.44            &       $ -0.10  $ \\
$U_{+0}^{-,{\rm Re}}$ &                 0.20            &   0.62       &     0.36           &       $  {\phantom -}0.50  $ \\
$U_{+0}^{+,{\rm Im}}$ &                 0.039           &   0.24      &     0.16           &       $  {\phantom -}0.51  $ \\
$U_{+0}^{+,{\rm Re}}$ &                 0.005          &   0.03     &      0.17            &       $  {\phantom -}0.05 $ \\
$U_{+-}^{-,{\rm Im}}$ &                 0.6            &   1.23       &     0.8           &       $ -0.27  $ \\
$U_{+-}^{-,{\rm Re}}$ &                 0.5            &   1.10       &     0.7           &       $  {\phantom -}0.15  $ \\
$U_{+-}^{+,{\rm Im}}$ &                 0.038           &   0.15      &     0.25            &       $ -0.22  $ \\
$U_{+-}^{+,{\rm Re}}$ &                 0.21            &   0.84      &      0.28           &       $ -0.21  $ \\
$U_+^-$               &                 0.021           &   0.23      &     0.06          &       $ -0.46  $ 
\label{tab:full1700Cov}
\end{tabular}
\end{center}
\end{table}

\subsection{$B$ Background Branching Fractions}

We account for uncertainties on the branching fractions for the various $B$ background classes by performing fits to data after increasing and decreasing the expected number of $B$ background events by $10\%$. The value of $10\%$ is an estimate of the degree of uncertainty on the expected number of events in the $B$ background classes with the largest contributions (see Table~\ref{tab:bbkgtab}). We calculate a systematic covariance matrix for the $U$ and $I$ parameters from these fits. Element $(i,j)$ is given by
 \begin{eqnarray}
	C_{ij} \equiv \frac{\delta_i}{2} \frac{\delta_j}{2},
\end{eqnarray}
\noindent where $\delta_i$ is the difference between the value of the $i{\rm th}$ $U$ or $I$ parameter in the fit with increased $B$ background contributions, and the fit with decreased contributions. The systematic errors ($\sigma^{i}_{\rm syst} = \sqrt{C_{ii}}$) associated with the $B$-background branching fractions are given in Table~\ref{tab:bfsyst}. The first column of numbers contains the systematic errors calculated from the square root of the diagonal elements of the covariance matrix while the second column of numbers contains the ratio of these uncertainties to the statistical uncertainties from the nominal fit to the full dataset.

\begin{table}
\caption{Square root of the diagonal elements of the covariance matrix associated with the uncertainties on the $U$ and $I$ parameters due to uncertainties on $B$-background branching fractions. The rightmost column contains the ratio of these systematic uncertainties to the statistical uncertainties on the $U$ and $I$ parameters as obtained in the nominal fit. }
\begin{center}
\begin{tabular}{ l | l  c l}
Parameter & $\sigma_{\rm syst}$ & $\sigma_{\rm syst} / \sigma_{\rm stat}$ \\ \hline
$I_0$                 			& 0.0035      &       0.09 \\
$I_-$                 			& 0.007      &       0.11 \\
$I_{-0}^{\rm Im}$     		& 0.012       &       0.03 \\
$I_{-0}^{\rm Re}$    		& 0.017       &       0.03 \\
$I_+$                 			& 0.0022      &       0.04 \\
$I_{+0}^{\rm Im}$     		& 0.025       &       0.07 \\
$I_{+0}^{\rm Re}$     		& 0.031       &       0.07 \\
$I_{+-}^{\rm Im}$     		& 0.034       &       0.05 \\
$I_{+-}^{\rm Re}$     		& 0.032       &       0.04 \\
$U_0^-$               		& 0.0011      &       0.02 \\
$U_0^+$               		& 0.007        &       0.25 \\
$U_{-0}^{-,{\rm Im}}$ 	& 0.14        &       0.33 \\
$U_{-0}^{-,{\rm Re}}$ 	& 0.043       &       0.12 \\
$U_{-0}^{+,{\rm Im}}$ 	& 0.07        &       0.35 \\
$U_{-0}^{+,{\rm Re}}$ 	& 0.031       &       0.18 \\
$U_-^-$               		& 0.009       &       0.09 \\
$U_-^+$               		& 0.0018        &       0.03 \\
$U_{+0}^{-,{\rm Im}}$ 	& 0.017       &       0.06 \\
$U_{+0}^{-,{\rm Re}}$ 	& 0.011       &       0.04 \\
$U_{+0}^{+,{\rm Im}}$ 	& 0.016       &       0.10 \\
$U_{+0}^{+,{\rm Re}}$ 	& 0.011       &       0.07 \\
$U_{+-}^{-,{\rm Im}}$ 	& 0.10       &       0.20 \\
$U_{+-}^{-,{\rm Re}}$ 	& 0.019       &       0.04 \\
$U_{+-}^{+,{\rm Im}}$ 	& 0.05       &       0.20 \\
$U_{+-}^{+,{\rm Re}}$ 	& 0.06       &       0.22 \\
$U_+^-$               		& 0.012       &       0.13 
\label{tab:bfsyst}
\end{tabular}
\end{center}
\end{table}

\subsection{\boldmath $\rho$ Lineshapes}

The systematic uncertainties associated with the $\rho(770)^\pm$, $\rho(770)^0$, and $\rho(1450)$ lineshapes are calculated by varying their masses and widths according to the uncertainties listed in Table~\ref{tab:lineshapeparams} and the correlation matrix in Table~\ref{tab:lineshapecorrelation}. These correlations and uncertainties were determined using a Gounaris-Sakurai model fit to data from $\tau^+\rightarrow{\overline \nu}_\tau\rho^+ (\rightarrow\pi^+\pi^0)$ decays and $e^+e^-\rightarrow\rho^0\rightarrow\pi^+\pi^-$ annihilation~\cite{ref:lineshapeCitation}. We use updated lineshape fits including new data from $e^+e^-$ annihilation~\cite{ref:cmd} and $\tau$ spectral functions~\cite{ref:aleph}.

The correlations between the three $\rho$ masses and widths as well as their uncertainties were used to calculate the corresponding covariance matrix. A six-dimensional correlated Gaussian was defined with means corresponding to the central values of the lineshape parameters. A set of 50 vectors of lineshape parameters was sampled randomly from the multi-dimensional Gaussian distribution and used to perform 50 fits. Using the randomly sampled masses and widths, and initializing all other parameter values to those from the best fit to the full on-resonance dataset with nominal parameters, we performed a fit with each resonance configuration, producing 50 final sets of $U$ and $I$ parameters. The covariances of these 50 sets of $U$ and $I$ parameter values were calculated using the set of final fit values from the nominal fit as expected values.

As a test of the fitting framework, the mean of the number of signal events from the 50 fits was extracted and compared to the nominal value. They were found to be in agreement with a difference of $0.6\sigma_{\rm stat}$ (where $\sigma_{\rm stat}$ is the statistical error on the number of signal events from the nominal fit to the full on-resonance dataset), exhibiting negligible bias. As a further comparison, the ratios of the systematic errors (taken from the square root of the diagonal of the covariance matrix) to the statistical errors (taken from the nominal fit) were calculated and found to be small as shown in Table~\ref{tab:lineshapesyst}.

\begin{table}
\caption{ $\rho$ lineshape parameters and uncertainties used in the fits and in evaluating the uncertainties on the $U$ and $I$ parameters~\cite{ref:lineshapeCitation}.}
\begin{center}
\begin{tabular}{ c | c }
Parameter & Value \cite{ref:lineshapeCitation} (${\rm MeV}/c^2$) \\ \hline
$m_{\rho(770)^\pm}$		& $775.5 \pm 0.6$ \\
$m_{\rho(770)^0}$			& $773.1 \pm 0.5$ \\
$\Gamma_{\rho(770)^\pm}$	& $148.2 \pm 0.8$ \\
$\Gamma_{\rho(770)^0}$		& $148.0 \pm 0.9$ \\
$m_{\rho(1450)}$			& $1409 \pm 12$ \\
$\Gamma_{\rho(1450)}$		& $500 \pm 37$ 
\label{tab:lineshapeparams}
\end{tabular}
\end{center}
\end{table}

\begin{table}
\caption{Correlations between $\rho$ lineshape parameters used in evaluating the uncertainties on the $U$ and $I$ parameters~\cite{ref:lineshapeCitation}.}
\begin{center}
\begin{tabular}{ c | r r r r r r  }
					& $m_{\rho(770)^\pm}$ 	&	$m_{\rho(770)^0}$   &  $\Gamma_{\rho(770)^\pm}$     &  $\Gamma_{\rho(770)^0}$    & $m_{\rho(1450)}$      &  $\Gamma_{\rho(1450)}$ \\ \hline
$m_{\rho(770)^\pm}$		& {\phantom -}1.000        &    {\phantom -}0.109    &      {\phantom -}0.315     &      $-$0.035    &      {\phantom -}0.017     &      $-$0.150 \\
$m_{\rho(770)^0}$		& {\phantom -}0.109      &     {\phantom -}1.000     &       {\phantom -}0.049     &      {\phantom -}0.290     &      {\phantom -}0.142     &      $-$0.065 \\
$\Gamma_{\rho(770)^\pm}$	& {\phantom -}0.315      &     {\phantom -}0.049   &        {\phantom -}1.000      &      {\phantom -}0.361     &      {\phantom -}0.133     &      {\phantom -}0.024 \\
$\Gamma_{\rho(770)^0}$	& $-$0.035     &     {\phantom -}0.290   &        {\phantom -}0.361   &        {\phantom -}1.000      &      {\phantom -}0.180     &      {\phantom -}0.083 \\
$m_{\rho(1450)}$			& {\phantom -}0.017       &    {\phantom -}0.142   &        {\phantom -}0.133   &        {\phantom -}0.180    &       {\phantom -}1.000      &      {\phantom -}0.779 \\
$\Gamma_{\rho(1450)}$	& $-$0.150      &    $-$0.065 &         {\phantom -}0.024   &        {\phantom -}0.083   &        {\phantom -}0.779   &        {\phantom -}1.000 
\label{tab:lineshapecorrelation}
\end{tabular}
\end{center}
\end{table}

\begin{table}
\caption{Square root of the diagonal elements of the covariance matrix associated with the uncertainties on the $U$ and $I$ parameters due to uncertainties on the $\rho$ lineshape.  The rightmost column contains the ratio of these systematic uncertainties to the statistical uncertainties on the $U$ and $I$ parameters as obtained in the nominal fit.}
\begin{center}
\begin{tabular}{ l | l  c l}
Parameter & $\sigma_{\rm syst}$ & $\sigma_{\rm syst} / \sigma_{\rm stat}$ \\ \hline
$I_0$                 & 0.0005  &   0.01 \\
$I_-$                 & 0.0013   &   0.02 \\
$I_{-0}^{\rm Im}$     & 0.009   &   0.02 \\
$I_{-0}^{\rm Re}$     & 0.025    &   0.04  \\
$I_+$                 & 0.0007  &   0.01 \\
$I_{+0}^{\rm Im}$     & 0.012    &   0.03 \\
$I_{+0}^{\rm Re}$     & 0.06    &   0.13  \\
$I_{+-}^{\rm Im}$     & 0.05      &   0.07 \\
$I_{+-}^{\rm Re}$     & 0.07    &   0.09 \\
$U_0^-$               & 0.0014   &   0.03 \\
$U_0^+$               & 0.0020   &   0.07 \\
$U_{-0}^{-,{\rm Im}}$ & 0.029    &   0.07 \\
$U_{-0}^{-,{\rm Re}}$ & 0.018    &   0.05 \\
$U_{-0}^{+,{\rm Im}}$ & 0.020     &   0.10 \\
$U_{-0}^{+,{\rm Re}}$ & 0.017    &   0.10  \\
$U_-^-$               & 0.0014   &   0.01 \\
$U_-^+$               & 0.0017   &   0.02 \\
$U_{+0}^{-,{\rm Im}}$ & 0.019    &   0.07 \\
$U_{+0}^{-,{\rm Re}}$ & 0.011    &   0.03 \\
$U_{+0}^{+,{\rm Im}}$ & 0.007   &   0.04 \\
$U_{+0}^{+,{\rm Re}}$ & 0.005   &   0.03 \\
$U_{+-}^{-,{\rm Im}}$ & 0.028    &   0.06 \\
$U_{+-}^{-,{\rm Re}}$ & 0.023    &   0.05 \\
$U_{+-}^{+,{\rm Im}}$ & 0.020    &   0.08 \\
$U_{+-}^{+,{\rm Re}}$ & 0.026    &   0.10   \\
$U_+^-$               & 0.0013   &   0.02 
\label{tab:lineshapesyst}
\end{tabular}
\end{center}
\end{table}

\subsection{Uniform Background Contributions}
\label{sec:nonresSystematic}

The systematic uncertainties associated with incoherent uniform background contributions in the DP are calculated in the same manner as for the $\rho(1700)$. The nominal fit with the uniform background component allocates only 69.4 events to the uniform component while the number of signal events decreases by 28.8 events or $0.3\sigma$. The changes in the $U$ and $I$ parameters across the fits to bootstrapped samples are not significant relative to the bootstrap-estimated uncertainties (see last column of Table~\ref{tab:nonresCov}). Also, as in the case of the $\rho(1700)$ systematic, the changes in the $U$ and $I$ parameters with and without the extra component are not significant relative to the statistical uncertainties on the $U$ and $I$ parameters (see middle column in Table~\ref{tab:nonresCov}).

\begin{table}
\caption{Magnitude of changes in $U$ and $I$ parameters between fits with and without the uniform background component, the ratio of these magnitudes relative to statistical uncertainties from the fit without the uniform component, the bootstrapped estimate of the uncertainty on changes in $U$ and $I$ parameters between fits with and without the uniform component, and the ratio of the mean change in the $U$ and $I$ parameters across all bootstrapped fits to their bootstrap-estimated uncertainties.}
\begin{center}
\begin{tabular}{ l | l c l c l l }
Parameter                       &       $|\Delta U|$ & $ |\Delta U| / \sigma_{\rm stat}\quad$ &       $\sigma_{\Delta U}^{\rm bs}$ &       $\left<\Delta U\right> / \sigma_{\Delta U}^{\rm bs}$ \\ \hline
$I_0$                 &         0.005      &       0.13      &         0.006    &        $-0.62$      \\
$I_-$                 &         0.006      &       0.10      &         0.007    &         ${\phantom -}0.55$       \\
$I_{-0}^{\rm Im}$     &         0.041       &       0.10     &         0.16      &         $-0.18$      \\
$I_{-0}^{\rm Re}$     &         0.09       &       0.16      &         0.25      &         ${\phantom -}0.34$       \\
$I_+$                 &         0.0013      &       0.02     &         0.006    &         ${\phantom -}0.22$       \\
$I_{+0}^{\rm Im}$     &         0.016       &       0.05     &         0.08      &         $-0.11$      \\
$I_{+0}^{\rm Re}$     &         0.029       &       0.06       &         0.11      &         $-0.07$     \\
$I_{+-}^{\rm Im}$     &         0.006       &       0.01    &         0.17      &         $-0.04$     \\
$I_{+-}^{\rm Re}$     &         0.07        &       0.09     &         0.21       &         $-0.15$      \\
$U_0^-$               &         0.000005        &       0.00    &         0.008    &         $-0.07$     \\
$U_0^+$               &         0.0036       &       0.12      &         0.0036    &         ${\phantom -}0.79$       \\
$U_{-0}^{-,{\rm Im}}$ &         0.14        &       0.33      &         0.18      &         $-0.58$      \\
$U_{-0}^{-,{\rm Re}}$ &         0.019       &       0.05     &         0.13      &         $-0.03$     \\
$U_{-0}^{+,{\rm Im}}$ &         0.06       &       0.31      &         0.06      &         ${\phantom -}0.64$       \\
$U_{-0}^{+,{\rm Re}}$ &         0.026       &       0.16      &         0.06     &         $-0.33$      \\
$U_-^-$               &         0.006       &       0.06     &         0.012     &         ${\phantom -}0.39$       \\
$U_-^+$               &         0.005        &       0.07     &         0.010    &         ${\phantom -}0.30$       \\
$U_{+0}^{-,{\rm Im}}$ &         0.0032      &       0.01     &         0.10     &        ${\phantom -}0.16$       \\
$U_{+0}^{-,{\rm Re}}$ &         0.031       &       0.10     &         0.11      &        ${\phantom -}0.27$       \\
$U_{+0}^{+,{\rm Im}}$ &         0.0026       &       0.02     &         0.041     &         $-0.08$      \\
$U_{+0}^{+,{\rm Re}}$ &         0.006      &       0.04     &         0.044     &         $-0.14$      \\
$U_{+-}^{-,{\rm Im}}$ &         0.045        &       0.09     &         0.16      &         $-0.38$      \\
$U_{+-}^{-,{\rm Re}}$ &         0.0032       &       0.01    &         0.18      &         $-0.05$     \\
$U_{+-}^{+,{\rm Im}}$ &         0.043       &       0.16      &         0.06     &         $-0.34$      \\
$U_{+-}^{+,{\rm Re}}$ &         0.05       &       0.21      &         0.13      &         $-0.23$      \\
$U_+^-$               &         0.007       &       0.07     &         0.012     &         $-0.47$      
\label{tab:nonresCov}
\end{tabular}
\end{center}
\end{table}

\subsection{Other Contributions}

In the 2007 {\babar} analysis~\cite{ref:matt2007}, several uncertainties were considered and found to provide only small contributions to the systematic uncertainty: 
\begin{itemize}
	\item{Uncertainties in $\Delta m_d$, the $B^0$ lifetime, each $\Delta t$ resolution parameter, tagging fractions, self-cross-feed fractions, $B$-background tagging fractions, and other minor systematics.}

	\item{Uncertainties on $\CP$ violation in the $B$ backgrounds (calculated by varying the parameters describing $\CP$ violation in each $B$ background class according to their uncertainties).}

	\item{Uncertainties caused by interference between $b \rightarrow c{\overline u}d$ and $b \rightarrow {\overline u} c {\overline d}$ on the tag side.}
	
\end{itemize}
These uncertainties are not expected to provide significantly different contributions in the current analysis. Therefore, these studies were not repeated, though their contributions to the systematic covariance matrix are included. The systematic uncertainties for these contributions, as calculated in the 2007 {\babar} analysis, are given in Table~\ref{tab:oldSystCov}.

\begin{table}
\caption{Systematic uncertainties for the three sources of systematic uncertainty not revisited for the present analysis. These values are taken from the 2007 {\babar} analysis and used without modification~\cite{ref:matt2007}. }
\begin{center}
\begin{tabular}{ l | c  c  c }
Parameter &	$B$ Bkgnd $\CP$	&      Tag Side Interference        &        ``Others'' \\ \hline
$I_0$                  		&       0.004	&	0.000	&	0.002	\\
$I_-$                  			&       0.013	&	0.003	&	0.007	\\
$I_{-0}^{\rm Im}$      		&       0.064	&	0.018	&	0.077	\\
$I_{-0}^{\rm Re}$      		&       0.083	&	0.006	&	0.059	\\
$I_+$                  		&       0.009	&	0.011	&	0.005	\\
$I_{+0}^{\rm Im}$      		&       0.050	&	0.003	&	0.065	\\
$I_{+0}^{\rm Re}$     		&       0.121	&	0.019	&	0.092	\\
$I_{+-}^{\rm Im}$      		&       0.168	&	0.033	&	0.133	\\
$I_{+-}^{\rm Re}$     		&       0.088	&	0.026	&	0.078	\\
$U_0^-$                		&       0.015	&	0.000	&	0.004	\\
$U_0^+$                		&       0.005	&	0.001	&	0.004	\\
$U_{-0}^{-,{\rm Im}}$  	&       0.052	&	0.007	&	0.016	\\
$U_{-0}^{-,{\rm Re}}$  	&       0.044	&	0.022	&	0.046	\\
$U_{-0}^{+,{\rm Im}}$  	&       0.038	&	0.022	&	0.011	\\
$U_{-0}^{+,{\rm Re}}$  	&       0.015	&	0.012	&	0.007	\\
$U_-^-$                		&       0.041	&	0.004	&	0.015	\\
$U_-^+$                		&       0.014	&	0.003	&	0.010	\\
$U_{+0}^{-,{\rm Im}}$  	&       0.073	&	0.028	&	0.038	\\
$U_{+0}^{-,{\rm Re}}$  	&       0.052	&	0.004	&	0.037	\\
$U_{+0}^{+,{\rm Im}}$  	&       0.042	&	0.001	&	0.032	\\
$U_{+0}^{+,{\rm Re}}$  	&       0.059	&	0.031	&	0.060	\\
$U_{+-}^{-,{\rm Im}}$  	&       0.055	&	0.031	&	0.045	\\
$U_{+-}^{-,{\rm Re}}$  	&       0.238	&	0.044	&	0.112	\\
$U_{+-}^{+,{\rm Im}}$  	&       0.028	&	0.031	&	0.012	\\
$U_{+-}^{+,{\rm Re}}$  	&       0.038	&	0.028	&	0.079	\\
$U_+^-$                		&       0.036	&	0.007	&	0.009	
\label{tab:oldSystCov}
\end{tabular}
\end{center}
\end{table}

\subsection{Total Systematic Uncertainties}

The square root of the diagonal elements of the total systematic covariance matrix for the $U$ and $I$ parameters are provided in Table~\ref{tab:finalSystCov}. These values include all systematic uncertainties described above, including the three sources of systematic uncertainty that were not recalculated in the present analysis. Table~\ref{tab:finalSystCov} also contains the ratio of the total systematic error to the statistical error from the final fit and the ratio of the total error (calculated by adding the systematic and statistical errors in quadrature) to the statistical error.

\begin{table}
\caption{The square root of the diagonal elements of the final systematic covariance matrix for the $U$ and $I$ parameters, the ratio of these total systematic errors to the statistical errors from the final fit, and the ratio of the total error (including statistical and systematic contributions) to the statistical error.}
\begin{center}
\begin{tabular}{ l | l c c }
Parameter & $\sigma_{\rm syst}$ &       $\sigma_{\rm syst}/\sigma_{\rm stat}$ & $\sqrt{\sigma_{\rm stat}^2 + \sigma_{\rm syst}^2}/\sigma_{\rm stat}$ \\ \hline
$I_0$                  &       0.022       &       0.59        &       1.16 \\
$I_-$                  &       0.028       &       0.45         &       1.10 \\
$I_{-0}^{\rm Im}$      &       0.5        &       1.08         &       1.47 \\
$I_{-0}^{\rm Re}$      &       0.8        &       1.34         &       1.67 \\
$I_+$                  &       0.029       &       0.46        &       1.10 \\
$I_{+0}^{\rm Im}$      &       0.43        &       1.21         &       1.57 \\
$I_{+0}^{\rm Re}$      &       0.7         &       1.57         &       1.86 \\
$I_{+-}^{\rm Im}$      &       0.9        &       1.26         &       1.61 \\
$I_{+-}^{\rm Re}$      &       1.0        &       1.30         &       1.64 \\
$U_0^-$                &       0.034       &       0.65        &       1.19 \\
$U_0^+$                &       0.020       &       0.68        &       1.21 \\
$U_{-0}^{-,{\rm Im}}$  &       0.5        &       1.19         &       1.56 \\
$U_{-0}^{-,{\rm Re}}$  &       0.37        &       1.05         &       1.45 \\
$U_{-0}^{+,{\rm Im}}$  &       0.24        &       1.16         &       1.53 \\
$U_{-0}^{+,{\rm Re}}$  &       0.18        &       1.10          &       1.49 \\
$U_-^-$                &       0.06        &       0.59        &       1.16 \\
$U_-^+$                &       0.05       &       0.72        &       1.23 \\
$U_{+0}^{-,{\rm Im}}$  &       0.5        &       1.59         &       1.88 \\
$U_{+0}^{-,{\rm Re}}$  &       0.38        &       1.19         &       1.55 \\
$U_{+0}^{+,{\rm Im}}$  &       0.17        &       1.09         &       1.48 \\
$U_{+0}^{+,{\rm Re}}$  &       0.19        &       1.26         &       1.61 \\
$U_{+-}^{-,{\rm Im}}$  &       0.8        &       1.68          &       1.95 \\
$U_{+-}^{-,{\rm Re}}$  &       0.8        &       1.70         &       1.98 \\
$U_{+-}^{+,{\rm Im}}$  &       0.26        &       1.02         &       1.43 \\
$U_{+-}^{+,{\rm Re}}$  &       0.33         &       1.29         &       1.63 \\
$U_+^-$                &       0.07       &       0.77        &       1.26 
\label{tab:finalSystCov}
\end{tabular}
\end{center}
\end{table}

Tables~\ref{tab:finalSystCovI}--\ref{tab:finalSystCovIII} contain the complete systematic correlation matrix for the $U$ and $I$ parameters. Because the three additional sources of systematic uncertainty taken from our previous analysis have small contributions, and only the diagonal elements of their covariance matrix are available to us, we only use the diagonal elements when creating the total systematic covariance matrix that is used, in turn, to generate the total correlation matrix.

\begin{table*} 
\caption{Systematic correlation matrix for the $U$ and $I$ parameters, including contributions from three sources of minor systematic uncertainties evaluated in the 2007 {\babar} analysis. Elements above the diagonal are redundant and not included in the table. (Continued in Tables \ref{tab:finalSystCovII} and \ref{tab:finalSystCovIII}.)}
\begin{center}
\begin{tabular}{ l | l  l l l l l l l}
                      & $I_{0}           $   & $I_{-}          $   & $I^{\rm Im}_{-0}$   & $I^{\rm Re}_{-0}$   & $I_{+}          $   & $I^{\rm Im}_{+0}$   & $I^{\rm Re}_{+0}$   & $I^{\rm Im}_{+-}$   \\ \hline
$I_{0}              $ &  ${\phantom -} 1.00$ &                     &                     &                     &                     &                     &                     &                     \\       
$I_{-}              $ &  $            -0.17$ & ${\phantom -} 1.00$ &                     &                     &                     &                     &                     &                     \\       
$I^{\rm Im}_{-0}    $ &  ${\phantom -} 0.13$ & $            -0.13$ & ${\phantom -} 1.00$ &                     &                     &                     &                     &                     \\       
$I^{\rm Re}_{-0}    $ &  $            -0.07$ & $            -0.00$ & $            -0.03$ & ${\phantom -} 1.00$ &                     &                     &                     &                     \\       
$I_{+}              $ &  $            -0.20$ & $            -0.13$ & $            -0.11$ & $            -0.01$ & ${\phantom -} 1.00$ &                     &                     &                     \\       
$I^{\rm Im}_{+0}    $ &  ${\phantom -} 0.04$ & $            -0.09$ & ${\phantom -} 0.06$ & ${\phantom -} 0.02$ & ${\phantom -} 0.00$ & ${\phantom -} 1.00$ &                     &                     \\       
$I^{\rm Re}_{+0}    $ &  ${\phantom -} 0.21$ & $            -0.05$ & ${\phantom -} 0.04$ & ${\phantom -} 0.08$ & $            -0.14$ & ${\phantom -} 0.16$ & ${\phantom -} 1.00$ &                     \\       
$I^{\rm Im}_{+-}    $ &  ${\phantom -} 0.13$ & $            -0.00$ & ${\phantom -} 0.09$ & ${\phantom -} 0.09$ & $            -0.08$ & ${\phantom -} 0.15$ & ${\phantom -} 0.26$ & ${\phantom -} 1.00$ \\ \hline
$I^{\rm Re}_{+-}    $ &  ${\phantom -} 0.06$ & ${\phantom -} 0.08$ & ${\phantom -} 0.20$ & ${\phantom -} 0.01$ & $            -0.31$ & $            -0.07$ & ${\phantom -} 0.16$ & ${\phantom -} 0.20$ \\       
$U^{-}_{0}          $ &  ${\phantom -} 0.09$ & ${\phantom -} 0.04$ & $            -0.13$ & ${\phantom -} 0.08$ & ${\phantom -} 0.10$ & $            -0.04$ & $            -0.09$ & $            -0.08$ \\       
$U^{+}_{0}          $ &  $            -0.11$ & ${\phantom -} 0.10$ & $            -0.07$ & ${\phantom -} 0.08$ & ${\phantom -} 0.09$ & $            -0.08$ & $            -0.05$ & $            -0.12$ \\       
$U^{-,{\rm Im}}_{-0}$ &  ${\phantom -} 0.08$ & $            -0.21$ & ${\phantom -} 0.10$ & $            -0.31$ & ${\phantom -} 0.01$ & ${\phantom -} 0.05$ & ${\phantom -} 0.01$ & $            -0.11$ \\       
$U^{-,{\rm Re}}_{-0}$ &  $            -0.03$ & $            -0.01$ & $            -0.41$ & ${\phantom -} 0.13$ & ${\phantom -} 0.09$ & $            -0.01$ & ${\phantom -} 0.03$ & $            -0.16$ \\       
$U^{+,{\rm Im}}_{-0}$ &  $            -0.18$ & ${\phantom -} 0.21$ & $            -0.04$ & ${\phantom -} 0.18$ & ${\phantom -} 0.02$ & $            -0.17$ & $            -0.11$ & $            -0.00$ \\       
$U^{+,{\rm Re}}_{-0}$ &  ${\phantom -} 0.19$ & $            -0.15$ & ${\phantom -} 0.32$ & $            -0.29$ & $            -0.09$ & ${\phantom -} 0.00$ & $            -0.03$ & ${\phantom -} 0.15$ \\       
$U^{-}_{-}          $ &  $            -0.12$ & ${\phantom -} 0.04$ & $            -0.08$ & ${\phantom -} 0.15$ & ${\phantom -} 0.05$ & $            -0.02$ & ${\phantom -} 0.11$ & $            -0.05$ \\       
$U^{+}_{-}          $ &  ${\phantom -} 0.01$ & $            -0.08$ & $            -0.23$ & ${\phantom -} 0.06$ & ${\phantom -} 0.24$ & ${\phantom -} 0.27$ & ${\phantom -} 0.09$ & $            -0.06$ \\ \hline
$U^{-,{\rm Im}}_{+0}$ &  $            -0.00$ & ${\phantom -} 0.07$ & $            -0.11$ & ${\phantom -} 0.21$ & ${\phantom -} 0.01$ & ${\phantom -} 0.04$ & ${\phantom -} 0.45$ & ${\phantom -} 0.14$ \\       
$U^{-,{\rm Re}}_{+0}$ &  $            -0.07$ & ${\phantom -} 0.11$ & $            -0.12$ & ${\phantom -} 0.12$ & $            -0.09$ & $            -0.17$ & ${\phantom -} 0.13$ & $            -0.11$ \\       
$U^{+,{\rm Im}}_{+0}$ &  ${\phantom -} 0.02$ & ${\phantom -} 0.04$ & $            -0.08$ & ${\phantom -} 0.11$ & $            -0.02$ & ${\phantom -} 0.01$ & ${\phantom -} 0.18$ & ${\phantom -} 0.09$ \\       
$U^{+,{\rm Re}}_{+0}$ &  ${\phantom -} 0.06$ & $            -0.01$ & ${\phantom -} 0.08$ & $            -0.02$ & $            -0.07$ & ${\phantom -} 0.11$ & $            -0.07$ & ${\phantom -} 0.08$ \\       
$U^{-,{\rm Im}}_{+-}$ &  ${\phantom -} 0.04$ & $            -0.01$ & ${\phantom -} 0.19$ & ${\phantom -} 0.04$ & $            -0.13$ & $            -0.28$ & ${\phantom -} 0.02$ & ${\phantom -} 0.19$ \\       
$U^{-,{\rm Re}}_{+-}$ &  $            -0.10$ & ${\phantom -} 0.07$ & ${\phantom -} 0.11$ & $            -0.04$ & $            -0.15$ & $            -0.25$ & $            -0.37$ & $            -0.30$ \\       
$U^{+,{\rm Im}}_{+-}$ &  ${\phantom -} 0.05$ & $            -0.08$ & ${\phantom -} 0.11$ & $            -0.20$ & $            -0.07$ & ${\phantom -} 0.08$ & $            -0.22$ & ${\phantom -} 0.05$ \\       
$U^{+,{\rm Re}}_{+-}$ &  ${\phantom -} 0.24$ & $            -0.11$ & ${\phantom -} 0.33$ & $            -0.26$ & $            -0.17$ & ${\phantom -} 0.15$ & ${\phantom -} 0.07$ & ${\phantom -} 0.16$ \\       
$U^{-}_{+}          $ &  ${\phantom -} 0.09$ & $            -0.04$ & ${\phantom -} 0.19$ & $            -0.07$ & $            -0.23$ & $            -0.15$ & $            -0.25$ & ${\phantom -} 0.03$ \\ \hline
\label{tab:finalSystCovI}
\end{tabular}
\end{center}
\end{table*}

\begin{table*}
\caption{(Continued from Table~\ref{tab:finalSystCovI}) Systematic correlation matrix for the $U$ and $I$ parameters, including contributions from three sources of minor systematic uncertainties evaluated in the 2007 {\babar} analysis. Elements above the diagonal are redundant and not included in the table. (Continued in Table \ref{tab:finalSystCovIII}.)}
\begin{center}
\begin{tabular}{ l | l  l l l l l l l l}
                      & $I^{\rm Re}_{+-}$   & $U^{-}_{0}      $   & $U^{+}_{0}      $   &$U^{-,{\rm Im}}_{-0}$&$U^{-,{\rm Re}}_{-0}$&$U^{+,{\rm Im}}_{-0}$&$U^{+,{\rm Re}}_{-0}$& $U^{-}_{-}      $   & $U^{+}_{-}      $   \\ \hline
$I^{\rm Re}_{+-}    $ & ${\phantom -} 1.00$ &                     &                     &                     &                     &                     &                     &                     &                     \\       
$U^{-}_{0}          $ & $            -0.22$ & ${\phantom -} 1.00$ &                     &                     &                     &                     &                     &                     &                     \\       
$U^{+}_{0}          $ & ${\phantom -} 0.07$ & ${\phantom -} 0.07$ & ${\phantom -} 1.00$ &                     &                     &                     &                     &                     &                     \\       
$U^{-,{\rm Im}}_{-0}$ & ${\phantom -} 0.05$ & $            -0.14$ & $            -0.17$ & ${\phantom -} 1.00$ &                     &                     &                     &                     &                     \\       
$U^{-,{\rm Re}}_{-0}$ & $            -0.18$ & ${\phantom -} 0.11$ & $            -0.04$ & $            -0.09$ & ${\phantom -} 1.00$ &                     &                     &                     &                     \\       
$U^{+,{\rm Im}}_{-0}$ & ${\phantom -} 0.02$ & ${\phantom -} 0.08$ & ${\phantom -} 0.35$ & $            -0.49$ & ${\phantom -} 0.04$ & ${\phantom -} 1.00$ &                     &                     &                     \\       
$U^{+,{\rm Re}}_{-0}$ & ${\phantom -} 0.36$ & $            -0.15$ & $            -0.03$ & ${\phantom -} 0.26$ & $            -0.34$ & $            -0.14$ & ${\phantom -} 1.00$ &                     &                     \\       
$U^{-}_{-}          $ & ${\phantom -} 0.01$ & $            -0.07$ & ${\phantom -} 0.04$ & $            -0.05$ & ${\phantom -} 0.07$ & ${\phantom -} 0.05$ & $            -0.22$ & ${\phantom -} 1.00$ &                     \\       
$U^{+}_{-}          $ & $            -0.58$ & ${\phantom -} 0.24$ & ${\phantom -} 0.01$ & $            -0.13$ & ${\phantom -} 0.24$ & $            -0.04$ & $            -0.40$ & ${\phantom -} 0.05$ & ${\phantom -} 1.00$ \\ \hline
$U^{-,{\rm Im}}_{+0}$ & ${\phantom -} 0.05$ & $            -0.14$ & ${\phantom -} 0.02$ & $            -0.12$ & ${\phantom -} 0.09$ & ${\phantom -} 0.03$ & $            -0.20$ & ${\phantom -} 0.18$ & ${\phantom -} 0.12$ \\       
$U^{-,{\rm Re}}_{+0}$ & $            -0.07$ & $            -0.04$ & ${\phantom -} 0.03$ & $            -0.12$ & ${\phantom -} 0.12$ & ${\phantom -} 0.10$ & $            -0.27$ & ${\phantom -} 0.16$ & ${\phantom -} 0.04$ \\       
$U^{+,{\rm Im}}_{+0}$ & $            -0.02$ & ${\phantom -} 0.11$ & $            -0.18$ & $            -0.12$ & ${\phantom -} 0.16$ & $            -0.00$ & $            -0.07$ & ${\phantom -} 0.02$ & $            -0.00$ \\       
$U^{+,{\rm Re}}_{+0}$ & $            -0.07$ & $            -0.12$ & $            -0.26$ & ${\phantom -} 0.06$ & $            -0.12$ & $            -0.21$ & $            -0.01$ & $            -0.02$ & $            -0.02$ \\       
$U^{-,{\rm Im}}_{+-}$ & ${\phantom -} 0.45$ & $            -0.14$ & ${\phantom -} 0.06$ & ${\phantom -} 0.08$ & $            -0.18$ & ${\phantom -} 0.08$ & ${\phantom -} 0.35$ & $            -0.13$ & $            -0.57$ \\       
$U^{-,{\rm Re}}_{+-}$ & ${\phantom -} 0.27$ & $            -0.06$ & ${\phantom -} 0.16$ & ${\phantom -} 0.08$ & $            -0.11$ & ${\phantom -} 0.12$ & ${\phantom -} 0.17$ & $            -0.06$ & $            -0.48$ \\       
$U^{+,{\rm Im}}_{+-}$ & ${\phantom -} 0.02$ & ${\phantom -} 0.02$ & $            -0.26$ & ${\phantom -} 0.21$ & $            -0.16$ & $            -0.27$ & ${\phantom -} 0.23$ & $            -0.20$ & $            -0.17$ \\       
$U^{+,{\rm Re}}_{+-}$ & ${\phantom -} 0.33$ & $            -0.15$ & $            -0.13$ & ${\phantom -} 0.34$ & $            -0.34$ & $            -0.29$ & ${\phantom -} 0.48$ & $            -0.19$ & $            -0.29$ \\       
$U^{-}_{+}          $ & ${\phantom -} 0.28$ & $            -0.14$ & $            -0.06$ & ${\phantom -} 0.20$ & $            -0.20$ & $            -0.02$ & ${\phantom -} 0.30$ & $            -0.23$ & $            -0.42$ \\ \hline
\label{tab:finalSystCovII}
\end{tabular}
\end{center}
\end{table*}

\begin{table*}
\caption{(Continued from Table~\ref{tab:finalSystCovII}) Systematic correlation matrix for the $U$ and $I$ parameters, including contributions from three sources of minor systematic uncertainties evaluated in the 2007 {\babar} analysis. Elements above the diagonal are redundant and not included in the table.}
\begin{center}
\begin{tabular}{ l | l  l l l l l l l l}
                      &$U^{-,{\rm Im}}_{+0}$&$U^{-,{\rm Re}}_{+0}$&$U^{+,{\rm Im}}_{+0}$&$U^{+,{\rm Re}}_{+0}$&$U^{-,{\rm Im}}_{+-}$&$U^{-,{\rm Re}}_{+-}$&$U^{+,{\rm Im}}_{+-}$&$U^{+,{\rm Re}}_{+-}$& $U^{-}_{+}$           \\ \hline
$U^{-,{\rm Im}}_{+0}$ & ${\phantom -} 1.00$ &                     &                     &                     &                     &                     &                     &                     &                       \\       
$U^{-,{\rm Re}}_{+0}$ & ${\phantom -} 0.15$ & ${\phantom -} 1.00$ &                     &                     &                     &                     &                     &                     &                       \\       
$U^{+,{\rm Im}}_{+0}$ & ${\phantom -} 0.06$ & $            -0.01$ & ${\phantom -} 1.00$ &                     &                     &                     &                     &                     &                       \\       
$U^{+,{\rm Re}}_{+0}$ & $            -0.04$ & $            -0.03$ & $            -0.10$ & ${\phantom -} 1.00$ &                     &                     &                     &                     &                       \\       
$U^{-,{\rm Im}}_{+-}$ & $            -0.08$ & $            -0.08$ & ${\phantom -} 0.07$ & $            -0.04$ & ${\phantom -} 1.00$ &                     &                     &                     &                       \\       
$U^{-,{\rm Re}}_{+-}$ & $            -0.25$ & ${\phantom -} 0.05$ & $            -0.05$ & $            -0.07$ & ${\phantom -} 0.18$ & ${\phantom -} 1.00$ &                     &                     &                       \\       
$U^{+,{\rm Im}}_{+-}$ & $            -0.29$ & $            -0.19$ & $            -0.01$ & ${\phantom -} 0.05$ & ${\phantom -} 0.02$ & ${\phantom -} 0.15$ & ${\phantom -} 1.00$ &                     &                       \\       
$U^{+,{\rm Re}}_{+-}$ & $            -0.16$ & $            -0.21$ & $            -0.13$ & ${\phantom -} 0.14$ & ${\phantom -} 0.16$ & ${\phantom -} 0.09$ & ${\phantom -} 0.28$ & ${\phantom -} 1.00$ &                       \\       
$U^{-}_{+}          $ & $            -0.19$ & $            -0.16$ & $            -0.05$ & ${\phantom -} 0.01$ & ${\phantom -} 0.42$ & ${\phantom -} 0.28$ & ${\phantom -} 0.24$ & ${\phantom -} 0.28$ & ${\phantom -} 1.00$  \\ \hline
\label{tab:finalSystCovIII}
\end{tabular}
\end{center}
\end{table*}

\section{RESULTS}
\label{sec:resultssec}

\begin{figure*}[!htb]
\begin{center}
\includegraphics[clip=true,trim=25 195 25 0,width=0.99\textwidth]{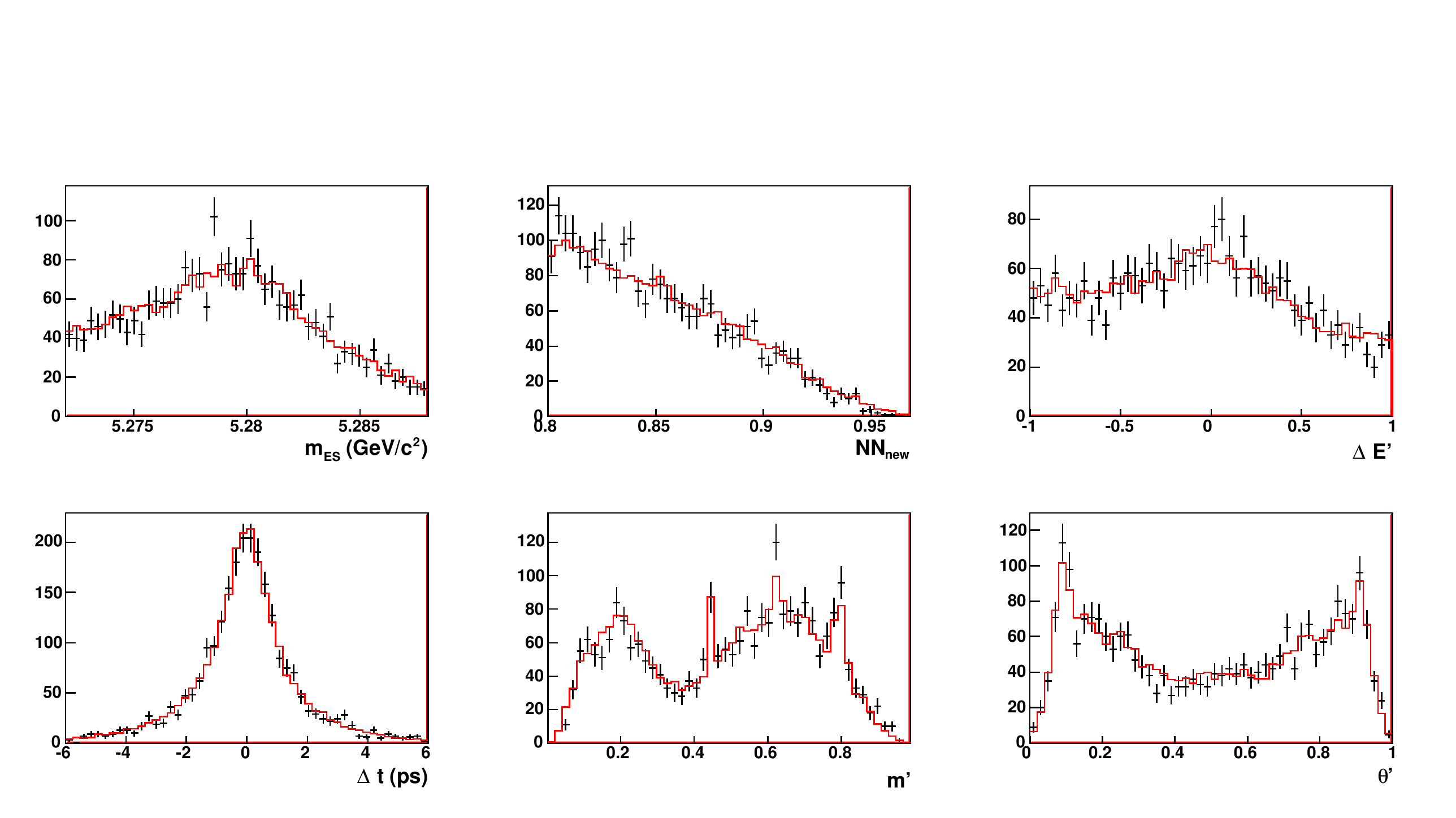}
\includegraphics[clip=true,trim=25 370 25 80,width=0.99\textwidth]{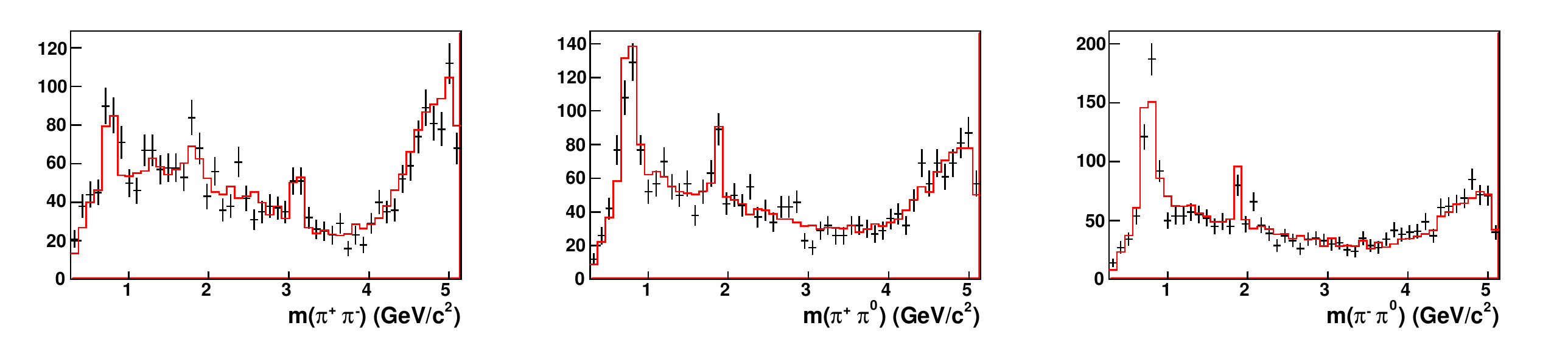}
\caption{(color online) Overlay plots of fit variable distributions in on-resonance data (points with error bars) and in a parameterized MC sample generated from the final fit results (red line) with 10 times the number of events in data. The MC histograms are scaled to have the same integral as the data histograms onto which they are overlaid. A restrictive selection criterion is applied to the NN output (${\tt NN}_{\rm new} > 0.8$) to enhance the signal component. }
\label{fig:overlayFigureEnhanced}
\end{center}
\end{figure*}

The final values of the $U$ and $I$ parameters extracted from the extended maximum likelihood fit to the full on-resonance dataset are provided in Table~\ref{tab:run16finalUI}. Tables~\ref{tab:finalStatCovI}--\ref{tab:finalStatCovIII} present the statistical correlation matrix for the $U$ and $I$ parameters in the fit. From an on-resonance dataset containing 53,084 events, the fit extracts 2,940$\pm100$ signal events and 46,750$ \pm 220$ continuum events. The goodness of fit is illustrated in Fig.~\ref{fig:overlayFigureEnhanced}, which shows overlaid distributions of fit variables in the data used in the final fit and in a parameterized MC sample generated using the results of the final fit and equivalent to 10 times the integrated luminosity of the data sample. The signal component of these plots is enhanced by a restrictive selection criterion on the NN variable. A study of the $U$ and $I$ parameters (see Appendix~\ref{sec:UIrobustness}) establishes that there is negligible bias in their extraction and good robustness in the presence of statistical fluctuations.

\begin{table}
\caption{$U$ and $I$ parameter values from the final fit to the complete on-resonance dataset, where the first error is statistical and the second is systematic.}
\begin{center}
\begin{tabular}{  l | l}
Parameter & Final Fit Value \\ \hline
$I_{0}              $  &       $           -0.042        \pm 0.038   \pm 0.022   $ \\
$I_{-}              $  &       $           -0.00         \pm 0.06    \pm 0.03    $ \\
$I^{\rm Im}_{-0}    $  &       $           -0.61         \pm 0.43    \pm 0.46    $ \\
$I^{\rm Re}_{-0}    $  &       ${\phantom -}0.4          \pm 0.6     \pm 0.8     $ \\
$I_{+}              $  &       ${\phantom -}0.05         \pm 0.06    \pm 0.03    $ \\
$I^{\rm Im}_{+0}    $  &       $           -0.04         \pm 0.36    \pm 0.43    $ \\
$I^{\rm Re}_{+0}    $  &       ${\phantom -}0.5          \pm 0.5     \pm 0.7     $ \\
$I^{\rm Im}_{+-}    $  &       $           -0.5          \pm 0.7     \pm 0.9     $ \\
$I^{\rm Re}_{+-}    $  &       $           -0.6          \pm 0.8     \pm 1.0     $ \\
$U^{-}_{0}          $  &       ${\phantom -}0.04         \pm 0.05    \pm 0.03    $ \\
$U^{+}_{0}          $  &       ${\phantom -}0.225        \pm 0.030   \pm 0.020   $ \\
$U^{-,{\rm Im}}_{-0}$  &       ${\phantom -}0.53         \pm 0.44    \pm 0.52    $ \\
$U^{-,{\rm Re}}_{-0}$  &       ${\phantom -}0.49         \pm 0.35    \pm 0.37    $ \\
$U^{+,{\rm Im}}_{-0}$  &       $           -0.39         \pm 0.20    \pm 0.24    $ \\
$U^{+,{\rm Re}}_{-0}$  &       $           -0.05         \pm 0.17    \pm 0.18    $ \\
$U^{-}_{-}          $  &       $           -0.27         \pm 0.10    \pm 0.06    $ \\
$U^{+}_{-}          $  &       ${\phantom -}1.22         \pm 0.07    \pm 0.05    $ \\
$U^{-,{\rm Im}}_{+0}$  &       ${\phantom -}0.10         \pm 0.29    \pm 0.45    $ \\
$U^{-,{\rm Re}}_{+0}$  &       ${\phantom -}0.30         \pm 0.32    \pm 0.38    $ \\
$U^{+,{\rm Im}}_{+0}$  &       ${\phantom -}0.41         \pm 0.16    \pm 0.17    $ \\
$U^{+,{\rm Re}}_{+0}$  &       ${\phantom -}0.01         \pm 0.15    \pm 0.19    $ \\
$U^{-,{\rm Im}}_{+-}$  &       ${\phantom -}1.1          \pm 0.5     \pm 0.8     $ \\
$U^{-,{\rm Re}}_{+-}$  &       $           -0.5          \pm 0.5     \pm 0.8     $ \\
$U^{+,{\rm Im}}_{+-}$  &       $           -0.07         \pm 0.26    \pm 0.26    $ \\
$U^{+,{\rm Re}}_{+-}$  &       $           -0.19         \pm 0.25    \pm 0.33    $ \\
$U^{-}_{+}          $  &       ${\phantom -}0.25         \pm 0.09    \pm 0.07    $ 
\label{tab:run16finalUI}
\end{tabular}
\end{center}
\end{table}

\begin{table*}
\caption{Statistical correlation matrix for the $U$ and $I$ parameters. Elements above the diagonal are redundant and not included in the table. (Continued in Tables \ref{tab:finalStatCovII} and \ref{tab:finalStatCovIII}.)}
\begin{center}
\begin{tabular}{ l | l  l l l l l l l}
                      & $I_{0}              $& $I_{-}          $   & $I^{\rm Im}_{-0}$   & $I^{\rm Re}_{-0}$   & $I_{+}          $   & $I^{\rm Im}_{+0}$   & $I^{\rm Re}_{+0} $  & $I^{\rm Im}_{+-} $   \\ \hline
$I_{0}              $ &  ${\phantom -} 1.00$ &                     &                     &                     &                     &                     &                     &                      \\       
$I_{-}              $ &  $            -0.04$ & ${\phantom -} 1.00$ &                     &                     &                     &                     &                     &                      \\       
$I^{\rm Im}_{-0}    $ &  ${\phantom -} 0.00$ & $            -0.06$ & ${\phantom -} 1.00$ &                     &                     &                     &                     &                      \\       
$I^{\rm Re}_{-0}    $ &  $            -0.05$ & ${\phantom -} 0.05$ & ${\phantom -} 0.05$ & ${\phantom -} 1.00$ &                     &                     &                     &                      \\       
$I_{+}              $ &  $            -0.02$ & $            -0.01$ & ${\phantom -} 0.00$ & $            -0.00$ & ${\phantom -} 1.00$ &                     &                     &                      \\       
$I^{\rm Im}_{+0}    $ &  $            -0.03$ & $            -0.01$ & ${\phantom -} 0.01$ & $            -0.01$ & ${\phantom -} 0.05$ & ${\phantom -} 1.00$ &                     &                      \\       
$I^{\rm Re}_{+0}    $ &  $            -0.12$ & $            -0.02$ & ${\phantom -} 0.02$ & $            -0.01$ & $            -0.09$ & ${\phantom -} 0.17$ & ${\phantom -} 1.00$ &                      \\       
$I^{\rm Im}_{+-}    $ &  ${\phantom -} 0.01$ & $            -0.06$ & ${\phantom -} 0.01$ & $            -0.01$ & $            -0.04$ & ${\phantom -} 0.01$ & ${\phantom -} 0.01$ & ${\phantom -} 1.00$  \\ \hline
$I^{\rm Re}_{+-}    $ &  ${\phantom -} 0.02$ & $            -0.08$ & ${\phantom -} 0.02$ & $            -0.04$ & $            -0.01$ & ${\phantom -} 0.03$ & ${\phantom -} 0.06$ & ${\phantom -} 0.04$  \\       
$U^{-}_{0}          $ &  $            -0.04$ & ${\phantom -} 0.01$ & $            -0.02$ & ${\phantom -} 0.10$ & ${\phantom -} 0.00$ & $            -0.05$ & $            -0.04$ & $            -0.00$  \\       
$U^{+}_{0}          $ &  $            -0.08$ & ${\phantom -} 0.01$ & $            -0.02$ & ${\phantom -} 0.01$ & $            -0.00$ & ${\phantom -} 0.03$ & ${\phantom -} 0.02$ & $            -0.00$  \\       
$U^{-,{\rm Im}}_{-0}$ &  $            -0.05$ & $            -0.06$ & ${\phantom -} 0.01$ & ${\phantom -} 0.06$ & $            -0.01$ & ${\phantom -} 0.05$ & ${\phantom -} 0.09$ & ${\phantom -} 0.01$  \\       
$U^{-,{\rm Re}}_{-0}$ &  $            -0.03$ & ${\phantom -} 0.02$ & $            -0.17$ & $            -0.01$ & $            -0.00$ & ${\phantom -} 0.00$ & ${\phantom -} 0.00$ & $            -0.01$  \\       
$U^{+,{\rm Im}}_{-0}$ &  ${\phantom -} 0.00$ & ${\phantom -} 0.03$ & ${\phantom -} 0.05$ & $            -0.05$ & ${\phantom -} 0.01$ & $            -0.05$ & $            -0.09$ & $            -0.01$  \\       
$U^{+,{\rm Re}}_{-0}$ &  ${\phantom -} 0.03$ & $            -0.07$ & ${\phantom -} 0.17$ & $            -0.11$ & $            -0.00$ & ${\phantom -} 0.03$ & ${\phantom -} 0.04$ & ${\phantom -} 0.01$  \\       
$U^{-}_{-}          $ &  $            -0.00$ & $            -0.02$ & ${\phantom -} 0.03$ & ${\phantom -} 0.02$ & ${\phantom -} 0.00$ & $            -0.01$ & $            -0.02$ & $            -0.00$  \\       
$U^{+}_{-}          $ &  $            -0.03$ & ${\phantom -} 0.02$ & $            -0.07$ & ${\phantom -} 0.06$ & $            -0.01$ & ${\phantom -} 0.01$ & $            -0.01$ & ${\phantom -} 0.00$  \\ \hline
$U^{-,{\rm Im}}_{+0}$ &  ${\phantom -} 0.01$ & $            -0.00$ & $            -0.00$ & ${\phantom -} 0.01$ & ${\phantom -} 0.01$ & ${\phantom -} 0.05$ & ${\phantom -} 0.26$ & ${\phantom -} 0.00$  \\       
$U^{-,{\rm Re}}_{+0}$ &  $            -0.08$ & ${\phantom -} 0.01$ & $            -0.01$ & ${\phantom -} 0.01$ & $            -0.06$ & $            -0.06$ & $            -0.02$ & $            -0.00$  \\       
$U^{+,{\rm Im}}_{+0}$ &  $            -0.01$ & $            -0.01$ & $            -0.00$ & ${\phantom -} 0.01$ & $            -0.05$ & ${\phantom -} 0.06$ & $            -0.01$ & $            -0.00$  \\       
$U^{+,{\rm Re}}_{+0}$ &  ${\phantom -} 0.05$ & $            -0.00$ & $            -0.00$ & ${\phantom -} 0.01$ & $            -0.03$ & ${\phantom -} 0.09$ & $            -0.09$ & $            -0.00$  \\       
$U^{-,{\rm Im}}_{+-}$ &  ${\phantom -} 0.01$ & $            -0.04$ & ${\phantom -} 0.00$ & ${\phantom -} 0.01$ & $            -0.04$ & ${\phantom -} 0.01$ & ${\phantom -} 0.03$ & ${\phantom -} 0.04$  \\       
$U^{-,{\rm Re}}_{+-}$ &  ${\phantom -} 0.01$ & $            -0.01$ & ${\phantom -} 0.00$ & $            -0.01$ & $            -0.00$ & ${\phantom -} 0.01$ & ${\phantom -} 0.01$ & $            -0.11$  \\       
$U^{+,{\rm Im}}_{+-}$ &  ${\phantom -} 0.02$ & $            -0.02$ & ${\phantom -} 0.02$ & $            -0.03$ & $            -0.01$ & ${\phantom -} 0.04$ & ${\phantom -} 0.06$ & $            -0.04$  \\       
$U^{+,{\rm Re}}_{+-}$ &  ${\phantom -} 0.02$ & $            -0.04$ & ${\phantom -} 0.02$ & $            -0.04$ & $            -0.03$ & ${\phantom -} 0.04$ & ${\phantom -} 0.08$ & ${\phantom -} 0.08$  \\       
$U^{-}_{+}          $ &  ${\phantom -} 0.01$ & $            -0.01$ & ${\phantom -} 0.00$ & $            -0.01$ & $            -0.05$ & $            -0.03$ & $            -0.04$ & $            -0.00$  \\ \hline
\label{tab:finalStatCovI}
\end{tabular}
\end{center}
\end{table*}

\begin{table*}
\caption{(Continued from Table~\ref{tab:finalStatCovI}) Statistical correlation matrix for the $U$ and $I$ parameters. Elements above the diagonal are redundant and not included in the table. (Continued in Table \ref{tab:finalStatCovIII}.)}
\begin{center}
\begin{tabular}{ l | l  l l l l l l l l}
                      & $I^{\rm Re}_{+-} $  & $U^{-}_{0}      $   & $U^{+}_{0}      $   &$U^{-,{\rm Im}}_{-0}$&$U^{-,{\rm Re}}_{-0}$&$U^{+,{\rm Im}}_{-0}$&$U^{+,{\rm Re}}_{-0}$&$U^{-}_{-}       $   &$U^{+}_{-}       $    \\ \hline
$I^{\rm Re}_{+-}    $ & ${\phantom -} 1.00$ &                     &                     &                     &                     &                     &                     &                     &                      \\       
$U^{-}_{0}          $ & $            -0.03$ & ${\phantom -} 1.00$ &                     &                     &                     &                     &                     &                     &                      \\       
$U^{+}_{0}          $ & $            -0.03$ & ${\phantom -} 0.07$ & ${\phantom -} 1.00$ &                     &                     &                     &                     &                     &                      \\       
$U^{-,{\rm Im}}_{-0}$ & ${\phantom -} 0.08$ & ${\phantom -} 0.02$ & ${\phantom -} 0.00$ & ${\phantom -} 1.00$ &                     &                     &                     &                     &                      \\       
$U^{-,{\rm Re}}_{-0}$ & $            -0.02$ & $            -0.05$ & ${\phantom -} 0.02$ & ${\phantom -} 0.09$ & ${\phantom -} 1.00$ &                     &                     &                     &                      \\       
$U^{+,{\rm Im}}_{-0}$ & $            -0.09$ & ${\phantom -} 0.02$ & ${\phantom -} 0.05$ & $            -0.26$ & $            -0.04$ & ${\phantom -} 1.00$ &                     &                     &                      \\       
$U^{+,{\rm Re}}_{-0}$ & ${\phantom -} 0.06$ & $            -0.04$ & $            -0.07$ & ${\phantom -} 0.06$ & ${\phantom -} 0.01$ & $            -0.03$ & ${\phantom -} 1.00$ &                     &                      \\       
$U^{-}_{-}          $ & ${\phantom -} 0.01$ & ${\phantom -} 0.00$ & $            -0.03$ & $            -0.08$ & $            -0.07$ & ${\phantom -} 0.06$ & $            -0.02$ & ${\phantom -} 1.00$ &                      \\       
$U^{+}_{-}          $ & $            -0.05$ & ${\phantom -} 0.05$ & ${\phantom -} 0.26$ & ${\phantom -} 0.02$ & ${\phantom -} 0.05$ & $            -0.05$ & $            -0.05$ & $            -0.12$ & ${\phantom -} 1.00$  \\ \hline
$U^{-,{\rm Im}}_{+0}$ & ${\phantom -} 0.01$ & ${\phantom -} 0.13$ & ${\phantom -} 0.05$ & ${\phantom -} 0.02$ & $            -0.01$ & $            -0.01$ & $            -0.00$ & $            -0.01$ & ${\phantom -} 0.01$  \\       
$U^{-,{\rm Re}}_{+0}$ & $            -0.02$ & $            -0.10$ & ${\phantom -} 0.04$ & $            -0.03$ & ${\phantom -} 0.02$ & ${\phantom -} 0.03$ & $            -0.02$ & ${\phantom -} 0.01$ & ${\phantom -} 0.03$  \\       
$U^{+,{\rm Im}}_{+0}$ & $            -0.01$ & ${\phantom -} 0.04$ & ${\phantom -} 0.15$ & ${\phantom -} 0.03$ & ${\phantom -} 0.02$ & $            -0.01$ & $            -0.00$ & $            -0.01$ & ${\phantom -} 0.07$  \\       
$U^{+,{\rm Re}}_{+0}$ & $            -0.01$ & ${\phantom -} 0.02$ & $            -0.09$ & $            -0.00$ & ${\phantom -} 0.01$ & $            -0.00$ & ${\phantom -} 0.01$ & $            -0.00$ & ${\phantom -} 0.02$  \\       
$U^{-,{\rm Im}}_{+-}$ & ${\phantom -} 0.05$ & ${\phantom -} 0.00$ & ${\phantom -} 0.02$ & ${\phantom -} 0.06$ & ${\phantom -} 0.03$ & $            -0.05$ & ${\phantom -} 0.01$ & ${\phantom -} 0.01$ & ${\phantom -} 0.06$  \\       
$U^{-,{\rm Re}}_{+-}$ & $            -0.04$ & $            -0.01$ & $            -0.01$ & ${\phantom -} 0.02$ & ${\phantom -} 0.00$ & $            -0.02$ & ${\phantom -} 0.01$ & $            -0.08$ & $            -0.04$  \\       
$U^{+,{\rm Im}}_{+-}$ & ${\phantom -} 0.01$ & $            -0.03$ & $            -0.01$ & ${\phantom -} 0.12$ & ${\phantom -} 0.01$ & $            -0.12$ & ${\phantom -} 0.06$ & ${\phantom -} 0.01$ & ${\phantom -} 0.02$  \\       
$U^{+,{\rm Re}}_{+-}$ & ${\phantom -} 0.14$ & $            -0.04$ & $            -0.03$ & ${\phantom -} 0.11$ & $            -0.01$ & $            -0.12$ & ${\phantom -} 0.07$ & $            -0.07$ & $            -0.06$  \\       
$U^{-}_{+}          $ & ${\phantom -} 0.04$ & $            -0.01$ & ${\phantom -} 0.01$ & ${\phantom -} 0.04$ & ${\phantom -} 0.01$ & $            -0.04$ & ${\phantom -} 0.01$ & $            -0.05$ & ${\phantom -} 0.04$  \\ \hline
\label{tab:finalStatCovII}
\end{tabular}
\end{center}
\end{table*}

\begin{table*}
\caption{(Continued from Table~\ref{tab:finalStatCovII}) Statistical correlation matrix for the $U$ and $I$ parameters. Elements above the diagonal are redundant and not included in the table.}
\begin{center}
\begin{tabular}{ l | l  l l l l l l l l}
                      &$U^{-,{\rm Im}}_{+0}$&$U^{-,{\rm Re}}_{+0}$&$U^{+,{\rm Im}}_{+0}$&$U^{+,{\rm Re}}_{+0}$&$U^{-,{\rm Im}}_{+-}$&$U^{-,{\rm Re}}_{+-}$&$U^{+,{\rm Im}}_{+-}$&$U^{+,{\rm Re}}_{+-}$& $U^{-}_{+}$           \\ \hline
$U^{-,{\rm Im}}_{+0}$ & ${\phantom -} 1.00$ &                     &                     &                     &                     &                     &                     &                     &                       \\       
$U^{-,{\rm Re}}_{+0}$ & $            -0.14$ & ${\phantom -} 1.00$ &                     &                     &                     &                     &                     &                     &                       \\       
$U^{+,{\rm Im}}_{+0}$ & ${\phantom -} 0.22$ & ${\phantom -} 0.07$ & ${\phantom -} 1.00$ &                     &                     &                     &                     &                     &                       \\       
$U^{+,{\rm Re}}_{+0}$ & ${\phantom -} 0.03$ & $            -0.01$ & $            -0.09$ & ${\phantom -} 1.00$ &                     &                     &                     &                     &                       \\       
$U^{-,{\rm Im}}_{+-}$ & $            -0.00$ & ${\phantom -} 0.00$ & ${\phantom -} 0.03$ & ${\phantom -} 0.00$ & ${\phantom -} 1.00$ &                     &                     &                     &                       \\       
$U^{-,{\rm Re}}_{+-}$ & ${\phantom -} 0.00$ & $            -0.00$ & $            -0.00$ & $            -0.00$ & ${\phantom -} 0.31$ & ${\phantom -} 1.00$ &                     &                     &                       \\       
$U^{+,{\rm Im}}_{+-}$ & ${\phantom -} 0.01$ & $            -0.02$ & ${\phantom -} 0.02$ & $            -0.00$ & ${\phantom -} 0.18$ & $            -0.06$ & ${\phantom -} 1.00$ &                     &                       \\       
$U^{+,{\rm Re}}_{+-}$ & ${\phantom -} 0.01$ & $            -0.03$ & ${\phantom -} 0.00$ & $            -0.00$ & $            -0.17$ & ${\phantom -} 0.06$ & ${\phantom -} 0.13$ & ${\phantom -} 1.00$ &                       \\       
$U^{-}_{+}          $ & ${\phantom -} 0.04$ & $            -0.01$ & $            -0.00$ & ${\phantom -} 0.02$ & $            -0.14$ & ${\phantom -} 0.01$ & ${\phantom -} 0.06$ & ${\phantom -} 0.01$ & ${\phantom -} 1.00$   \\ \hline
\label{tab:finalStatCovIII}
\end{tabular}
\end{center}
\end{table*}

\subsection{Quasi-Two-Body Parameters}

The $U$ and $I$ parameters and associated correlations can be used to extract the values of the Q2B $B^0 ({\overline B^0}) \rightarrow \rho^\pm \pi^\mp$ $\CP$-violation parameters in the time-dependent decay rate defined in Ref.~\cite{ref:cpv}:
\begin{align}
f^{\rho^\pm \pi^\mp}_{Q_{\rm tag}}(\Delta t) &= (1 \pm {\mathcal A}_{\rho\pi}) \frac{e^{-|\Delta t|/\tau}}{4\tau} \nonumber \\
	&\times \left[ 1 + Q_{\rm tag} ({\mathcal S} \pm \Delta{\mathcal S}) \sin(\Delta m_d \Delta t) \right. \nonumber \\
	&\left. - Q_{\rm tag} ({\mathcal C} \pm \Delta{\mathcal C})\cos(\Delta m_d \Delta t) \right] ,
\end{align}
where $Q_{\rm tag} = +1\ (-1)$ when the tag-side $B$ meson is a $B^0$ (${\overline B}^0$). The time- and flavor-integrated charge asymmetry ${\mathcal A}_{\rho\pi}$ quantifies direct $\CP$ violation, while ${\mathcal S}$ and ${\mathcal C}$ parameterize mixing-induced $\CP$ violation related to the angle $\alpha$, and flavor-dependent direct $\CP$ violation, respectively. The parameter $\Delta {\mathcal C}$ describes the asymmetry between the rates $\Gamma (B^0 \rightarrow \rho^+ \pi^-) + \Gamma ( {\overline B}^0 \rightarrow \rho^- \pi^+) $ and $\Gamma (B^0 \rightarrow \rho^- \pi^+) + \Gamma ( {\overline B}^0 \rightarrow \rho^+ \pi^-) $, while $\Delta {\mathcal S}$ relates to the strong-phase difference between the different amplitudes involved in the decay $B^0 \rightarrow \rho \pi$. The $U$ and $I$ parameters are related to the Q2B parameters through the relations of Eqs.~\eqref{q2bequation}--\eqref{lastQ2B}.

We can also use the $U$ and $I$ parameters and associated correlations to extract the $B^0\rightarrow\rho^0\pi^0$ $\CP$-violation parameters and decay fraction:
\begin{align}
	{\mathcal C}_{00} &= \frac{U^-_0}{U^+_0}, \\
	{\mathcal S}_{00} &= \frac{2 I_0}{U^+_0}, \\
	f_{00} &= \frac{U^+_0}{U^+_+ + U^+_- + U^+_0}.
\end{align}
These eight Q2B parameters related to $\rho^\pm\pi^\mp$ and $\rho^0\pi^0$ decays are extracted using a $\chi^2$ minimization technique that accounts for the statistical and systematic correlations between the $U$ and $I$ parameters. In each step of the minimization process, the current values of the free parameters are used to calculate the corresponding values of the eight $U$ and $I$ parameters on which they depend and a vector $V_{\rm diff}$ is constructed from the differences between these $U$ and $I$ parameter values and the values obtained in our final fit to the full on-resonance dataset. The $\chi^2$ is then calculated as 
\begin{eqnarray}
	\chi^2_{\rm Q2B} =  V_{\rm diff}^T (C^{\rm data})^{-1} V_{\rm diff},
\end{eqnarray}
\noindent where $C^{\rm data}$ is the $8 \times 8$ covariance matrix for the relevant $U$ and $I$ parameters from the fit to data. Table~\ref{tab:q2b} presents the Q2B parameters extracted from the full fit to on-resonance data along with their statistical and systematic errors. In a study of correlations between $\sigma_{\Delta t}$ and DP position we find a small contribution to the systematic uncertainty on $f_{00}$, which is included in Table~\ref{tab:q2b}. The correlation matrix for the Q2B parameters is provided in Table~\ref{tab:q2bCor}. A study of the Q2B parameters (see Appendix~\ref{sec:q2brobustness}) establishes that there is negligible bias in their extraction and good robustness in the presence of statistical fluctuations.

\begin{table}
\caption{Quasi-two-body parameter values and uncertainties corresponding to the fit to the complete on-resonance dataset. }
\begin{center}
\begin{tabular}{ l | l  l l }
Param       &     Value        	    &   $\sigma_{\rm stat}$   &  $\sigma_{\rm syst}$ \\ \hline
${\mathcal A}_{\rho \pi}   $   &                $-0.100$      &   0.029         &  0.021       \\
${\mathcal C}              $   &     ${\phantom -}0.016$      &   0.059	      &  0.036	     \\
$\Delta{\mathcal C}        $   &     ${\phantom -}0.234$      &   0.061	      &  0.048	     \\
${\mathcal S}              $   &     ${\phantom -}0.053$      &   0.081 	      &  0.034	     \\
$\Delta{\mathcal S}        $   &     ${\phantom -}0.054$      &   0.082	      &  0.039	     \\
${\mathcal C}_{00}                    $   &     ${\phantom -}0.19$       &   0.23	      &  0.15	     \\
${\mathcal S}_{00}                    $   &                $-0.37$       &   0.34	      &  0.20	     \\
$f_{00}                    $   &     ${\phantom -}0.092$      &   0.011         &  0.009       
\label{tab:q2b}
\end{tabular}
\end{center}
\end{table}

\begin{table*}
\caption{Combined statistical and systematic correlation matrix for the quasi-two-body parameters corresponding to the fit to the complete on-resonance dataset. Values above the diagonal are redundant and omitted for clarity.}
\begin{center}
\begin{tabular}{l | l  l l l l l l l}
                         & ${\phantom -}{\mathcal A}_{\rho\pi}$ &    ${\phantom -}{\mathcal C}$      & ${\phantom -}\Delta{\mathcal C}$ &   ${\phantom -}{\mathcal S}$      &  ${\phantom -}\Delta{\mathcal S}$ & ${\phantom -}{\mathcal C}_{00}$              & ${\phantom -}{\mathcal S}_{00}$              & ${\phantom -}f_{00}$              \\ \hline
${\mathcal A}_{\rho\pi}$ & ${\phantom -}1.000$      &                        &                      &                       &                       &                       &                       &                       \\
${\mathcal C}          $ & ${\phantom -}0.035$      &    ${\phantom -}1.000$ &			    &			    &			    &			    &			    &			    \\
$\Delta{\mathcal C}    $ & ${\phantom -}0.154$      &    ${\phantom -}0.213$ &	${\phantom -}1.000$ &			    &			    &			    &			    &			    \\
${\mathcal S}          $ & $	       -0.040$      &    $	     -0.065$ &	$	    -0.070$ & 	${\phantom -}1.000$ &			    &			    &			    &			    \\
$\Delta{\mathcal S}    $ & $	       -0.041$      &    $	     -0.038$ &	$	    -0.060$ & 	${\phantom -}0.199$ & 	${\phantom -}1.000$ &			    &			    &			    \\
${\mathcal C}_{00}                $ & $	       -0.088$      &    $	     -0.041$ &	$	    -0.034$ & 	${\phantom -}0.026$ & 	${\phantom -}0.011$ & 	${\phantom -}1.000$ &			    &			    \\
${\mathcal S}_{00}                $ & $	       -0.005$      &    ${\phantom -}0.007$ &	${\phantom -}0.044$ & 	$	    -0.081$ & 	$	    -0.007$ & 	${\phantom -}0.002$ & 	${\phantom -}1.000$ &			    \\
$f_{00}                $ & ${\phantom -}0.074$      &    ${\phantom -}0.009$ &	${\phantom -}0.016$ &   ${\phantom -}0.029$ & 	$	    -0.016$ & 	$	    -0.062$ & 	${\phantom -}0.062$ &	${\phantom -}1.000$ \\ \hline
\label{tab:q2bCor}
\end{tabular}
\end{center}
\end{table*}

\begin{figure}[!htb]
\begin{center}
\includegraphics[width=0.45\textwidth]{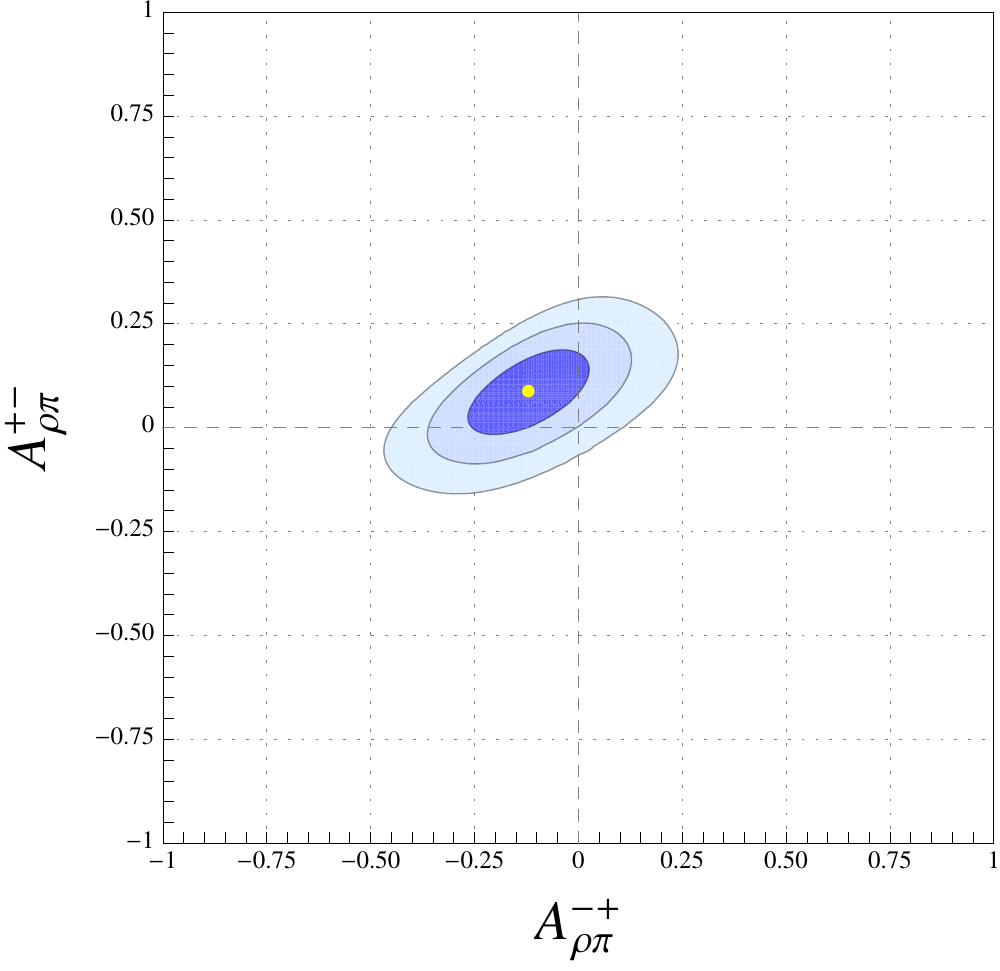}
\caption{(color online) Combined statistical and systematic two-dimensional likelihood scan of  ${\mathcal A}^{+-}_{\rho\pi}$ versus ${\mathcal A}^{-+}_{\rho\pi}$ with $68.3\%$, $95.5\%$, and $99.7\%$ confidence-level contours ($\Delta\chi^2=\{2.30,6.18,11.83\}$). The yellow dot inside the contours indicates the central value.}
\label{fig:contours}
\end{center}
\end{figure}

The parameters ${\mathcal A}_{\rho\pi}$ and ${\mathcal C}$ can be transformed into the direct $\CP$-violation parameters ${\mathcal A}^{+-}_{\rho\pi}$ and ${\mathcal A}^{-+}_{\rho\pi}$ where
\begin{align}
	{\mathcal A}^{+-}_{\rho\pi} & \equiv \frac{\Gamma({\overline B}^0 \rightarrow \rho^-\pi^+) - \Gamma( B^0 \rightarrow \rho^+\pi^-)}{\Gamma({\overline B}^0 \rightarrow \rho^-\pi^+) + \Gamma( B^0 \rightarrow \rho^+\pi^-)},\\
	{\mathcal A}^{-+}_{\rho\pi} & \equiv \frac{\Gamma({\overline B}^0 \rightarrow \rho^+\pi^-) - \Gamma( B^0 \rightarrow \rho^-\pi^+)}{\Gamma({\overline B}^0 \rightarrow \rho^+\pi^-) + \Gamma( B^0 \rightarrow \rho^-\pi^+)},
\end{align}
using the relations
\begin{align}
	{\mathcal A}^{+-}_{\rho\pi} & = - \frac{{\mathcal A}_{\rho\pi} + {\mathcal C} + {\mathcal A}_{\rho\pi} \Delta {\mathcal C}}{1+ \Delta {\mathcal C} + {\mathcal A}_{\rho\pi} {\mathcal C}},\\
	{\mathcal A}^{-+}_{\rho\pi} & = \frac{{\mathcal A}_{\rho\pi} - {\mathcal C} - {\mathcal A}_{\rho\pi} \Delta {\mathcal C}}{1- \Delta {\mathcal C} - {\mathcal A}_{\rho\pi} {\mathcal C}}.
\end{align}
We extract the central values and uncertainties for these parameters using a $\chi^2$ minimization procedure in the two-dimensional plane corresponding to ${\mathcal A}^{+-}_{\rho\pi}$ versus ${\mathcal A}^{-+}_{\rho\pi}$. At each point in the plane, the values of ${\mathcal A}^{+-}_{\rho\pi}$ and ${\mathcal A}^{-+}_{\rho\pi}$ are fixed and used in combination with $\Delta {\mathcal C}$ (which is free to vary) to determine the corresponding values of ${\mathcal A}_{\rho\pi}$ and ${\mathcal C}$. These values are then used in combination with  $\Delta {\mathcal C}$ and the five other Q2B parameters to calculate a $\chi^2$ value as described above. From this two-dimensional scan, we find
\begin{align}
	{\mathcal A}^{+-}_{\rho\pi} & = {\phantom -} 0.09 ^{+0.05}_{-0.06} \pm 0.04, \\
	{\mathcal A}^{-+}_{\rho\pi} & = -0.12 \pm 0.08 ^{+0.04}_{-0.05},
\end{align}
with a correlation of 0.55 evaluated from the $1\sigma$ contour for statistical and systematic errors combined. A two-dimensional likelihood scan with combined statistical and systematic uncertainties and $68.3\%$, $95.5\%$, and $99.7\%$ confidence-level contours is provided in Fig.~\ref{fig:contours}. The origin, corresponding to no direct $\CP$ violation, lies on the $96.0\%$ confidence-level contour ($\Delta\chi^2=6.42$), corresponding to a $p=4.0\%$ probability, under the assumption of no direct $\CP$ violation, of obtaining a result that deviates from the origin at least as much as ours.

\subsection{\boldmath Alpha Scan}
\label{sec:scan}

In order to extract likely values of $\alpha$ from the $U$ and $I$ parameters obtained in our final fit, we perform a scan of $\alpha$ from $0^\circ$ to $180^\circ$. At each scan point, a $\chi^2$-minimization fit is performed using the goodness of fit measure:
\begin{align}
	\chi^2(\alpha) =&  \left[ V^{\rm data} - V^{\rm scan} \right]^T (C^{\rm data})^{-1} \left[ V^{\rm data} - V^{\rm scan} \right] \nonumber \\
				&+ \left( \frac{1 - U^{+\ {\rm scan}}_+}{\epsilon} \right)^2,
\end{align}
\noindent where $C^{\rm data}$ is the $26 \times 26$ covariance matrix for $U$s and $I$s from our fit to data and  $V^{\rm data}$ and $V^{\rm scan}$ are vectors of the 26 $U$ and $I$ parameters from our fit to data and the current iteration of the minimization, respectively. The last term is a Gaussian constraint that restricts $U^+_+$ in the scan to lie within $\epsilon$ of 1 (the value to which it is fixed in the fits to data); $\epsilon$ is set to 0.0001. Because the overall scaling of the $U$ and $I$ parameters is not physically meaningful, $U^+_+$ is allowed to be exactly 1 in the fit and this constraint term does not significantly contribute to the total $\chi^2$. In the $\chi^2$-minimizations, the actual free parameters are the tree ($T^{\pm,0}$) and penguin ($P^{\pm}$) amplitudes, which are related to the $\rho$-resonance amplitudes through the formulae
\begin{eqnarray}
	\label{AsDefFirst}
	A^+ &=& T^+ e^{-i \alpha} + P^+, \\
	A^- &=& T^- e^{-i \alpha} + P^-, \\
	A^0 &=& T^0 e^{-i \alpha} + P^0, \\
	{\overline A}^+ &=& T^- e^{+i \alpha} + P^-, \\
	{\overline A}^- &=& T^+ e^{+i \alpha} + P^+, \\
	{\overline A}^0 &=& T^0 e^{+i \alpha} + P^0,
	\label{AsDefLast}
\end{eqnarray}
where
\begin{eqnarray}
	\label{amplitudeFirst}
	A^{+} &\equiv& A(B^0 \rightarrow \rho^+\pi^-),\\
	A^{-} &\equiv& A(B^0 \rightarrow \rho^-\pi^+),\\
	A^{0} &\equiv& A(B^0 \rightarrow \rho^0\pi^0),\\
	{\overline A}^{+} &\equiv& \frac{q}{p} A({\overline B}^0 \rightarrow \rho^+\pi^-),\\
	{\overline A}^{-} &\equiv& \frac{q}{p} A({\overline B}^0 \rightarrow \rho^-\pi^+),\\
	{\overline A}^{0} &\equiv& \frac{q}{p} A({\overline B}^0 \rightarrow \rho^0\pi^0).
	\label{amplitudeLast}
\end{eqnarray}
Note that due to $SU(2)$ flavor symmetry, the third penguin amplitude can be calculated from the other two using the relation $P^0 = -\frac{1}{2}\left( P^+ + P^- \right)$ (see Refs.~\cite{ref:flavorsymmone} and \cite{ref:flavorsymmtwo}). At each step in the minimization process, the current values of the tree and penguin amplitudes as well as the current fixed value of $\alpha$ are used to calculate the $\rho$-resonance amplitudes. These $\rho$ amplitudes are then used to calculate the $U$ and $I$ parameters that comprise the vector $V^{\rm scan}$, using Eqs.~\eqref{UsIsDefFirst} through \eqref{UsIsDefLast}, which relate the $U$ and $I$ parameters to the $\rho$ amplitudes. In the fits, we take advantage of a global phase that is not physically meaningful to fix the phase of $T^+$ to 0.

As the scan proceeds, a minimum $\chi^2$ value is extracted from the fit at each value of $\alpha$. We convert these $\chi^2$ values to ``$\Sigma$'' values by calculating the $\chi^2$ probability of each value according to
\begin{eqnarray}
	\Sigma \equiv \int_a^\infty f(x;1) dx,
\end{eqnarray}
\noindent where $a$ is the difference between the $\chi^2$ at the current scan point and the minimum $\chi^2$ for all the scan points, and $f(x;1)$ is a $\chi^2$ distribution with one degree of freedom. The variable ``$\Sigma$'' corresponds to what is commonly referred to as ``1$-$Confidence Level'' (1$-$C.L.) and is simply the $p$-value of a $\chi^2$ test at each scan point.

\subsubsection{Incorporating Information from Charged $B$ Decays}

Following the methods employed in Belle's 2007 $B^0\rightarrow\rho\pi$ analysis~\cite{ref:belle2007} and described in Ref.~\cite{ref:snyderquinnusis}, we perform a further $\alpha$ scan that makes use of measurements from the charged decays $B^\pm\rightarrow\rho^{\pm,0}\pi^{0,\pm}$. Amplitudes for these modes can be related to amplitudes in the neutral $B$ modes through isospin relations. These relations result in four ``constraint'' equations while introducing only two new free parameters in the fit (the real and imaginary parts of a tree amplitude, $T^{+0}$). The charged $B$ measurements of interest are the branching fractions and asymmetries:
\begin{eqnarray}
	\label{BFFirst}
	\mathcal{B}(\rho^+\pi^0) &=& c (|A^{+0}|^2 + |A^{-0}|^2) \tau_{B^+},\\
	\mathcal{B}(\rho^0\pi^+) &=& c (|A^{0+}|^2 + |A^{0-}|^2) \tau_{B^+},\\
	\mathcal{A}(\rho^+\pi^0) &=& \frac{|A^{-0}|^2 - |A^{+0}|^2}{|A^{-0}|^2 + |A^{+0}|^2},\\
	\mathcal{A}(\rho^0\pi^+) &=& \frac{|A^{0-}|^2 - |A^{0+}|^2}{|A^{0-}|^2 + |A^{0+}|^2},
	\label{BFLast}
\end{eqnarray}
where $c$ is a constant and
\begin{eqnarray}
	\label{secAmpFirst}
	A^{\pm 0} &=& \frac{q}{p} A(B^\pm \rightarrow \rho^{\pm}\pi^0),\\
	A^{0 \pm} &=& \frac{q}{p} A(B^\pm \rightarrow \rho^0\pi^{\pm}).
	\label{secAmpLast}
\end{eqnarray}
In the fit, we fix $c=1$ and no longer require $U^+_+=1$. This is equivalent to letting $c$ be a free parameter in the fit and setting $U^+_+=1$. Due to this convention, it is necessary to divide all the current values of the $U$ and $I$ parameters during the minimization process by the current value of $U^+_+$ before using them to calculate the current $\chi^2$ value.

According to $SU(2)$ isospin symmetry, several relations hold between the amplitudes in Eqs.~\eqref{amplitudeFirst}--\eqref{amplitudeLast} and \eqref{secAmpFirst}--\eqref{secAmpLast}:
\begin{align}
	A^+ + A^- + 2A^0 &= e^{-2 i \alpha} \left(\overline{A}^+ + \overline{A}^- + 2 \overline{A}^0\right)\\
				    &= \sqrt{2} (A^{+0} + A^{0+})\\
				    &= \sqrt{2} e^{-2 i \alpha} (A^{-0} + A^{0-}),
\end{align}
\begin{align}
	A^{+0} - A^{0+} - \sqrt{2} (A^+ - A^-) &= e^{-2 i \alpha} \left[ A^{-0} - A^{0-} {\phantom {\sqrt{A^2}}} \right. \nonumber \\
				     & \left. - \sqrt{2} (A^- - A^+)\right].
\end{align}
Based on these relations, we can parameterize the charged $B$ amplitudes according to
\begin{eqnarray}
	\label{ConstraintFirst}
	\sqrt{2} A^{+0} &=& e^{-i \alpha} T^{+0} + P^+ - P^-, \\	
	\sqrt{2} A^{0+} &=& e^{-i \alpha} (T^+ + T^- +2T^0 - T^{+0}) \nonumber \\
				& &- P^+ + P^-, \\
	\sqrt{2} A^{-0} &=& e^{+i \alpha} T^{+0} + P^+ - P^-, \\
	\sqrt{2} A^{0-} &=& e^{+i \alpha} (T^+ + T^- +2T^0 - T^{+0}) \nonumber \\
				& & - P^+ + P^-.
	\label{ConstraintLast}
\end{eqnarray}
When combined with Eqs.~\eqref{AsDefFirst}--\eqref{AsDefLast}, which parameterize the neutral $B$ amplitudes, this yields a parameterization that implicitly incorporates the isospin relations between the different charged and neutral $B$ modes. Because the global phase is physically irrelevant, we fix it by requiring that ${\rm Im} T^+=0$.

The isospin ``constrained'' and ``unconstrained'' scans are performed identically, except that Eqs.~\eqref{ConstraintFirst}--\eqref{ConstraintLast} are included as Gaussian constraints in the $\chi^2$ calculation. The system of equations~\eqref{BFFirst}--\eqref{BFLast} is used to express the magnitudes of $A^{\pm0}$ and $A^{0\pm}$ in terms of the branching fractions and asymmetries for the charged $B$ modes $B^\pm \rightarrow \rho^\pm \pi^0$ and $B^\pm \rightarrow \rho^0 \pi^\pm $. Using world average measurements from~\cite{ref:rpp2011}, we calculate the value of each of these magnitudes as well as their uncertainties. At each step in the minimization process, and for each of Eqs.~\eqref{ConstraintFirst}--\eqref{ConstraintLast}, a term is added to the $\chi^2$:
\begin{eqnarray}
	\left( \frac{|A_{\rm iter.}|-|A_{\rm meas.}|}{\sigma_{|A_{\rm meas.}|}} \right)^2,
\end{eqnarray} 
where $|A_{\rm iter.}|$ is the magnitude of the relevant $A^{\pm 0}$ or $A^{0 \pm}$ parameter for the current iteration of the minimization process, $|A_{\rm meas.}|$ is the magnitude of the amplitude based on branching fractions and asymmetry measurements, and $\sigma_{|A_{\rm meas.}|}$ is the uncertainty in the value of the magnitude due to measurement uncertainties for the branching fractions and asymmetries. For those branching fractions and asymmetries that have asymmetric uncertainties, we choose whether to use the upper or lower error for the calculation in a given iteration by ascertaining whether the value of the branching fraction or asymmetry corresponding to the tree and penguin parameters in the current iteration of the minimization process is less than or greater than the experimental value, respectively.

\subsubsection{Results of $\alpha$ Scan}

\begin{figure}[!htb]
		\includegraphics[width=0.49\textwidth]{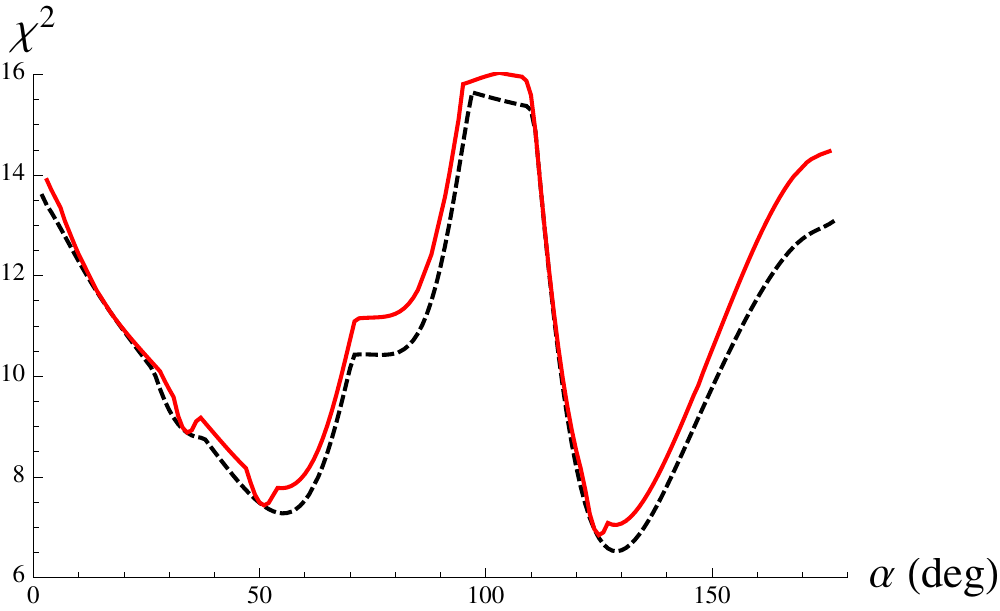}
		\caption{(color online) Isospin-constrained (solid red) and unconstrained (dashed black) scans of minimum $\chi^2$ values as a function of $\alpha$. The scans are based on the fit to the full on-resonance dataset and include contributions from both statistical and systematic uncertainties. Note that the origin is suppressed.}
		\label{fig:chi}
\end{figure}

Plots of the $\chi^2$ values from our final $\alpha$ scans with isospin constraints (solid red) and without isospin constraints (dashed black) are shown in Fig.~\ref{fig:chi}. The corresponding $\Sigma$ distributions are presented in Fig.~\ref{fig:1mcl}. As indicated by our robustness studies (see Appendix), the $\Sigma$ scan is not robust and cannot be interpreted in terms of Gaussian statistics.

\begin{figure}[!htb]
	\begin{center}
		\includegraphics[width=0.49\textwidth]{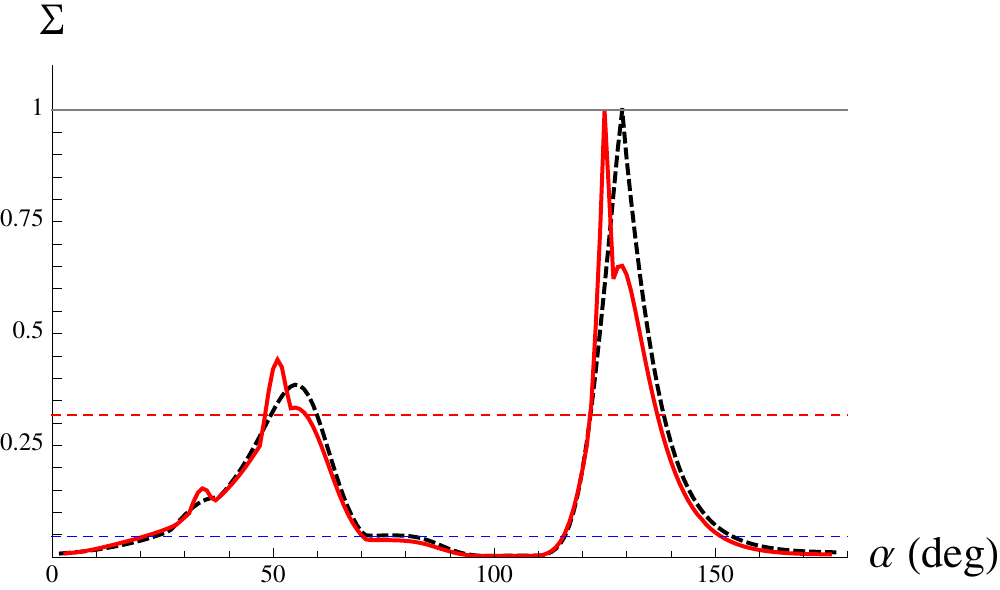}
		\caption{(color online) Isospin-constrained (solid red) and unconstrained (dashed black) scans of $\Sigma$ values as a function of $\alpha$. The scans are based on the fit to the full on-resonance dataset and include contributions from both statistical and systematic uncertainties. The upper and lower horizontal dashed lines correspond to $\Sigma=0.05$ and $0.32$, respectively.}
		\label{fig:1mcl}
	\end{center}
\end{figure}

\section{CONCLUSION}
\label{sec:conclusion}

We have performed a time-dependent Dalitz plot analysis of the mode $B^0 \rightarrow (\rho \pi)^0$ in which we extract 26 $U$ and $I$ parameter values describing the physics involved, as well as their full correlation matrix in an extended unbinned maximum likelihood fit. From these fit results, we extract the Q2B parameters ${\mathcal A}_{\rho\pi}$, ${\mathcal C}$, $\Delta{\mathcal C}$, ${\mathcal S}$, $\Delta{\mathcal S}$, ${\mathcal C}_{00}$, ${\mathcal S}_{00}$, and $f_{00}$, with values given in Table~\ref{tab:q2b}. These Q2B values are consistent with the results of the 2007 {\Babar} analysis~\cite{ref:matt2007} as well as the results obtained by the Belle collaboration~\cite{ref:belle2007}. We also perform a two-dimensional likelihood scan of the direct $\CP$-violation asymmetry parameters for $B^0\rightarrow\rho^\pm\pi^\mp$ decays, finding the change in $\chi^2$ between the minimum and the origin (corresponding to no direct $\CP$-violation) to be $\Delta \chi^2=6.42$. Finally, we perform one-dimensional likelihood-scans of the unitarity angle $\alpha$ (see Figure~\ref{fig:chi}) both with (solid red) and without (dashed black) isospin constraints. However, as indicated by our robustness studies (Appendix~\ref{sec:robustness}), the extraction of $\alpha$ with our current sample size is not robust. Maximum likelihood estimators are known to be Gaussian in general only in the limit of large data sets. This analysis would benefit greatly from increased sample sizes available at higher-luminosity experiments.

\section{ACKNOWLEDGMENTS}
\label{sec:Acknowledgments} 

\noindent
We are grateful for the extraordinary contributions of our PEP-II colleagues in achieving the excellent luminosity and machine conditions that have made this work possible. The success of this project also relies critically on the expertise and dedication of the computing organizations that support {\babar}. The collaborating institutions wish to thank  SLAC for its support and the kind hospitality extended to them. This work is supported by the US Department of Energy and National Science Foundation, the Natural Sciences and Engineering Research Council (Canada), the Commissariat \`a l'Energie Atomique and Institut National de Physique Nucl\'eaire et de Physique des Particules (France), the Bundesministerium f\"ur Bildung und Forschung and Deutsche Forschungsgemeinschaft (Germany), the Istituto Nazionale di Fisica Nucleare (Italy), the Foundation for Fundamental Research on Matter (The Netherlands), the Research Council of Norway, the Ministry of Education and Science of the Russian Federation,  Ministerio de Ciencia e Innovaci\'on (Spain), and the Science and Technology Facilities Council (United Kingdom). Individuals have received support from the Marie-Curie IEF program (European Union) and the A. P. Sloan Foundation (USA).

\appendix

\section{Robustness Studies}

We assess the robustness with which the fit framework extracts statistically accurate values and uncertainties for the $U$ and $I$ parameters, the Q2B parameters, and $\alpha$, by employing MC-simulated samples generated with a parameterized detector simulation and with signal and background contributions corresponding to those expected in the full on-resonance dataset. The simulated samples are generated using physical $U$ and $I$ parameters based on specific tree and penguin amplitudes and $\alpha=89^\circ$ (approximately the world average). Each simulated dataset is generated with the same parameter values, but a different random-number seed. By examining the results of the fits to each of these simulated datasets, we assess the the fit robustness.

For these studies, we use 25 samples generated with different seeds, so that the uncertainty on the bias is negligible compared to the other uncertainties.

\subsection{U and I Parameter Robustness Studies}
\label{sec:UIrobustness}

For each of the 26 $U$ and $I$ parameters extracted in the fits to the MC samples, we calculate the RMS (across the 25 MC samples) of the differences (measured in units of the statistical uncertainty on the extracted values of the parameters) between the generated and extracted values of the parameters. We also calculate the average across the 25 MC samples of these differences for each of the $U$ and $I$ parameters.
The RMS difference, averaged across the 26 $U$ and $I$ parameters, is 1.17, while the difference in units of statistical uncertainty, averaged across the $U$ and $I$ parameters, is 0.04. This demonstrates that the $U$ and $I$ parameter values are extracted robustly and with negligible bias.

\subsection{Quasi-Two-Body Robustness Studies}
\label{sec:q2brobustness}

The results of the robustness study of the Q2B parameters are provided in Table~\ref{tab:q2bRobust}. In this table, the first column of numbers is the ratio of the square root of the variance of the parameter across 25 MC samples divided by the mean error on the fit to the parameter across all MC samples. As one would expect, the square root of the variance of each parameter value is approximately equal to the mean statistical uncertainty on the variable as extracted from the fits to MC samples. The second column is the average difference (measured in units of statistical uncertainty, $\sigma$, on the extracted values of the parameters) between the extracted value of the parameter and its generated value, across all MC samples. These values are all within $0.28\sigma$ of 0, indicating negligible bias.  The third column is  the same as the second, except that the absolute value of each difference is taken before averaging. These values are no greater than $0.93\sigma$, indicating that the extracted values of the parameters are reliably close to the generated values. Taken together, these results indicate that the extraction of the Q2B parameters is robust to possible non-Gaussian fluctuations, and unbiased.

\begin{table}
\caption{Results of robustness study of quasi-two-body parameters. The first column of numbers is the ratio of the square root of the variance divided by the mean error on the fit to the parameter across 25 MC samples. The second column is the average difference (measured in units of statistical uncertainty, $\sigma$, on the extracted values of the parameters) between the extracted value of the parameter and its generated value, across all MC samples. The third column is  the same as the second, except that the absolute value of each difference is taken before averaging. }
\begin{center}
\begin{tabular}{ l | c c c }
Param	&	$\sqrt{\rm Variance}/\left<\sigma\right>$ &  Avg \#$\sigma$        & Avg Abs \#$\sigma$ \\
           	&			   			                         &  Diff From                    & Diff From \\
		&								      &	  Gen Val                      & Gen Val   \\ \hline
${\mathcal A}_{\rho\pi}$	&	0.94		&	$-0.13$					&	0.76\\
${\mathcal C}          $  	&	1.15		&	${\phantom -}0.06$		&	0.90\\
$\Delta{\mathcal C}    $	&	0.94		&	${\phantom -}0.04$		&	0.75\\
${\mathcal S}          $		&	1.11		&	${\phantom -}0.03$		&	0.92\\
$\Delta{\mathcal S}    $	&	1.02		&	$-0.20$					&	0.82\\
${\mathcal C}_{00}                $		&	1.15		&	$-0.10$					&	0.89\\
${\mathcal S}_{00}                $		&	1.13		&	${\phantom -}0.23$		&	0.92\\
$f_{00}                $		&	1.08		&	${\phantom -}0.28$		&	0.93
\label{tab:q2bRobust}
\end{tabular}
\end{center}
\end{table}

\subsection{\boldmath $\alpha$ Robustness Studies}
\label{sec:robustness}

\begin{table}
\caption{Results of parameterized MC study of robustness of $\alpha$ scans. The generated value $\alpha$ is $\alpha_{\rm gen}=89^\circ$ and the columns are described in the text.}
\begin{center}
\begin{tabular}{ c | c c c c c c }
 $\alpha$ Scan & Upper & Lower &  Mean & Min  & $\sqrt{2 \Delta \chi^2}$ Dist     & $\#\sigma$ Between \\
Peak		       & Error       & Error       & Error	    & $\chi^2$			   &  From & Gen and Fit \\
 & & & & & $\chi^2(\alpha_{\rm gen})$ & Peaks \\ \hline 
43    &	$+5$ 	&	$-4$ 		&	5 		&	27.2	      	&			2.6		 	&	$-9.0$ \\
44    &	$+5$ 	&	$-5$ 		&	5	     	&	18.7	 	&			5.3		 	&	$-8.5$ \\
48    &	$+5$ 	&	$-5$ 		&	5	 	&	21.0	 	&			2.3		 	&	$-8.2$ \\
49    &	$+5$ 	&	$-5$ 		&	5 		&	24.2 		&			2.9		 	&	$-8.7$ \\
52    &	$+5$ 	&	$-5$ 		&	5	 	&	16.1 		&			2.4		 	&	$-8.2$ \\
53    &	$+5$ 	&	$-21$ 	&	13	 	&	23.2	 	&			1.5		 	&	$-7.2$ \\
60    &	$+14$ 	&	$-8$ 		&	11	 	&	16.5	 	&			3.3		 	&	$-2.1$ \\
74    &	$+7$ 	&	$-6$ 		&	7 		&	21.2	 	&			0.7		 	&	$-2.1$ \\
74    &	$+5$ 	&	$-13$ 	&	9 		&	15.9	 	&			4.1		 	&	$-2.9$ \\
75    &	$+10$ 	&	$-8$ 		&	9	 	&	21.3	 	&			1.8		 	&	$-1.5$ \\
76    &	$+8$ 	&	$-18$ 	&	13	 	&	21.3	 	&			2.5		 	&	$-1.7$ \\
80    &	$+5$ 	&	$-6$ 		&	6 		&	30.0	 	&			2.3		 	&	$-1.7$ \\
83    &	$+6$ 	&	$-7$ 		&	7 		&	24.0	 	&			1.3     	 	&	$-0.9$ \\
84    &	$+6$ 	&	$-6$ 		&	6	 	&	26.4 		&			1.3	   	 	&	$-0.9$ \\
84    &	$+7$ 	&	$-6$ 		&	7	 	&	30.1	   	&			1.0		 	&	$-0.7$ \\
87    &	$+7$ 	&	$-7$ 		&	7	 	&	22.9	 	&			0.4		 	&	$-0.3$ \\
88    &	$+7$ 	&	$-6$ 		&	7	 	&	10.9	 	&			0.1			 &	$-0.1$ \\
89    &	$+9$ 	&	$-8$ 		&	8 		&	23.4	 	&			0.00	 		&	${\phantom -}0$ \\
91    &	$+9$ 	&	$-9$ 		&	9 		&	33.1	 	&			0.3			 &	${\phantom -}0.2$ \\
91    &	$+4$ 	&	$-5$ 		&	5 		&	63.3	 	&			0.6		 	&	${\phantom -}0.4$ \\
92    &	$+8$ 	&	$-6$ 		&	7 		&	39.2	 	&			0.7		 	&	${\phantom -}0.5$ \\
94    &	$+7$ 	&	$-6$ 		&	6	 	&	10.0	 	&			1.3		 	&	${\phantom -}0.9$ \\
112 &	$+5$ 	&	$-6$ 		&	5 		&	19.0 		&			3.0		 	&	${\phantom -}3.9$ \\
115 &	$+5$ 	&	$-6$ 		&	5 		&	23.3	 	&			2.2		 	&	${\phantom -}4.6$ \\
124 &	$+28$ 	&	$-15$ 	&	22	 	&	25.6	 	&			1.4		 	&	${\phantom -}2.3$
\label{tab:robustness}
\end{tabular}
\end{center}
\end{table}

\begin{figure}[!htb]
\begin{center}
\includegraphics[width=0.48\textwidth]{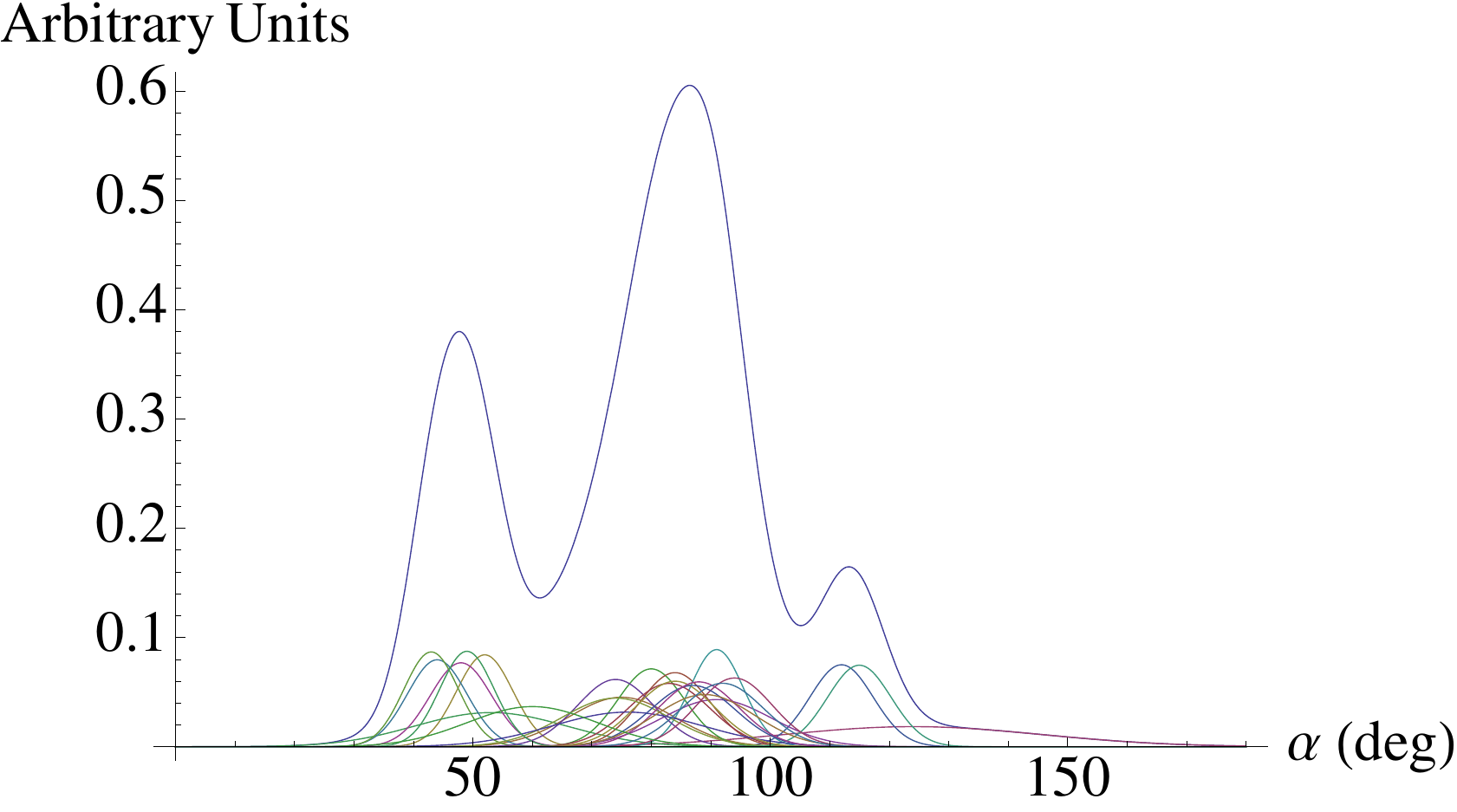}
\caption{(color online) Plot of the sum of 25 normalized Gaussians with means and widths determined by the peak positions and mean errors extracted from the 25 $\alpha$ scans of parameterized MC generated with $\alpha=89^\circ$. Also plotted are the individual Gaussians that contribute to the total PDF.}
\label{fig:gaussianRobustness}
\end{center}
\end{figure}

\begin{figure}[!htb]
\begin{center}
\includegraphics[width=0.48\textwidth]{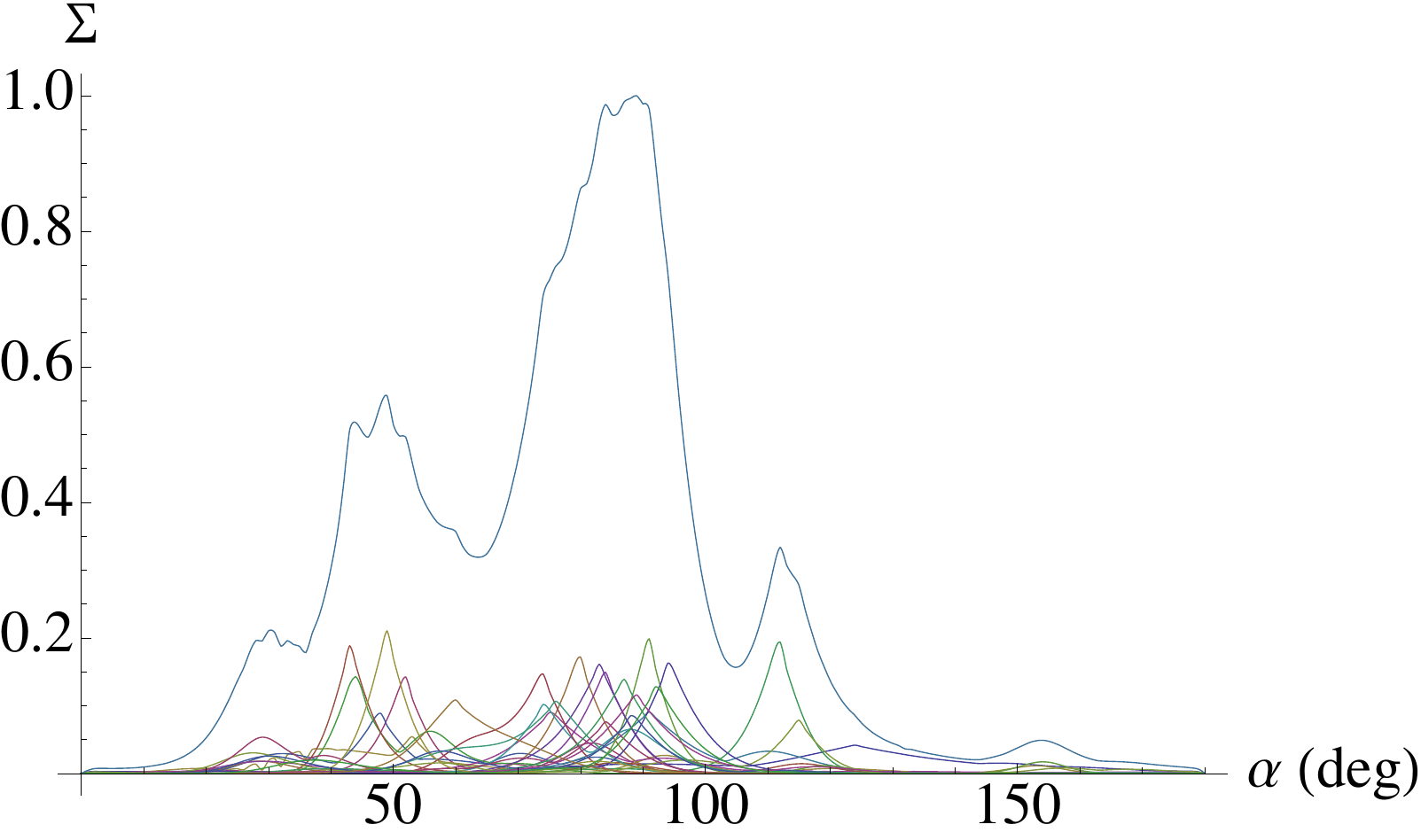}
\caption{(color online) Plot of the total $\Sigma$ distribution obtained by summing all 25 $\Sigma$ $\alpha$ scans of parameterized MC generated with $\alpha=89^\circ$ after normalizing their areas to 1. The total distribution is scaled so that it peaks at 1. Also plotted are the individual scaled $\Sigma$ scans that contribute to the total distribution.}
\label{fig:AprilRobustnessScans}
\end{center}
\end{figure}

The results of the robustness scans of $\alpha$ are provided in Table~\ref{tab:robustness}, sorted by the absolute difference between the extracted $\alpha$-scan peak position and the generated value of $89^\circ$. The first column lists the position of the most favored value of $\alpha$ in the scan. The second and third columns list the upper and lower errors, respectively, which are calculated as the number of degrees to either side of the $\alpha$ scan peak position at which the $\Sigma$ value drops to 0.32. The fourth column lists the mean of the upper and lower errors while the fifth column lists the value of the minimum $\chi^2$ obtained at the $\alpha$ scan peak position. The second to last column gives a measure of the consistency between the likelihoods for the peak $\alpha$ position and the generated $\alpha$ position based on the change in $\chi^2$, and the last column is the distance in $\sigma$ between the generated and extracted $\alpha$ peak positions (where the upper or lower error is used as appropriate). For 17 of the 25 scans, the extracted value of $\alpha$ lies within $3\sigma$ of the generated value. Examining the individual $\alpha$ scans reveals three distinct solutions for $\alpha$ that tend to be favored (including the generated value of $89^\circ$) and each scan tends to include at least one secondary peak in addition to the primary peak. Figure~\ref{fig:gaussianRobustness} illustrates the three solutions for alpha by providing the sum of 25 normalized Gaussians with means and widths determined by the peak positions and symmetric errors extracted from the 25 $\alpha$ scans. Also plotted are the individual Gaussians that contribute to the total PDF. Because the errors are not truly Gaussian, Figure~\ref{fig:gaussianRobustness} provides an incomplete picture of the scan results. A better illustration is provided by Figure~\ref{fig:AprilRobustnessScans}, which displays the total $\Sigma$ distribution obtained by summing all 25 $\Sigma$ $\alpha$ scans after normalizing each to the same area. The total distribution is scaled so that it peaks at 1. Also plotted are the individual scaled $\Sigma$ scans that contribute to the total distribution. The final PDF closely resembles that obtained by naively summing Gaussian distributions, though it exhibits more fine features. Again, the distribution indicates three distinct solutions for $\alpha$, with the generated value of $89^\circ$ being favored. At the $1\sigma$ level ($\Sigma$=0.32), the total scan distribution allows both the central and left peak. The presence of these secondary solutions indicates that with the current signal sample size and background levels, there is still a significant possibility that the favored value of $\alpha$ in a particular scan will correspond to a secondary solution. 

\end{document}

%% file: authors_sep2012.tex
%
\author{J.~P.~Lees}
\author{V.~Poireau}
\author{V.~Tisserand}
\affiliation{Laboratoire d'Annecy-le-Vieux de Physique des Particules (LAPP), Universit\'e de Savoie, CNRS/IN2P3,  F-74941 Annecy-Le-Vieux, France}
\author{E.~Grauges}
\affiliation{Universitat de Barcelona, Facultat de Fisica, Departament ECM, E-08028 Barcelona, Spain }
\author{A.~Palano$^{ab}$ }
\affiliation{INFN Sezione di Bari$^{a}$; Dipartimento di Fisica, Universit\`a di Bari$^{b}$, I-70126 Bari, Italy }
\author{G.~Eigen}
\author{B.~Stugu}
\affiliation{University of Bergen, Institute of Physics, N-5007 Bergen, Norway }
\author{D.~N.~Brown}
\author{L.~T.~Kerth}
\author{Yu.~G.~Kolomensky}
\author{G.~Lynch}
\affiliation{Lawrence Berkeley National Laboratory and University of California, Berkeley, California 94720, USA }
% ...
\collaboration{The \babar\ Collaboration}
\noaffiliation

%% file: abstract.tex
\def\CP{\ensuremath{C\!P}\xspace}
\def\Babar{\slshape B\kern-0.1em{\footnotesize A}\kern-0.1em B\kern-0.10em{\footnotesize A\kern-0.20em R}}

\begin{abstract} 
\noindent
We present results for a time-dependent Dalitz plot measurement of $\CP$-violating asymmetries in the mode $B^0 \rightarrow \pi^+ \pi^- \pi^0$. The dataset is derived from the complete sample of $471 \times 10^6$ $B{\overline B}$ meson pairs collected with the {\Babar} detector at the PEP-II asymmetric-energy $e^+e^-$ collider at the SLAC National Accelerator Laboratory operating on the $\Upsilon(4S)$ resonance. We extract parameters describing the time-dependent $B^0\rightarrow\rho\pi$ decay probabilities and $\CP$ asymmetries, including ${\mathcal C} = 0.016\pm0.059\pm0.036$, $\Delta{\mathcal C} = 0.234\pm0.061\pm0.048$, ${\mathcal S} = 0.053\pm0.081\pm0.034$, and $\Delta{\mathcal S} = 0.054\pm0.082\pm0.039$, where the uncertainties are statistical and systematic, respectively. We perform a two-dimensional likelihood scan of the direct $\CP$-violation asymmetry parameters for $B^0\rightarrow\rho^\pm\pi^\mp$ decays, finding the change in $\chi^2$ between the minimum and the origin (corresponding to no direct $\CP$ violation) to be $\Delta \chi^2=6.42$. We present information on the $\CP$-violating parameter $\alpha$ in a likelihood scan that incorporates $B^\pm\rightarrow\rho\pi$ measurements.

%% file: paper.bbl
\begin{thebibliography}{99}
\bibitem{ref:conj}
Throughout this paper, whenever a mode is given, the charge conjugate is also implied unless indicated otherwise.
\end{thebibliography}
